\documentclass[aps,prc,twocolumn,showpacs,preprintnumbers,
               nofootinbib,float,superscriptaddress]{revtex4-1}

\usepackage{epsfig}
\usepackage{color}
\usepackage[dvipsnames]{xcolor}
\usepackage{float,amsmath,amssymb}
\usepackage{graphicx}
\usepackage{url}
\usepackage{bbold}
\usepackage[utf8]{inputenc}
\newcommand{\rt}{{\mathbf{r}_\perp}}

\newcommand{\xt}{{\mathbf{x}_\perp}}

\newcommand{\bt}{{\mathbf{b}_\perp}}

\usepackage{hyperref}

\begin{document}

\title{Running the gamut of high energy nuclear collisions}

\author{Bj\"orn Schenke}
\affiliation{Physics Department, Brookhaven National Laboratory, Upton, NY 11973, USA}

\author{Chun Shen}
\affiliation{Department of Physics and Astronomy, Wayne State University, Detroit, Michigan 48201, USA}
\affiliation{RIKEN BNL Research Center, Brookhaven National Laboratory, Upton, NY 11973, USA}

\author{Prithwish Tribedy}
\affiliation{Physics Department, Brookhaven National Laboratory, Upton, NY 11973, USA}

\begin{abstract}
We present calculations of bulk properties and multiparticle correlations in a large variety of collision systems within a hybrid formalism consisting of IP-Glasma initial conditions, \textsc{Music} viscous relativistic hydrodynamics, and UrQMD microscopic hadronic transport. In particular, we study heavy ion collisions at the Large Hadron Collider (LHC), including Pb+Pb, Xe+Xe, and O+O collisions, and Au+Au, U+U, Ru+Ru, Zr+Zr, and O+O collisions at the Relativistic Heavy Ion Collider (RHIC). We further study asymmetric systems, including p+Au, d+Au, $^3$He+Au, and p+Pb collisions at various energies as well as p+p collisions at 0.5 and 13 TeV. We describe experimental observables in all heavy ion systems well with one fixed set of parameters, validating the energy and system dependence of the framework. Many observables in the smaller systems are also well described, although they test the limits of the model. Calculations of O+O collisions provide predictions for potential future runs at RHIC and LHC. 
\end{abstract}

\maketitle


\section{Introduction}
The application of relativistic hydrodynamics to the description of particle production in heavy ion collisions has a long history of success \cite{Heinz:2013th,Gale:2013da}. Especially since the inclusion of event-by-event fluctuations of the initial conditions \cite{Socolowski:2004hw,Andrade:2006yh,Broniowski:2007ft,Takahashi:2009na,Andrade:2009em,Hirano:2009bd,Alver:2010gr,Werner:2010aa,Holopainen:2010gz,Alver:2010dn,Petersen:2010cw,Schenke:2010rr,Gale:2012rq} a wide range of observables, including anisotropic flow harmonics, measured in heavy ion collisions performed at the top energy of the Relativistic Heavy Ion Collider (RHIC) and at the Large Hadron Collider (LHC) could be well described. This, in combination with observations of jet quenching and developments in computing the equation of state of nuclear matter from first principles \cite{Philipsen:2012nu,Borsanyi:2013bia,Bazavov:2014pvz}, has established that the quark gluon plasma (QGP), a novel state of matter, is being formed in high energy nuclear collisions, and that it behaves like an almost perfect fluid.
Further improvement to the hydrodynamic framework can be achieved by describing the low temperature regime of the fireball using microscopic hadronic cascades (see \cite{Petersen:2014yqa} for a review). This is particularly important for observables involving identified hadrons, such as the anisotropic flow of protons and anti-protons.

The state-of-the-art for comprehensive simulations of heavy ion collisions now involves (i) fluctuating initial states that either simply model the geometry in the transverse plane or ideally are derived from first principles calculations, (ii) second order viscous relativistic hydrodynamics including temperature dependent expressions for shear ($\eta/s$) and bulk ($\zeta/s$) viscous coefficients, and (iii) a microscopic hadronic cascade stage using UrQMD \cite{Bass:1998ca,Bleicher:1999xi}, JAM \cite{Hirano:2005xf,Hirano:2007ei}, or SMASH \cite{Weil:2016zrk}.

In this work we apply the hybrid framework, consisting of the color glass condensate (CGC) \cite{McLerran:1994ni,McLerran:1994ka,Iancu:2003xm} based IP-Glasma initial state model \cite{Schenke:2012wb,Schenke:2012hg}, the viscous relativistic hydrodynamic simulation \textsc{Music} \cite{Schenke:2010nt,Schenke:2010rr,Schenke:2011bn}, and UrQMD microscopic transport, to the study of collisions of different heavy ions (Ru+Ru, Zr+Zr, Xe+Xe, Au+Au, Pb+Pb, U+U) at center of mass energies $\sqrt{s_{\rm NN}} = 200\,{\rm GeV}$ and above, collisions of light ions with heavy ions, such as p+Au, d+Au, $^3$He+Au, and p+Pb, as well as p+p and O+O collisions. 

We demonstrate that a wide range of observables in all systems larger than p+p, and most successfully in the heavy ion systems, are well described using only one fixed set of parameters, which is determined in 200 GeV Au+Au collisions. We only change the collision systems and center of mass energies to match the corresponding experiment.

The main purpose of this paper is to present all details of the model framework and establish its usefulness in producing predictions for the bulk observables in heavy ion collisions and smaller systems, as well as to demonstrate where the model begins to fail. This establishes a baseline for detailed future investigations of more complex observables and a wide variety of physics questions, including understanding of the phase structure of quantum chromodynamics (QCD), the values of QCD transport coefficients in the QGP and hadron resonance gas, thermalization, jet quenching, production of electromagnetic probes and heavy flavors, the chiral magnetic effect, and more. 

We further present predictions for systems for which experimental data is not yet available or data has not been taken yet (500 GeV p+p and 200 GeV O+O, Ru+Ru, and Zr+Zr, and 5.02 TeV O+O collisions). 

This paper is organized as follows. In Section \ref{sec:framework} we present the employed framework, discussing the relevant physics and providing all details necessary to reproduce our results using the publicly available software. In Section \ref{sec:mult} we present results for the multiplicity distributions of charged hadrons and identified particles. In Section \ref{sec:pt} we show the mean transverse momentum of charged hadrons and identified particles, and in Section \ref{sec:vn} various observables measuring momentum anisotropies, both integrated over and differential in transverse momentum.
We close with conclusions in Section \ref{sec:conclusions}. We further provide Appendix A to lay out numerical treatments when viscous corrections are large in the hydrodynamic simulations, Appendix \ref{sec:centrality} to discuss how the centrality selection in this work differs from our previous studies, Appendix \ref{sec:flowAnalysis} to detail how flow cumulants are computed, and Appendix \ref{sec:deltaf} to discuss the effect of off-equilibrium corrections to the thermal distribution functions on the switching surface from hydrodynamics to microscopic transport.

\section{Framework}\label{sec:framework}
All components of the employed hybrid framework are publicly available \cite{iEBE-MUSIC, ipglasma, music, iSS, urqmd, afterburner_toolkit}. In the following we give a detailed description of the framework, specifying all parameters and settings necessary to reproduce the results presented in this work.

\subsection{IP-Glasma - Initial state}
The impact parameter dependent Glasma (IP-Glasma) initial state model \cite{Schenke:2012wb,Schenke:2012hg} is based on the color glass condensate effective theory \cite{McLerran:1994ni,McLerran:1994ka,Iancu:2003xm} and uses the classical description of gluon production first introduced in \cite{Kovner:1995ja,Kovchegov:1997ke}, employing numerical methods pioneered in \cite{Krasnitz:1998ns,Krasnitz:1999wc,Krasnitz:2000gz,Lappi:2003bi}.
The impact parameter dependence is derived from the impact parameter dependent dipole saturation model (IPSat) \cite{Bartels:2002cj,Kowalski:2003hm}, which was originally developed to describe deeply inelastic scattering experiments of electrons on protons at HERA. In particular the application to diffractive vector meson production in e+p collisions allowed for constraints on the spatial gluon distribution. 

The IPSat model is an extension of the Golec-Biernat– W\"usthoff dipole model \cite{GolecBiernat:1998js,Golec-Biernat:1999qd}, giving the correct perturbative result in the limit that the dipole size $|\rt|$ goes to zero.  The model parametrizes the dipole-proton scattering cross section as a function of Bjorken $x$, dipole separation $\rt$, and impact parameter $\bt$:
\begin{equation}\label{eq:dipxsec}
    \frac{d\sigma^p_{\rm dip}}{d^2\bt}(x,\rt,\bt) = 2[1-\exp(-F(x,\rt,\bt))]\,,
\end{equation}
with the function 
\begin{equation}
    F(x,\rt,\bt) = \frac{\pi^2}{2N_c}\mathbf{r}_\perp^2 \alpha_s(\tilde{\mu}) x g(x,\tilde{\mu}^2) T_p(\bt)\,,
\end{equation}
with $N_c=3$, where the scale 
\begin{equation}
    \tilde{\mu}^2 = \frac{4}{\mathbf{r}_\perp^2}+\tilde{\mu}_0^2\,,
\end{equation}
and the running coupling is given by the leading order expression
\begin{equation}
    \alpha_s(\tilde{\mu}) = \frac{2\pi}{(11-2N_f/3)\ln(\tilde{\mu}/\Lambda_{\rm QCD})}\,,
\end{equation}
with $N_f=3$ the number of flavors\footnote{Here, we do not vary $N_f$ as a function of the energy scale, which can be done \cite{Mantysaari:2018nng}.}, and $\Lambda_{\rm QCD}$ the scale of quantum chromo dynamics (QCD).
$xg(x,\tilde{\mu}^2)$ is the gluon density for a given value of $x$, evolved from $\tilde{\mu}_0^2=1.51\,{\rm GeV}^2$ \cite{Rezaeian:2012ji} to the scale $\tilde{\mu}^2$ using leading order DGLAP evolution without quarks. The initial condition for this evolution is given by
\begin{equation}
    x g(x,\tilde{\mu}_0^2) = A_g x^{-\lambda_g}(1-x)^{5.6}\,,
\end{equation}
where $A_g=2.308$ and $\lambda_g=0.058$ are parameters determined from fits to HERA data \cite{Rezaeian:2012ji}.
Finally, a component of great importance for our purposes is the spatial dependence introduced via the proton thickness function $T_p(\bt)$.
In this work we use sub-nucleonic fluctuations, introducing three hot-spots per nucleon, such that
\begin{equation}\label{eq:Tp}
     T_p(\bt) = \sum_{i=1}^{3} T_q(\bt-\mathbf{b}_\perp^i)\,,
\end{equation}
where each hot-spot is parametrized by a Gaussian in the transverse plane
\begin{equation}\label{eq:2DGauss}
     T_q(\bt) = \frac{1}{2\pi B_{q}} e^{-\mathbf{b}_\perp^2/2B_q}\,,
\end{equation}
with $B_q=0.3\,{\rm GeV}^{-2}$, constrained in \cite{Mantysaari:2016ykx}. Each hot-spot center $\mathbf{b}_\perp^i$ is also sampled from a 2D-Gaussian distribution of the same form as \eqref{eq:2DGauss}, with width parameter $B_{qc}=4\,{\rm GeV}^{-2}$ \cite{Mantysaari:2016ykx}.\footnote{In this work we shift the center of the nucleon back to $\mathbf{b}_\perp^i$ after sampling the hot spots. The values quoted for $B_{qc}$ and $B_{q}$ are determined in \cite{Mantysaari:2016ykx}, where the nucleon is also recentered.}

Further fluctuations of the normalization of each $T_q$ improve agreement with experimentally measured multiplicity distributions \cite{McLerran:2015qxa} as well as exclusive vector meson production data in deep inelastic scattering \cite{Mantysaari:2016ykx}. We thus follow  \cite{McLerran:2015qxa} by introducing fluctuations of the logarithm of $Q_s^2$, or, in practice, the logarithm of the normalization of $T_q$, and sample from the log-normal distribution
\begin{equation}\label{eq:Qsfluctuations}
    P\Big(\ln\Big(\frac{T_q(\bt)}{\langle T_q(\bt)\rangle}\Big)\Big)=\frac{1}{\sqrt{2\pi}\sigma} e^{\big(-\frac{\ln^2(T_q(\bt)/\langle T_q(\bt)\rangle)}{2\sigma^2}\big)}
\end{equation}
where $\langle T_q(\bt)\rangle$ is given by Eq.\,\eqref{eq:2DGauss}. We then make sure that the original average $\langle T_q \rangle$ is recovered, by dividing $T_q$ by $\exp(\sigma^2/2)$. In this work we use $\sigma = 0.6$.

In \cite{Schenke:2013dpa} the parameter $r_{\rm max}$ was introduced to set the thickness function of a nucleus $T_A(\bt)$, whose construction is discussed in more detail below, to zero once it reaches a value as small as that of a nucleon a distance $r_{\rm max}$ from its center. Here we use a very large $r_{\rm max}=10\,{\rm fm}$, such that its effect is negligible.

From the dipole amplitude $\mathcal{N}(x,\rt,\bt) = ( d\sigma^p_{\rm dip}/d^2\bt ) (x,\rt,\bt)/2$ given by Eq.\,\eqref{eq:dipxsec}, we can extract a saturation scale $Q_s(x)$ by using the definition that $Q_s^2=2/R_s^2$, with $R_s$ defined via $\mathcal{N}(x,R_s)=1-\exp(-1/2)$. Note that $\mathcal{N}$ and $Q_s$ also depend on the thickness function $T_A$, and in practice we tabulate $Q_s(x,T_A)$ for use in the following.

Having determined the $x$ and thickness function dependence of $Q_s$ we can now construct the initial color charge distributions of the nuclei we are about to collide. For proton projectiles we simply use the description above, generating $T_p$ according to Eq.~\eqref{eq:Tp}. For large nuclei, we first sample nucleon positions from a Woods-Saxon distribution 
\begin{equation}\label{eq:WS}
    \rho(r,\theta) = \frac{\rho_0}{1+\exp[(r-R'(\theta))/a]}\,,
\end{equation}
with $R'(\theta)=R[1+\beta_2 Y_2^0(\theta)+\beta_4 Y_4^0(\theta)]$, and $\rho_0$ the nuclear density at the center of the nucleus. $R$ is the radius parameter, $a$ the skin depth. The spherical harmonic functions $Y_l^m(\theta)$ and the parameters $\beta_2$ and $\beta_4$ account for the possible deformation from a spherical shape. Parameters for all nuclei for which we employ the Woods-Saxon form are given in Table \ref{tab:WS}. We further impose a minimal distance of $d_{\rm min}=0.9\,{\rm fm}$ between nucleons when sampling in three dimensions. When a nucleon is added and violates the minimum distance criterion with one or more already sampled nucleons, we resample its angular coordinates (not the radial position). For the deformed nuclei, we only resample the azimuthal angle $\phi$ to keep the distributions of radial distances and polar angles unchanged \cite{Moreland:2014oya}.

\begin{table}[hb]
\begin{center}  
    \begin{tabular}{ | c | c | c | c | c |}
    \hline  
    Nucleus & $R$ [fm] & $a$ [fm] & $\beta_2$ & $\beta_4$ \\ \hline
    $^{238}$U  & 6.81  & 0.55  & 0.28  & 0.093 \\ \hline
    $^{208}$Pb & 6.62  & 0.546 & 0 & 0 \\ \hline
    $^{197}$Au & 6.37  & 0.535 & -0.13 & -0.03 \\ \hline
    $^{129}$Xe & 5.42 & 0.57  & 0.162 & -0.003 \\ \hline 
    $^{96}$Ru & 5.085 & 0.46  & 0.158 & 0     \\ \hline 
    $^{96}$Zr & 5.02  & 0.46  & 0     & 0     \\ 
    \hline
    \end{tabular}
\end{center}
\caption{Parameter values used for nuclei described with the Woods-Saxon parametrization \eqref{eq:WS} \cite{Filip:2009zz,Masui:2009qk,Hirano:2012kj,Shen:2014vra,Schenke:2014tga,Pritychenko:2013gwa,Goldschmidt:2015qya,Goldschmidt:2015kpa,Moller:2015fba}. \label{tab:WS}}
\end{table}

Smaller nuclei, such as $^{16}$O, and $^3$He are described using a variational Monte-Carlo method (VMC) using the Argonne v18 (AV18) two-nucleon potential +UIX interactions \cite{Carlson:1997qn}. In practice we use the $^3$He and $^{16}$O configurations available in the PHOBOS Monte-Carlo Glauber distribution \cite{Loizides:2014vua,tglauber}.

For the results we will show involving the deuteron, we employ a simple Hulthen wave function of the form \cite{Miller:2007ri}
\begin{equation}
    \phi(d_{\rm pn}) =\frac{\sqrt{a_H b_H(a_H+b_H)}}{b_H-a_H}\frac{e^{-a_H d_{\rm pn}}-e^{-b_H d_{\rm pn}}}{\sqrt{2\pi} d_{\rm pn}}\,,
\end{equation}
where $d_{\rm pn}$ is the separation between the proton and the neutron, and the parameters are experimentally determined to be $a_H=0.228\,{\rm fm}^{-1}$ and $b_H=1.18\,{\rm fm}^{-1}$.

Once all nucleon centers are sampled, the nucleon substructure is sampled as in the case of the proton, then $T_A(\bt)$ is determined by summing all nucleons' $T_p(\bt)$. With that information, we can self-consistently determine $Q_s(x,\bt)$, using an iterative procedure, where $x=x(\bt)=Q_s(x,\bt)/\sqrt{s_{\rm NN}}$, with $\sqrt{s_{\rm NN}}$ the center of mass energy of the collision. Note that we are interested in mid-rapidity observables, so we assume rapidity $y=0$ for the determination of $x$. The results will be boost-invariant. For ways to go beyond boost-invariance in the IP-Glasma framework see \cite{Schenke:2016ksl,McDonald:2018wql}.

Color charges can now be sampled using the assumption of local Gaussian correlations as in the McLerran-Venugopalan (MV) model \cite{McLerran:1994ka,McLerran:1994vd}
\begin{equation}
\langle \rho_i^a(\bt) \rho_i^b(\xt) \rangle = g^2 \mu_i^2 (x,\bt) \delta^{ab} \delta^{(2)}(\bt-\xt)\,,
\end{equation}
where $g^2\mu_i(x,\bt) = c Q_s^i(x,\bt)$, with a proportionality constant $c$ that can be determined numerically. Here we use $c=1.25$, which is close to the value determined in \cite{Lappi:2007ku}. 
The coupling $g$ scales out of the classical calculation, and we will set it to 1 in all calculations. Its actual value will only affect the normalizations (e.g. of produced gluon number or energy density) and will be absorbed in a universal normalization constant for the energy momentum tensor $T^{\mu\nu}$ in the end.

The index $i$ labels the nucleus, taking values $P$ or $T$ for projectile or target, respectively.
From the color charges in both projectile and target we can construct the color currents, generated by the static charges moving with the speed of light in the positive or negative $z$-direction, respectively. Using the gauge where $A^-$ or $A^+$ are zero, respectively, the currents can be written as
\begin{equation}\label{eq:currents}
    J^\nu_i = \delta^{\nu\pm} \rho_i(x^{\mp},\xt) = \delta^{\nu\pm} \rho^a_i (x^{\mp},\xt)t^a \,,
\end{equation}
where $t^a$ are SU(3) generators in the fundamental representation, with $a$ running from 1 to 8. Upper signs refer to the right moving nucleus, lower signs to the left moving one. Here, we work in lightcone coordinates, where the components are $(+,-,1,2)$ with 1,2 labeling the transverse coordinates. Lightcone coordinates are defined as $v^\pm = (v^0 \pm v^3 )/\sqrt{2}$. Later we switch to Milne coordinates $(\tau, \eta, x^1, x^2)$, where proper time can be expressed using $x^+$ and $x^-$ as $\tau=\sqrt{2 x^+ x^-}$, and the spatial rapidity coordinate reads $\eta = 0.5 \ln(x^+/x^-)$. The metric in Milne coordinates reads $g_{\mu\nu} = {\rm diag}(1,-\tau^2,-1,-1)$. 

The currents \eqref{eq:currents} form the sources in the Yang-Mills equation (omitting nucleus labels)
\begin{equation}\label{eq:YM} 
    [D_\mu,F^{\mu\nu}] = J^\nu\,,
\end{equation}
where $D_\mu=\partial_\mu - igA_\mu$, and $F^{\mu\nu} = \frac{i}{g}[D^\mu,D^\nu] = \partial^\mu A^\nu - \partial^\nu A^\mu - ig[A^\mu,A^\nu]$ is the field strength tensor, with the gluon fields $A^\mu=A^\mu_a t^a$.

Eq.\,\eqref{eq:YM} can be solved in Lorentz gauge, where $\partial_\mu A^\mu = 0$, resulting in 
\begin{equation}\label{eq:Apm}
    A^\pm_i = -\frac{\rho_i(x^\mp,\xt)}{\boldsymbol{\nabla}_\perp^2-m^2}\,,
\end{equation}
where we introduced the infrared regulator $m$ in the denominator, which incorporates confinement effects and tames otherwise appearing Coulomb tails.\footnote{Note that the sign of $m^2$ in \eqref{eq:Apm} was misquoted in \cite{Schenke:2012wb,Schenke:2012hg,Schenke:2013dpa}.}  We choose $m$ to be close to the QCD scale $\Lambda_{\rm QCD}$  and in this work use $m=0.2\,{\rm GeV}$.\footnote{Note that the values for $B_q$ and $B_{qc}$ in \cite{Mantysaari:2016ykx} were determined for $m=0.4\,{\rm GeV}$. Here, we have checked explicitly that the combination of $m=0.2\,{\rm GeV}$ and the $B_q$ and $B_{qc}$ quoted above, along with the additional normalization fluctuations \eqref{eq:Qsfluctuations}, also provide a good fit to diffractive vector meson production data from HERA.} Alternatively, color neutrality on the nucleon size scale can be modeled explicitly \cite{Krasnitz:2002mn}, however, introducing $m$ is simpler and leads to very similar results.

In order to find a solution for the gluon fields after the collision, it will be convenient to convert \eqref{eq:Apm} to lightcone gauge, where we set $A^+$ or $A^-$ to zero, depending on the direction the nucleus is moving. In that gauge, the other components of the gluon field are $A^-=0$ ($A^+=0$), and the sum of the left and right moving nucleus leads to \cite{McLerran:1994ni,McLerran:1994ka,McLerran:1994vd,Kovner:1995ts,Kovner:1995ja,Kovchegov:1996ty,Jalilian-Marian:1997xn}
\begin{align}\label{eq:Aj}
    A^j(\xt) = &\theta(x^-)\theta(-x^+)\frac{i}{g} V_P(\xt)\partial^j V_P^\dag(\xt) \notag \\
    &+\theta(x^+)\theta(-x^-)\frac{i}{g} V_T(\xt)\partial^j V_T^\dag(\xt)\,,
\end{align}
where the path-ordered exponentials, or Wilson lines, are given by
\begin{equation}\label{eq:Wilson}
    V_i(\xt) = P \exp\Bigg(-ig \int dx^\mp \frac{\rho_i(x^\mp,\xt)}{\boldsymbol{\nabla}_\perp^2-m^2}\Bigg)\,,
\end{equation}
where again upper symbols are for right moving ($i=P$) and lower symbols for left moving ($i=T$) nuclei.

The gauge choice discussed above, where $A^+=0$ for one and $A^-=0$ for the other nucleus, can be summarized as the Fock-Schwinger gauge condition $A^\tau = (x^+ A^- + x^- A^+) = 0$. In this gauge the solution for the gauge fields in the forward light cone (after the collision) at time $\tau=0^+$ can be derived by requiring that $[D_\mu,F^{\mu j}]=0$ and $[D_\mu,F^{\mu \pm}]=J^\pm$ have no singular terms as $\tau\rightarrow 0$. The solution at time $\tau=0^+$ has the simple form \cite{Kovner:1995ts,Kovner:1995ja}
\begin{align}
    A^j &= A^j_P + A^j_T \,, \\
    A^\eta &= -\frac{ig}{2} \Big[ A_{Pj} , A^j_T \Big]\,,\\
    \partial_\tau A^j &= 0\,,\\
    \partial_\tau A^\eta &= 0\,.
\end{align}

We need to solve for the fields in the forward lightcone numerically, which is best done using the Wilson lines directly. First, the path-ordered exponential can be evaluated numerically by discretizing the longitudinal direction in $N_y$ slices, sampling color charges in each slice and evaluating \cite{Lappi:2007ku}
\begin{equation}
    V_i(\xt) = \prod_{k=1}^{N_y} \exp \Big( -ig \frac{\rho_i^k(\xt)}{\boldsymbol{\nabla}^2-m^2}\Big)\,,
\end{equation}
where the $\rho_i^k$ satisfy
\begin{equation}
\langle \rho_{i,k}^a(\bt) \rho_{i,l}^b(\xt) \rangle = \frac{g^2 \mu_i^2 (x,\bt)}{N_y} \delta^{ab}\delta^{kl} \delta^{(2)}(\bt-\xt)\,.
\end{equation}

For each nucleus $i$, we can then assign to each lattice site $s$ the pure gauge configuration \eqref{eq:Aj} expressed using the corresponding Wilson line
\begin{equation}
    U_i^j(s) = V_i(s) V^\dag_i(s+\hat{e}_j)\,, 
\end{equation}
where $\hat{e}_j$ is a shift by one lattice site in the $j \in (1,2)$ direction.

To obtain the Wilson lines $U^j(s)$ at site $s$ after the collision, we need to solve \cite{Krasnitz:1998ns}
\begin{align}\label{eq:initU}
    {\rm Tr} \Big\{ t^a \Big[ & \Big( U_P^j(s)+U_T^j(s) \Big) (1+U^{j\dag}(s)) \notag \\ 
        &- (1+U^j(s))\Big( U_P^{j\dag}(s) + U_T^{j\dag}(s)\Big) \Big] \Big\} = 0\,,
\end{align}
which is a set of $N_c^2-1$ equations that we solve iteratively for $N_c=3$.

Because we are assuming boost-invariance, the  rapidity component of the gluon field becomes an adjoint representation scalar $\phi = A_\eta = -\tau^2 A^\eta$.
The longitudinal electric field is then given by $\pi = E^\eta = \dot{\phi}/\tau$.\footnote{Please note that $E_\eta$ in \cite{Schenke:2012hg} should be $E^\eta$.}
Its initial condition can be written in terms of the solution to Eq.\,\eqref{eq:initU}:
\begin{align}\label{eq:Einit}
    E^\eta&(s) = \notag\\
        &\frac{i}{4g} \sum_{j=1,2} \Big[\Big(U_P^j(s)-U_T^j(s) \Big)(U^{j\dag}(s)-1) \notag\\ 
        & ~~~~~~~~~~~ - {\rm h.c} - (U^{j\dag}(s-\hat{e}_j)-1)\notag\\
        & ~~~~~~~~~~~\times
        \Big( U_P^j(s-\hat{e}_j)-U_T^j(s-\hat{e}_j) \Big)\notag\\
         & ~~~~~~~~~~~+{\rm h.c.}\Big]\,,
\end{align}
where the sum is over the transverse directions $j\in(1,2)$.

Besides the initial condition for $U^j$ determined in Eq.\,\eqref{eq:initU} and $\pi=E^\eta$ from \eqref{eq:Einit}, we have $E^j=0$ and $\phi=A_\eta=0$.

The equations of motion for the fields on the lattice can be obtained from the lattice Hamiltonian
\begin{align}\label{eq:Ham}
    aH = \sum_s \Big[ &\frac{g^2 a}{\tau} E^j(s)E^j(s) + \frac{2\tau}{g^2a} (N_c-{\rm Re} {\rm Tr} U_{(1,2)}(s)) \notag\\
    & +\frac{\tau}{a} {\rm Tr} \pi^2(s) +\frac{a}{\tau}\sum_j(\phi(s)-\tilde{\phi}_j(s))^2 \Big]\,,
\end{align}
where the sum is over all sites $s$, $a$ is the lattice spacing, and the parallel transported scalar field in cell $s$ is given by
\begin{equation}
    \tilde{\phi}_j(s) = U_j(s) \phi(s+\hat{e}_j) U_{j}^{\dag}(s)\,.
\end{equation}
The plaquette in the transverse plane is defined as
\begin{equation}
    U_{(1,2)}(s) = U_1(s) U_2(s+\hat{e}_1) U_1^{\dag}(s+\hat{e}_2)U_2^{\dag}(s)\,.
\end{equation}
We can identify the terms in \eqref{eq:Ham} as the energy in the transverse electric fields, the longitudinal magnetic fields, the longitudinal electric fields, and the transverse magnetic fields, respectively. 
At the initial time $\tau=0^+$ we only have longitudinal electric and magnetic contributions. 

The equations of motion are obtained by taking the Poisson bracket of the fields with the Hamiltonian and read
\begin{align}
    \dot{U}_j &= i \frac{g^2}{\tau}E^j U_j ~~({\rm no~sum~over~}j)\label{eq:eom}\\
    \dot{\phi} &= \tau \pi\\
    \dot{E}^1 &= \frac{i\tau}{2g^2} [U_{(1,2)} + 
    U_{(1,-2)}-U^\dag_{(1,2)}-U^\dag_{(1,-2)}-T_1]\notag\\
    & ~~~~~+\frac{i}{\tau}[\tilde{\phi}_1,\phi]\\
    \dot{E}^2 &= \frac{i\tau}{2g^2} [U_{(2,1)} + 
    U_{(2,-1)}-U^\dag_{(2,1)}-U^\dag_{(2,-1)}-T_2]\notag\\
    & ~~~~~+\frac{i}{\tau}[\tilde{\phi}_2,\phi]\\
    \dot{\pi} &= \frac{1}{\tau}\sum_j \big[ \tilde{\phi}_j + \tilde{\phi}_{-j} - 2\phi\big]\,,\label{eq:eom5}
\end{align}
where the subtraction of traces
$T_1=\pmb{\mathbb{1}}{\rm Tr} [U_{(1,2)} + U_{(1,-2)}-U^\dag_{(1,2)}-U^\dag_{(1,-2)}]/N_c$ and $T_2=\pmb{\mathbb{1}}{\rm Tr} [U_{(2,1)} + U_{(2,-1)}-U^\dag_{(2,1)}-U^\dag_{(2,-1)}]/N_c$ ensures that all components of $F^{\mu\nu}$ remain traceless. 

Eqs.\,\eqref{eq:eom}-\eqref{eq:eom5} can be solved using a leapfrog algorithm on a 2D lattice. For collisions of the larger nuclei we use a lattice size of $L=28\,{\rm fm}$ and a lattice spacing of $a=0.04\,{\rm fm}$. For O+O collisions we used $L=14\,{\rm fm}$, keeping the number of lattice sites the same. For p+p and p+A collisions we use $a=0.04\,{\rm fm}$, but a smaller $L=10\,{\rm fm}$. When initializing the hydrodynamic simulation, we transfer our results to a grid of size $512^2$ with length $L=34\,{\rm fm}$ for the larger A+A collisions, $L=28\,{\rm fm}$ for O+O, and $L=20\,{\rm fm}$ for the p+A and p+p collisions.

To transfer initial state information to the hydrodynamic simulation, we compute all components of the energy momentum tensor 
$T^{\mu\nu} = -g^{\mu\alpha}g^{\nu\beta} g^{\gamma\delta} F_{\alpha\gamma} F_{\beta\delta} +\frac{1}{4} g^{\mu\nu}g^{\alpha\gamma}g^{\beta\delta}
  F_{\alpha\beta}F_{\gamma\delta}$ 
of the Yang-Mills system at a given time $\tau_{\rm switch}$, in the following chosen to be $\tau_{\rm switch}=0.4\,{\rm fm}$. 

Naturally, $T^{\mu\nu}$ has to be evaluated on the lattice, involving the color electric field $E^j$, the scalar $\phi$, as well as the plaquettes representing the color magnetic fields. Here we show the $(\tau\tau)$ component of the lattice $T^{\mu\nu}$ as an example:
\begin{align}
    T^{\tau\tau}(s) &= \frac{g^2}{2\tau^2} {\rm Tr} [(E^1)^{2}(s)+(E^1)^{2}(s+\hat{e}_2)\notag\\
        &~~~~~~~~~~~~+(E^2)^{2}(s)+(E^2)^{2}(s+\hat{e}_1)]\notag\\
    &+\frac{1}{2\tau^2} {\rm Tr}\big[\big(\phi(s)-\tilde{\phi}_1(s)\big)^2\notag\\
        &~~~~~~~~~~~~+\big(\phi(s+\hat{e}_2)-\tilde{\phi}_1(s+\hat{e}_2)\big)^2\notag\\
        &~~~~~~~~~~~~+\big(\phi(s)-\tilde{\phi}_2(s)\big)^2\notag\\
        &~~~~~~~~~~~~+\big(\phi(s+\hat{e}_1)-\tilde{\phi}_2(s+\hat{e}_1)\big)^2\big]\notag\\
    &+\frac{1}{4} {\rm Tr} \big[\pi^2(s) + \pi^2(s+\hat{e}_1) \notag\\
        &~~~~~~~~~+ \pi^2(s+\hat{e}_2) + \pi^2(s+\hat{e}_1+\hat{e}_2) \big]\notag\\
    &+\frac{2}{g^2}\Big(N_c-{\rm Re}{\rm Tr}[U_{(1,2)}(s)]\Big)\,.
\end{align}
Note that with this definition $T^{\tau\tau}$ is technically defined in the center of the plaquette $U_{(1,2)}(s)$, but we still label it by site $s$.\footnote{This definition is different from \cite{Schenke:2012hg} but represents precisely what is used in the publicly available IP-Glasma code \cite{ipglasma}.}

\subsection{\textsc{Music} - Hydrodynamics}\label{sec:music}
The evolution of the energy momentum tensor continues under the assumption that the system remains near local thermal equilibrium, where an equation of state can be used to close the hydrodynamic equations given by
\begin{equation}\label{eq:hydro}
\partial_\mu T^{\mu\nu} = 0 \,.
\end{equation}
In the hydrodynamic phase, the energy momentum tensor can be decomposed as
\begin{equation}
T^{\mu\nu} = e u^\mu u^\nu - (P + \Pi) \Delta^{\mu\nu} + \pi^{\mu\nu}\,,
\end{equation}
with the energy density $e$, pressure $P$, flow velocity $u^\mu$, bulk viscous pressure $\Pi$, and shear viscous tensor $\pi^{\mu\nu}$. We defined $\Delta^{\mu\nu} = g^{\mu\nu}-u^\mu u^\nu$. 

As discussed in the previous section, the initial conditions are given by the energy momentum tensor of the Yang-Mills system at every position in the transverse plane. Let us call it $T_{\rm YM}^{\mu\nu}(\xt)$ in this section. 
As we have set the coupling constant $g=1$ above, we still need to apply an overall normalization factor, which mainly accounts for a more realistic strong coupling constant. In this work we multiply the initial $T^{\mu\nu}_{\rm YM}$ with a factor 0.235, which corresponds to $g \simeq 2.06$.
We extract the initial condition for the energy density and flow velocities using the relation $e u^\mu = T^{\mu\nu}_{\rm YM} u_\nu$. The initial value for the shear 
stress tensor then follows as \cite{Mantysaari:2017cni,Schenke:2019ruo,Mantysaari:2019hkq}
\begin{equation}
\pi^{\mu\nu}=T^{\mu\nu}_{\rm YM}-\frac{4}{3}e u^\mu u^\nu + \frac{e}{3}g^{\mu\nu}\,,
\end{equation}
where we used that the Yang-Mills system is conformal (has zero bulk viscosity) and obeys an ideal equation of state $P=e/3$.

In the hydrodynamic simulation we are going to use a more realistic QCD equation of state \cite{Moreland:2015dvc}, obtained from a combination of lattice QCD calculations \cite{Bazavov:2014pvz} and a hadron resonance gas model in the low temperature regime. This poses a problem as there will be a mismatch between the pressures on the Yang-Mills side $P_{\rm YM}$ and the one determined by the QCD equation of state on the hydrodynamic side, $P$, at the switching time $\tau_{\rm switch}$. There is no unique way to handle the mismatch as it is a result of missing physics in our description of how the initial gluon dominated non-equilibrium system evolves towards a thermalized quark gluon plasma. Here we adopt the same method as in \cite{Schenke:2019ruo}, and absorb the difference in pressures into an effective initial bulk viscous term $\Pi = e/3-P$, which ensures that we initialize with the exact $T^{\mu\nu}_{\rm YM}$ given by the IP-Glasma model. This way, the system should approach the QCD pressure on a time scale given by the bulk relaxation time.
We note that our auxiliary initial $\Pi$ is positive, leading to an additional outward push \cite{Schenke:2019ruo}, which is opposite to the usual behavior of a bulk viscous pressure.

To continue our discussion of the hydrodynamic evolution equations, we present the second order constitutive relations for the shear and bulk viscous parts. We use the expressions derived in \cite{Denicol:2012cn,Denicol:2014vaa}, given by
\begin{eqnarray}
	\tau_{\Pi} \dot{\Pi} + \Pi
	& = & - \zeta\, \theta
	- \delta_{\Pi \Pi} \Pi \, \theta
	+ \lambda_{\Pi \pi} \pi^{\mu\nu} \sigma_{\mu\nu}\label{eq:bulk} \\ 
	\tau_{\pi} \dot{\pi}^{\langle \mu\nu \rangle}
	+ \pi^{\mu\nu}
	& = & 2\eta\,\sigma^{\mu\nu}
	- \delta_{\pi \pi} \pi^{\mu\nu} \theta
	+ \varphi_7 \pi_{\alpha}^{\langle \mu} \pi^{\nu\rangle \alpha} \nonumber \\
	& & - \tau_{\pi \pi} \pi_{\alpha}^{\langle \mu} \sigma^{\nu\rangle \alpha}
	+ \lambda_{\pi \Pi} \Pi\, \sigma^{\mu\nu}\,,\label{eq:shear}
\end{eqnarray}
where $A^{\langle \cdot  \cdot \rangle}$ indicates symmetrized and traceless projections, $\theta = \nabla_{\mu} u^{\mu}$ is the expansion rate, and 
\begin{equation}
\sigma^{\mu\nu}
= \frac{1}{2}\left[ \nabla^{\mu} u^{\nu} + \nabla^{\nu} u^{\mu}
- \frac{2}{3} \Delta^{\mu\nu} (\nabla_{\alpha} u^{\alpha}) \right]
\end{equation}
the shear tensor, with $\nabla_{\mu} = ( g_{\mu\nu} - u_{\mu} u_{\nu} ) \partial^{\nu}$.

The first-order transport coefficients $\eta$ and $\zeta$ are the shear and bulk viscosities, respectively. They are determined by choosing the quantities $\eta/s$ and $\zeta/s$ such that the experimental data is best described.

The shear and bulk relaxation time, $\tau_{\pi}$ and $\tau_{\Pi}$ are then given by
\begin{eqnarray}
	\tau_{\pi} & = & \frac{5 \, \eta}{e + P } \, , \\
	\tau_{\Pi} & = & \frac{\zeta}{15 \,
		\left( \frac{1}{3} - c_s^2 \right)^2 ( e + P ) } \,,\label{eq:bulkrelax}
\end{eqnarray}
where $c_s$ is the speed of sound, which is determined by the equation of state. These choices for the relaxation times fulfill the linear/static causality conditions. \cite{Pu:2009fj, Huang:2010sa}.

These second-order transport coefficients appearing in Eqs.\,\eqref{eq:bulk} and \eqref{eq:shear} can be expressed in terms of the shear and bulk relaxation times and are given in Table \ref{tab:coeff}. The second order transport coefficients depend on the assumptions they are derived under and present a source of systematic uncertainty. We estimated this uncertainty by running simulations with all quantities listed in Table \ref{tab:coeff} set to zero. The change in multiplicities and mean transverse momenta was negligible. Flow harmonics were slightly reduced (typically by 5-10\% - maximally 15\% in 80-90\% central Au+Au collisions) when neglecting the second order transport coefficients.

\begin{table}
\begin{center}
    \begin{tabular}{ |@{\hspace{0.05cm}}@{}c@{}@{\hspace{0.05cm}}|@{\hspace{0.05cm}}@{}c@{}@{\hspace{0.05cm}}|@{\hspace{0.05cm}}@{}c@{}@{\hspace{0.05cm}}|@{\hspace{0.05cm}}@{}c@{}@{\hspace{0.05cm}}|@{\hspace{0.05cm}}@{}c@{}@{\hspace{0.05cm}}|@{\hspace{0.05cm}}@{}c@{}@{\hspace{0.05cm}}|}
    \hline
    $\tau_{\pi\pi} [\tau_\pi]$ & $\delta_{\pi\pi} [\tau_\pi]$ & $\varphi_7 P$ &
    $\lambda_{\pi\Pi} [\tau_\pi]$ & $\lambda_{\Pi\pi}[\tau_\Pi]$ & $\delta_{\Pi\Pi} [\tau_\Pi]$  \\ \hline
    10/7 & 4/3 & 9/70 & 6/5 & $8(1/3-c_s^2)/5$ & 2/3 \\
    \hline
    \end{tabular}
\caption{Transport coefficients appearing in the equations for the shear stress tensor $\pi^{\mu\nu}$ and the bulk viscous term $\Pi$. \label{tab:coeff}}
\end{center}
\end{table}

\begin{figure}[htb]
  \centering
  \includegraphics[width=0.48\textwidth]{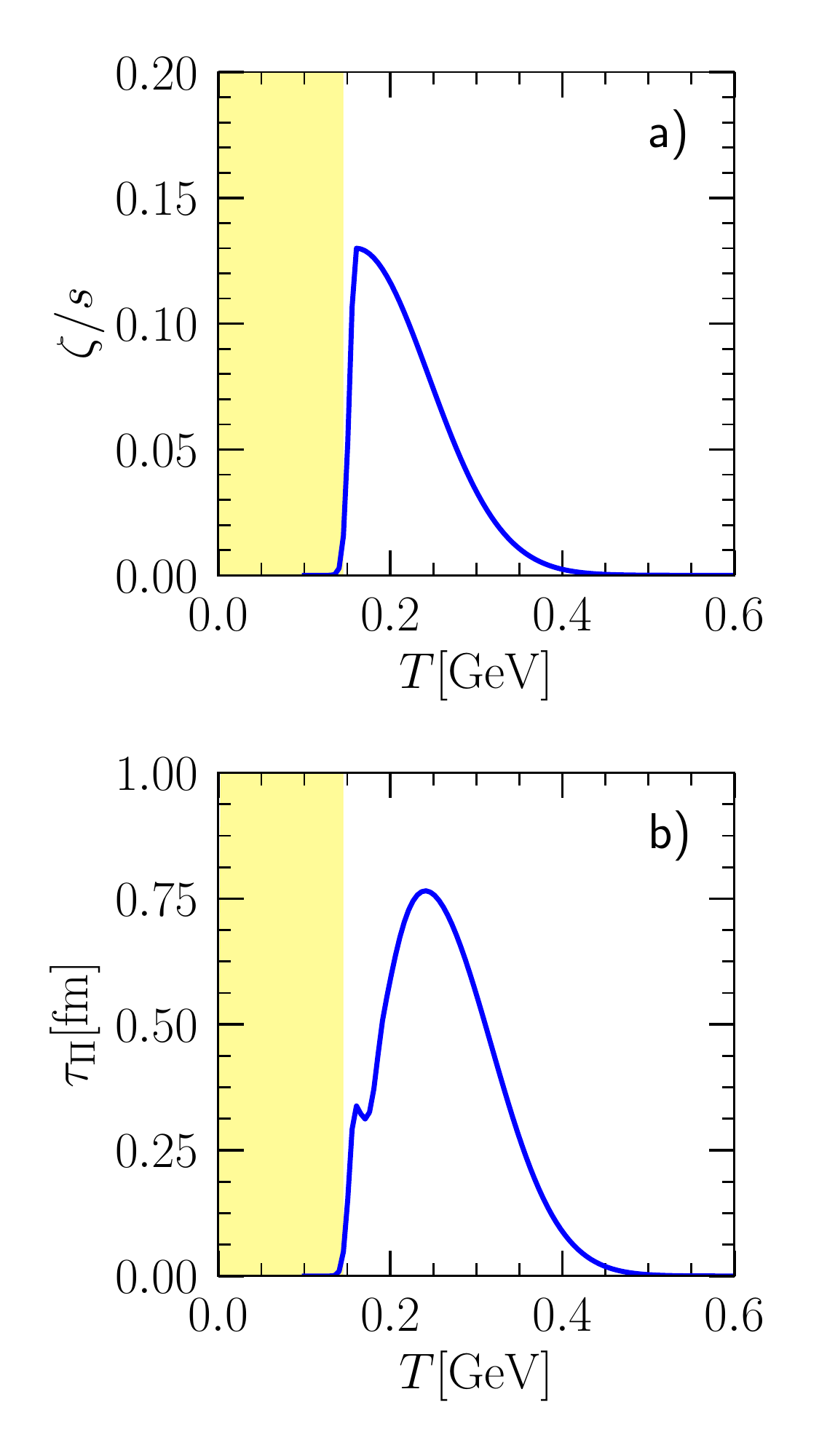}
  \caption{Temperature dependence of a) the bulk viscosity to entropy density ratio $\zeta/s$, and b) the bulk relaxation time $\tau_\Pi$. The shaded region indicates the range of temperature where we employ the hadronic transport simulation instead of hydrodynamics. \label{fig:coefficients}}
\end{figure}

In this work we will use a constant $\eta/s=0.12$ and a temperature dependent $(\zeta/s)(T)$, given by the parametrization
\begin{equation}
    \frac{\zeta}{s} (T) = \left\{ \begin{array}{cc}
        B_\mathrm{norm} \exp \left(- \frac{(T - T_\mathrm{peak})^2}{B_1^2} \right) & \quad T < T_\mathrm{peak} \\
        B_\mathrm{norm} \exp \left(- \frac{(T - T_\mathrm{peak})^2}{B_2^2} \right) & \quad T > T_\mathrm{peak} 
    \end{array} , \right.
\end{equation}
where $B_\mathrm{norm} = 0.13$, $B_1 = 0.01$\,GeV, $B_2 = 0.12$\,GeV, and $T_\mathrm{peak} = 0.16$\,GeV.

The used $(\zeta/s)(T)$ is shown in Fig.\,\ref{fig:coefficients}\,a). It is constructed to avoid large bulk viscous corrections on the switching surface (of fixed energy density $e_{\rm sw} = 0.18\,{\rm GeV/fm}^3$), because the corrections to the distribution functions to be discussed below in Sec.\,\ref{sec:sampling} are not well constrained. This explains the quick fall-off of $(\zeta/s)(T)$ as $T$ decreases below the peak. 
The bulk relaxation time follows from Eq.\,\eqref{eq:bulkrelax} and is shown in Fig.\,\ref{fig:coefficients}\,b) for our choice of $(\zeta/s)(T)$. We choose a Gaussian parametrization for the temperature dependent $(\zeta/s)(T)$ at $T > T_\mathrm{peak}$, which ensures the bulk relaxation time remains smaller than 1 fm. A previous parametrization using a Lorentzian form led to large $\tau_\Pi$ and Knudsen numbers at high temperatures \cite{Schenke:2019ruo, Schenke:2020unx}.
A summary of recent calculations of $(\zeta/s)(T)$ in the hadronic phase is shown in \cite{Rose:2020lfc}.

The simulation \textsc{Music} solves the hydrodynamic equations \eqref{eq:hydro}, \eqref{eq:bulk}, and \eqref{eq:shear} using the Kurganov-Tadmor (KT) algorithm \cite{Kurganov:2000} as described in detail in \cite{Schenke:2010nt,Jeon:2015dfa} for the ideal (non-viscous) case.\footnote{The viscous case is briefly discussed in \cite{Schenke:2010rr}.} The KT algorithm has a low numerical viscosity and can deal particularly well with shocks. 
One aspect of the simulation that was not previously discussed in detail is the regulation of viscous corrections that become very large compared to the ideal parts of $T^{\mu\nu}$. We present the details of this procedure in Appendix \ref{sec:regulator}.

\subsection{iSS - Particle sampling} \label{sec:sampling}
When the medium local energy density drops to the switching energy density $e_{\rm sw}=0.18\,{\rm GeV}/{\rm fm}^3$, the fluid is converted to particles by first computing the particle spectra according to the Cooper-Frye formula \cite{Cooper:1974mv}, using equilibrium distributions $f_0$ with viscous corrections $\delta f$, given in  \cite{Dusling:2009df,Bozek:2009dw,Paquet:2015lta} for shear and bulk viscous terms. The probability distribution of particle species $i$ emitted from a surface element $d^3 \sigma$ located on a constant energy density hyper-surface $\Sigma(x^\mu)$ can be computed as
\begin{equation}
    E \frac{d N^i}{d^3p}(x^\mu) = \frac{g_i}{(2\pi)^3} p^\mu d^3 \sigma_\mu [f^i_0(x^\mu, p^\mu) + \delta f^i(x^\mu, p^\mu)].
\end{equation}
Here $g_i$ is the degeneracy of the particle species $i$. Assuming longitudinal boost-invariance, the infinitesimal surface normal vector $d^3 \sigma_\mu = (\cosh(\eta_s), -\partial \tau/\partial x, -\partial \tau/\partial y, -\sinh(\eta_s)) \tau dx dy d\eta_s$. We choose the following viscous corrections $\delta f$ for shear (14-moment method \cite{Grad:1949}) and bulk viscosity (Chapman-Enskog method \cite{Chapman:1990})
\begin{equation}
    \delta f^i_\mathrm{shear}(x^\mu, p^\mu) = f^i_0 (1 \pm f^i_0) \frac{\pi^{\mu\nu} p_\mu p_\nu}{2 T^2 (e + P)}\,,
    \label{eq:sheardf}
\end{equation}
and
\begin{eqnarray}
    \delta f^i_\mathrm{bulk}(x^\mu, p^\mu) &=& f^i_0 (1 \pm f^i_0) \left( - \frac{\Pi}{\hat{\zeta}} \right) \frac{(p \cdot u)}{T} \nonumber \\ && \times \left[\frac{m_i^2}{3 (p \cdot u)^2} - \left(\frac{1}{3} - c_s^2 \right) \right]\,.
\end{eqnarray}
Here the coefficient $\hat{\zeta}$ is the following thermodynamic integral,
\begin{eqnarray}
    \hat{\zeta} &=& \frac{1}{3 T} \sum_i \frac{g_i}{(2\pi)^3} m_i^2 \int \frac{d^3 k}{E} f^i_0 (1 \pm f^i_0) \nonumber \\
    && \qquad \qquad \qquad \qquad \times E \left[\frac{m_i^2}{3 E^2} - \left(\frac{1}{3} - c_s^2 \right)  \right]\,,
\end{eqnarray}
where $E=p\cdot u$.

Should the value of any $\delta f$ be larger than that of the equilibrium distribution, we set the distribution to zero, avoiding negative values. For a detailed discussion of off-equilibrium corrections and different choices for $\delta f$, we refer the reader to \cite{McNelis:2019auj}.

To generate Monte-Carlo samples from individual fluid cells, we need to determine the average numbers of particles emitted from each fluid cell. The particle yield of a species $i$ from a surface fluid cell has the following analytic expression
\begin{equation}
    N^i(x^\mu) = \int \frac{d^3 p}{E} E \frac{d N^i}{d^3p}(x^\mu) = N^i_0 + N^i_\mathrm{bulk} + N^i_\mathrm{shear}.
\end{equation}
Here the equilibrium distribution function gives
\begin{eqnarray}
  N^i_0 = \frac{g_i}{2 \pi^2} d^3 \sigma_{\mu} u^{\mu}  m_i^2 T \sum_{n = 1}^{\infty} \frac{(\pm1)^{n - 1}}{n} K_2 \left(n \frac{m_i}{T} \right).
\end{eqnarray}
The bulk viscous corrections also contribute to the particle yield:
\begin{eqnarray}
  N^i_\mathrm{bulk} & = & \frac{g_i}{2 \pi^2} d^3 \sigma_{\mu}
  u^{\mu} \left( - \frac{\Pi}{\hat{\zeta}} \right) \left\{ - \left( \frac{1}{3} - c_s^2 \right)
  m_i^2 T \right. \nonumber \\
  && \times \sum_{n = 1} (\pm1)^{n - 1} \left[ \frac{m_i}{T}
  K_1 \left(n \frac{m_i}{T}\right) + \frac{3}{n} K_2 \left(n \frac{m_i}{T} \right) \right]
   \nonumber\\
  && \left. + \frac{m_i^3}{3} \sum_{n = 1}^{\infty} (\pm1)^{n - 1} K_1 \left(n \frac{m_i}{T}\right) \right\}.
\end{eqnarray}
Finally, the shear viscous correction does not modify the particle yield, and $N^i_\mathrm{shear} = 0$.

In this work, we use an open-source particle sampler \texttt{iSS} \cite{iSS, Shen:2014vra, Denicol:2018wdp} to perform the numerical sampling of particles from hydrodynamic hypersurfaces event-by-event. To achieve enough statistics, each individual hydrodynamic hypersurface is oversampled until we reach at least 100,000 thermal $\pi^+$ per unit of rapidity. The averaged number of oversampled events is about 100 in central collisions and  it can reach up to more than 10,000 in the most peripheral centrality bin.
\begin{figure}[tb]
  \centering
  \includegraphics[width=0.48\textwidth]{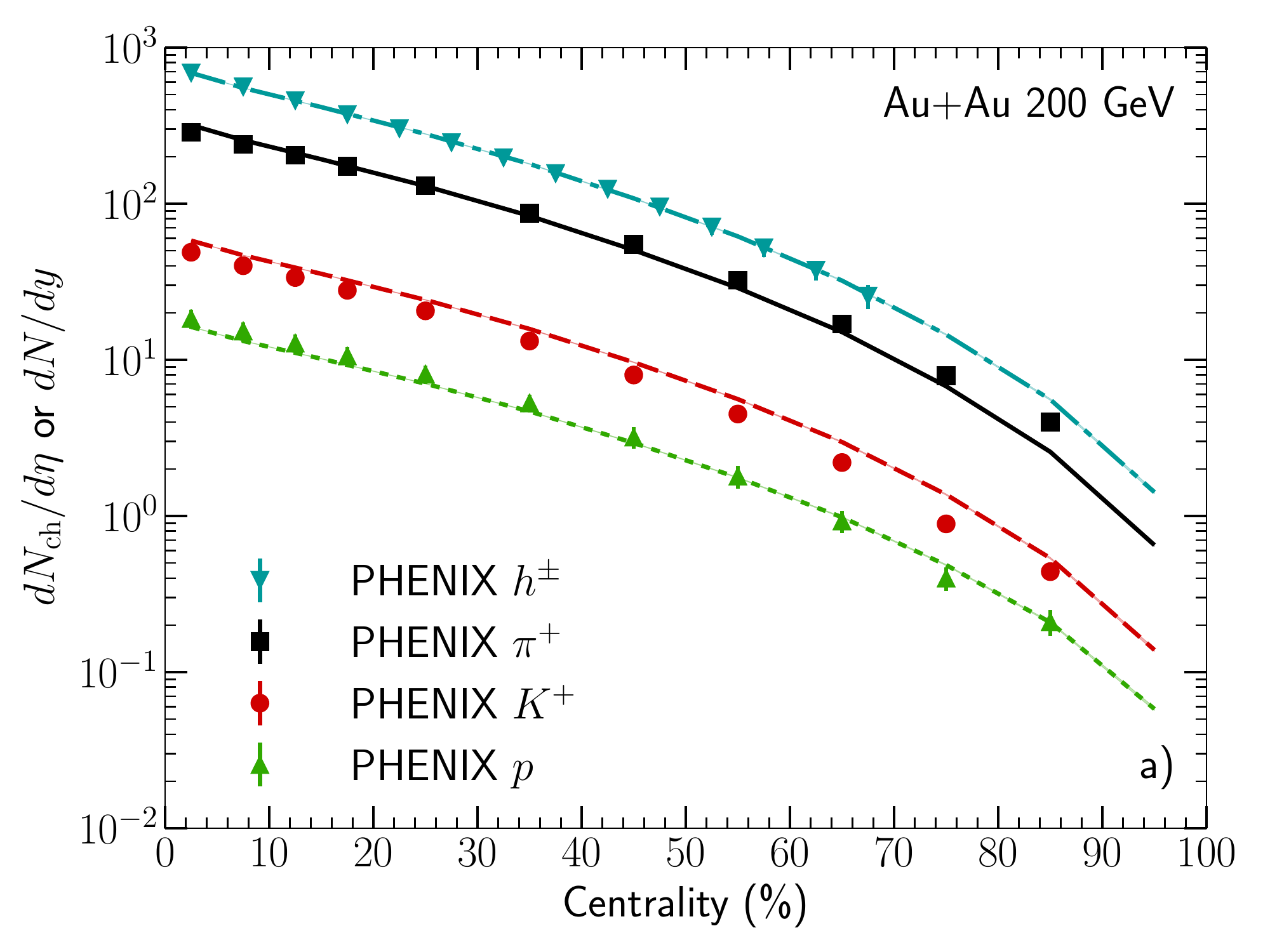}
  \includegraphics[width=0.48\textwidth]{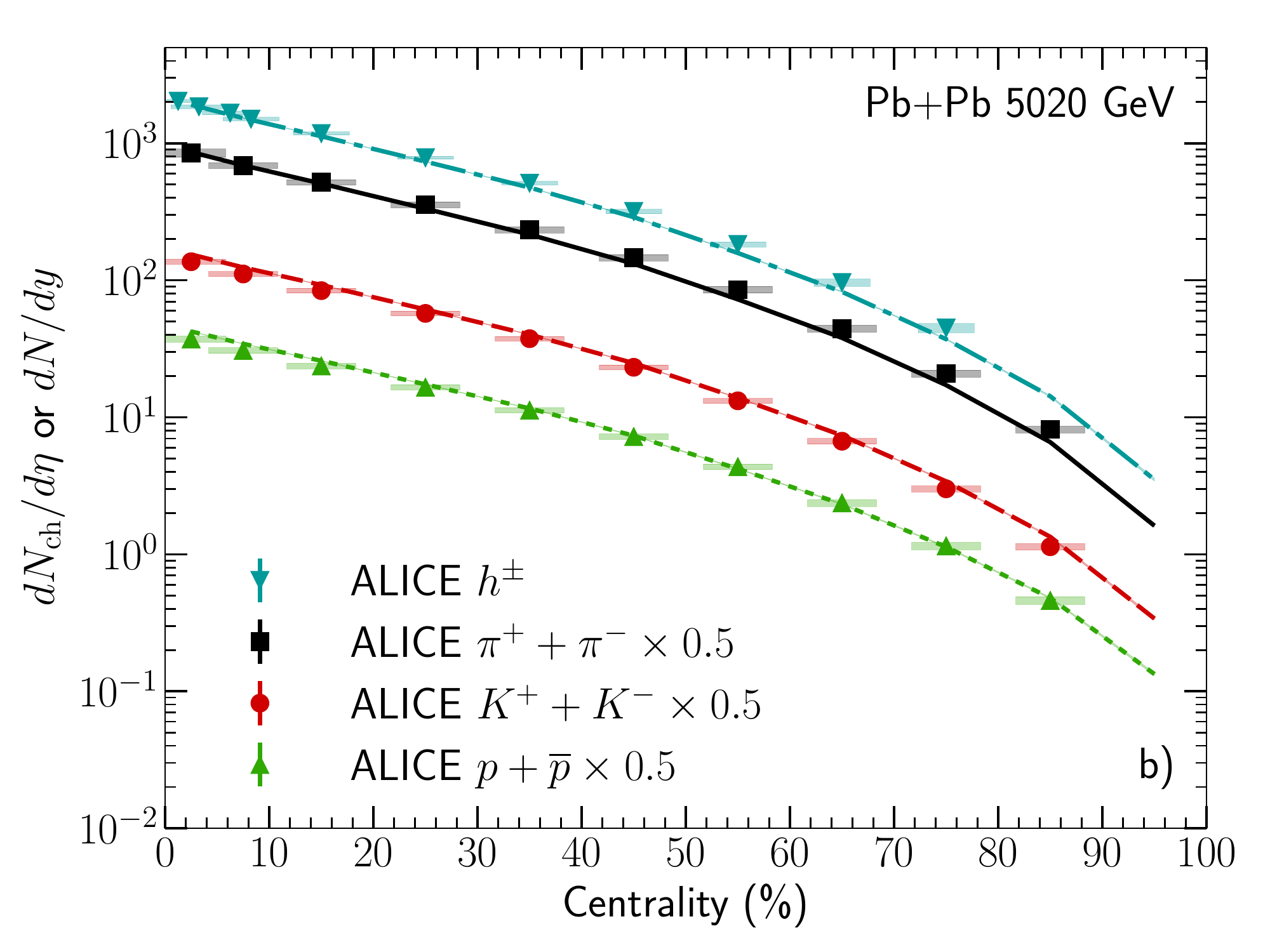}
  \caption{Charged hadron and identified particle multiplicities vs.~centrality for 200 GeV Au+Au collisions at RHIC (a) and 5.02 TeV Pb+Pb collisions at LHC (b). Experimental data from the PHENIX \cite{Adler:2003cb} and ALICE \cite{Adam:2015ptt,Acharya:2019yoi} Collaborations. \label{fig:NPID}}
\end{figure}

\begin{figure}[tb]
  \centering
    \includegraphics[width=0.48\textwidth]{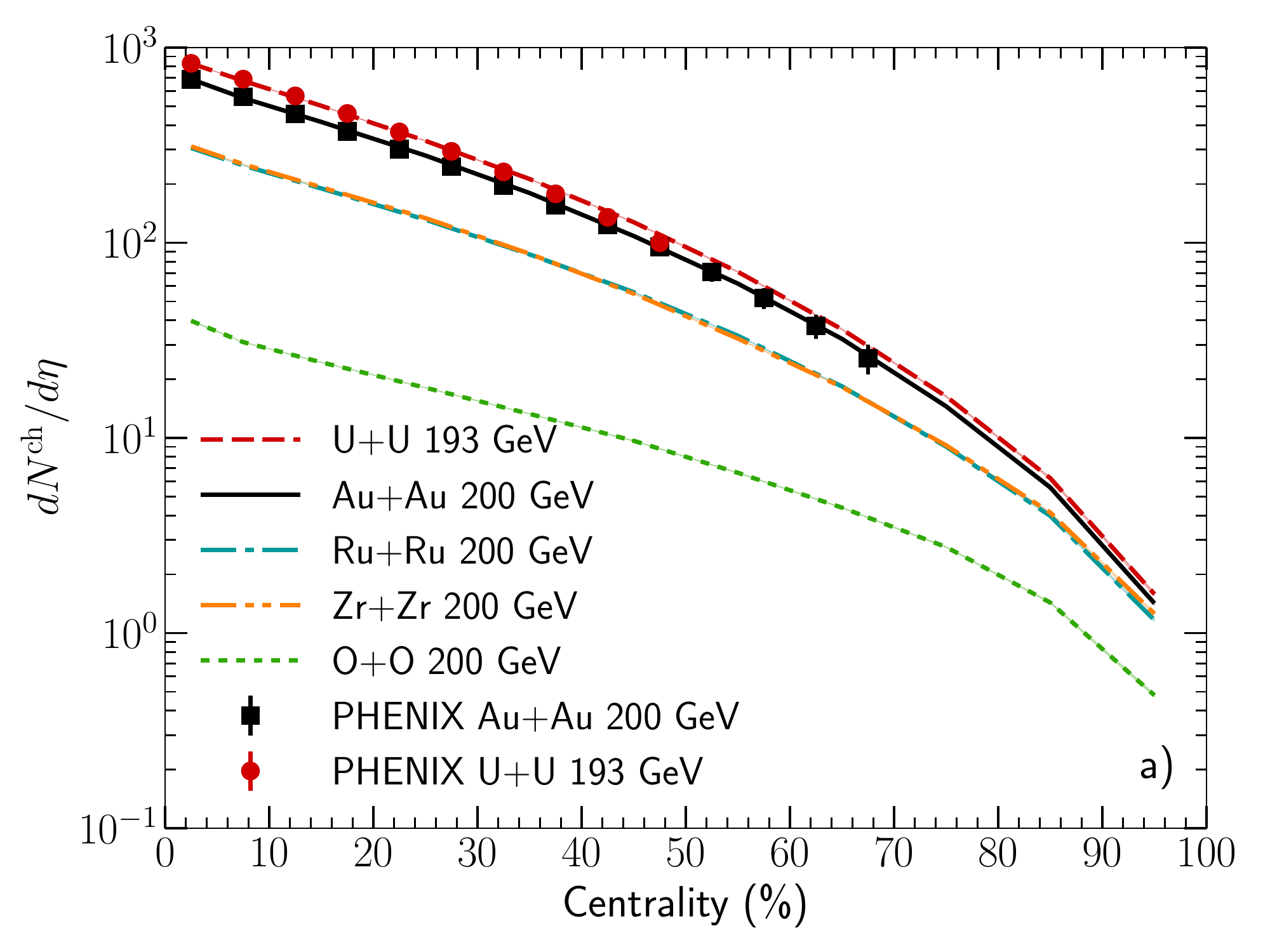}
    \includegraphics[width=0.48\textwidth]{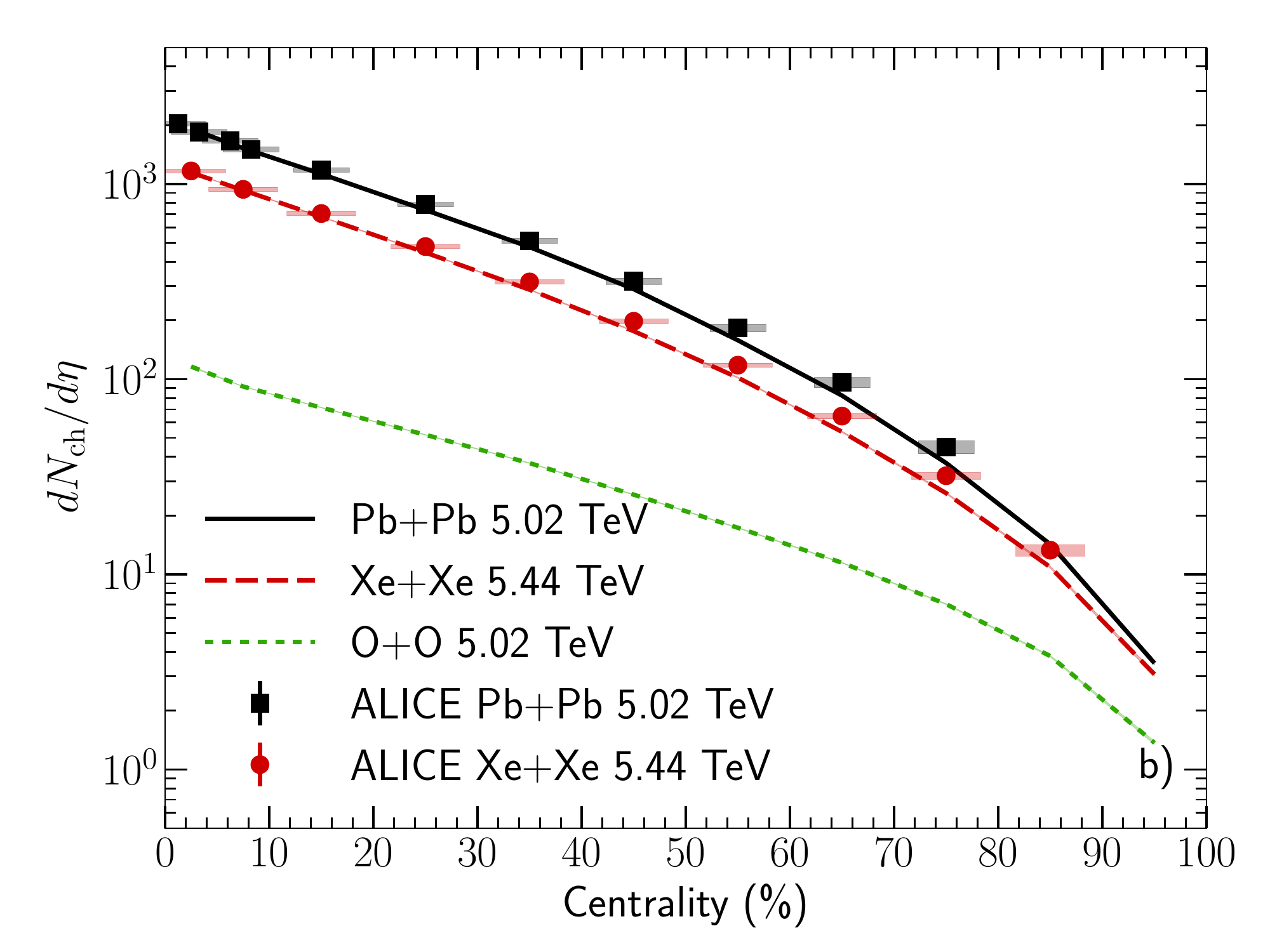}
  \caption{Charged hadron multiplicities as a function of centrality in 200 GeV O+O, Ru+Ru, Zr+Zr, Au+Au, and 193 GeV U+U collisions at RHIC (a) and in 5.02 TeV O+O, 5.44 TeV Xe+Xe, and 5.02 TeV Pb+Pb collisions at LHC (b). Experimental data from the PHENIX \cite{Adler:2003cb} and ALICE \cite{Adam:2015ptt,Acharya:2018hhy} Collaborations.\label{fig:dNdeta}}
\end{figure}

\subsection{UrQMD - Hadronic cascade}
The sampled particles are further propagated in the microscopic transport simulations using the \texttt{UrQMD} model \cite{Bass:1998ca,Bleicher:1999xi} with default parameters. \texttt{UrQMD} simulates scatterings among hadrons and their excited resonance states with masses up to 2.25 GeV as well as all strong decay processes. The interactions in \texttt{UrQMD} include elastic and inelastic scatterings as well as $B\bar{B}$ annihilations. Having \texttt{UrQMD} as an afterburner allows for a more realistic dynamical freeze-out for different species of hadrons in the final stage of the collisions.

\section{Charged hadron multiplicity}\label{sec:mult}
We begin our discussion of results with the most straight forward observable, namely the number of all charged hadrons as well as that of identified hadrons per unit rapidity (around mid-rapidity) as a function of collision centrality. In all results we show, we have performed the centrality selection using the final $dN/d\eta$ of charged hadrons.
In previous works we have instead used the gluon number distribution obtained directly in the IP-Glasma model \cite{Schenke:2013dpa}. Naturally, using the final, observable, charged hadron numbers is the right thing to do when comparing to experimental data. In Appendix \ref{sec:centrality} we demonstrate the differences between the two methods. 

In Fig.\,\ref{fig:NPID} a) we show the charged particle number per unit pseudo-rapidity $dN_{\rm ch}/d\eta$ as a function of centrality for $\sqrt{s_{\rm NN}}=200\,{\rm GeV}$ Au+Au collisions compared to experimental data from the PHENIX Collaboration \cite{Adler:2003cb}. Agreement with the experimental data is very good. We further show results for identified particles, namely positive pions, kaons, and protons. Again, agreement is good, with kaons slightly overestimated for all centrality classes in our model.

In Fig.\,\ref{fig:NPID} b) we show the same observables for $\sqrt{s_{\rm NN}}=5.02\,{\rm TeV}$ Pb+Pb collisions and compare to experimental data from the ALICE Collaboration \cite{Acharya:2019vdf} (we divide the identified particle spectra, which are given by the experiment as a sum of positive and negative particles by 2 for clarity of the plot\footnote{Our calculation does not include any baryon chemical potential, which is a good approximation for large energies, such that there is no difference between baryons and anti-baryons.}). As in the case of RHIC, agreement with experimental data is very good. We note that there are no additional free parameters once parameters have been fixed using Au+Au collisions at the top RHIC energy. The multiplicity evolution in the IP-Glasma model and entropy production from viscous hydrodynamics give a quantitative prediction of final charged hadron yields at 5.02 TeV. The deviations from the experimental measurements are within 10\%. This predictive power is a unique feature of our hybrid framework. All changes in the studied observables, as we modify the collision system or center of mass energy, are predictions of the model.

We extend our study to more systems in Figs.\,\ref{fig:dNdeta} a) for RHIC and b) for LHC. At RHIC we study, in addition to 200 GeV Au+Au collisions, U+U collisions with $\sqrt{s_{\rm NN}}=193\,{\rm GeV}$, Zr+Zr, Ru+Ru, and O+O collisions at 200 GeV. We compare Au+Au and U+U collisions to experimental data from the PHENIX Collaboration \cite{Adler:2003cb}. Both systems' $dN_{\rm ch}/d\eta$ as a function of centrality are well described within our model. The other results are predictions for the isobar run as well as for a potential future O+O run. 

At LHC, we repeat the result from Fig.\,\ref{fig:NPID} b) for Pb+Pb collisions and add results for Xe+Xe collisions at $\sqrt{s_{\rm NN}}=5.44\,{\rm TeV}$ and O+O collisions at $\sqrt{s_{\rm NN}}=5.02\,{\rm TeV}$. Experimental data is available for Xe+Xe, and we compare well with the ALICE result \cite{Acharya:2018hhy}, slightly underestimating the multiplicity in centrality classes $>20-30\%$, similar to the case of Pb+Pb collisions.

\section{Average transverse momentum}\label{sec:pt}
\begin{figure}[tb]
  \centering
   \includegraphics[width=0.48\textwidth]{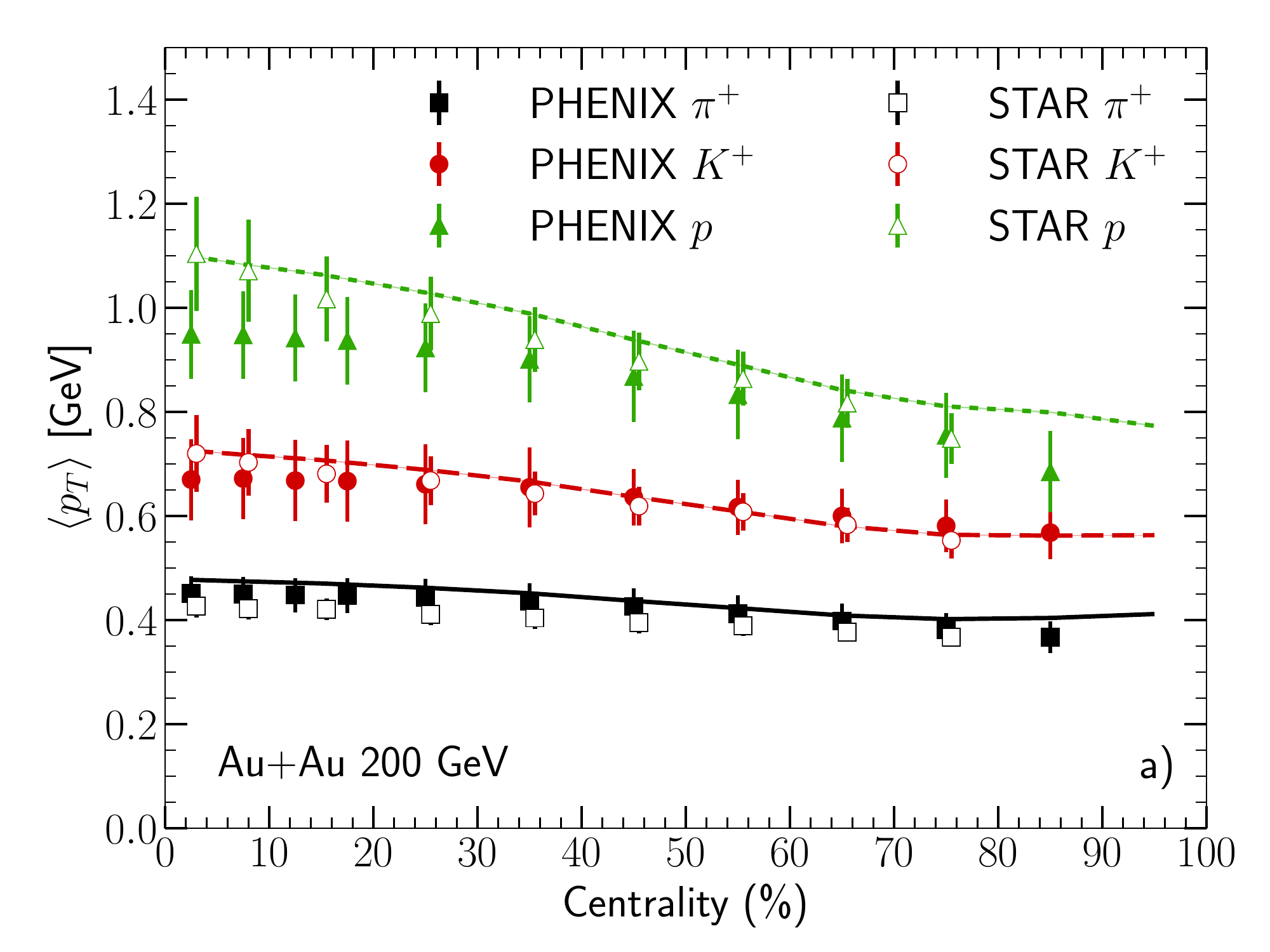}
     \includegraphics[width=0.48\textwidth]{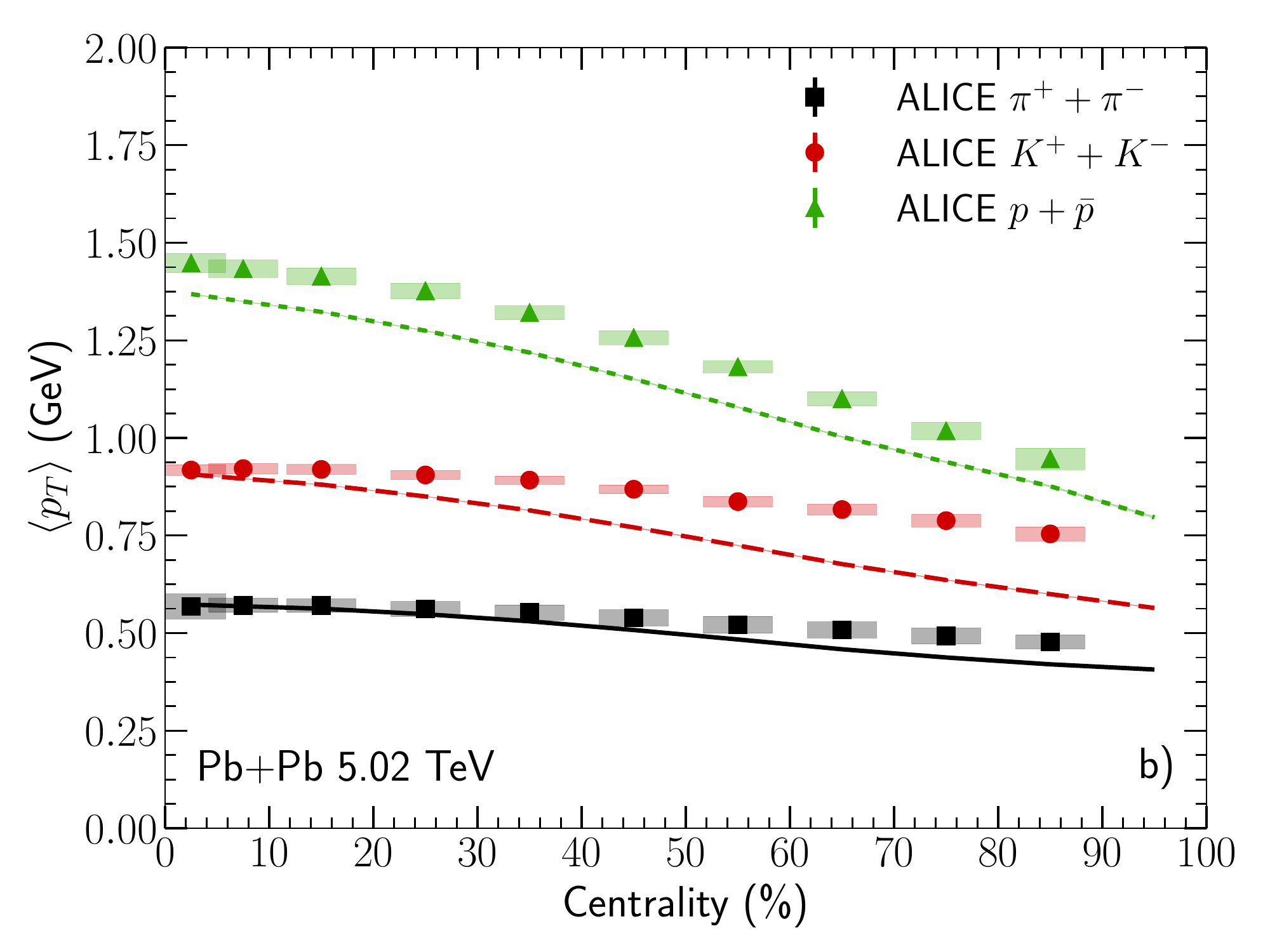}
  \caption{Identified particle mean transverse momentum vs.~centrality in a) 200 GeV Au+Au collisions at RHIC and b) 5.02 TeV Pb+Pb collisions at LHC. Experimental data from the PHENIX \cite{Adler:2003cb}, STAR \cite{Abelev:2008ab}, and ALICE \cite{Acharya:2019yoi} Collaborations. \label{fig:PIDmeanpT}}
\end{figure}

\begin{figure}[tb]
  \centering
  \includegraphics[width=0.48\textwidth]{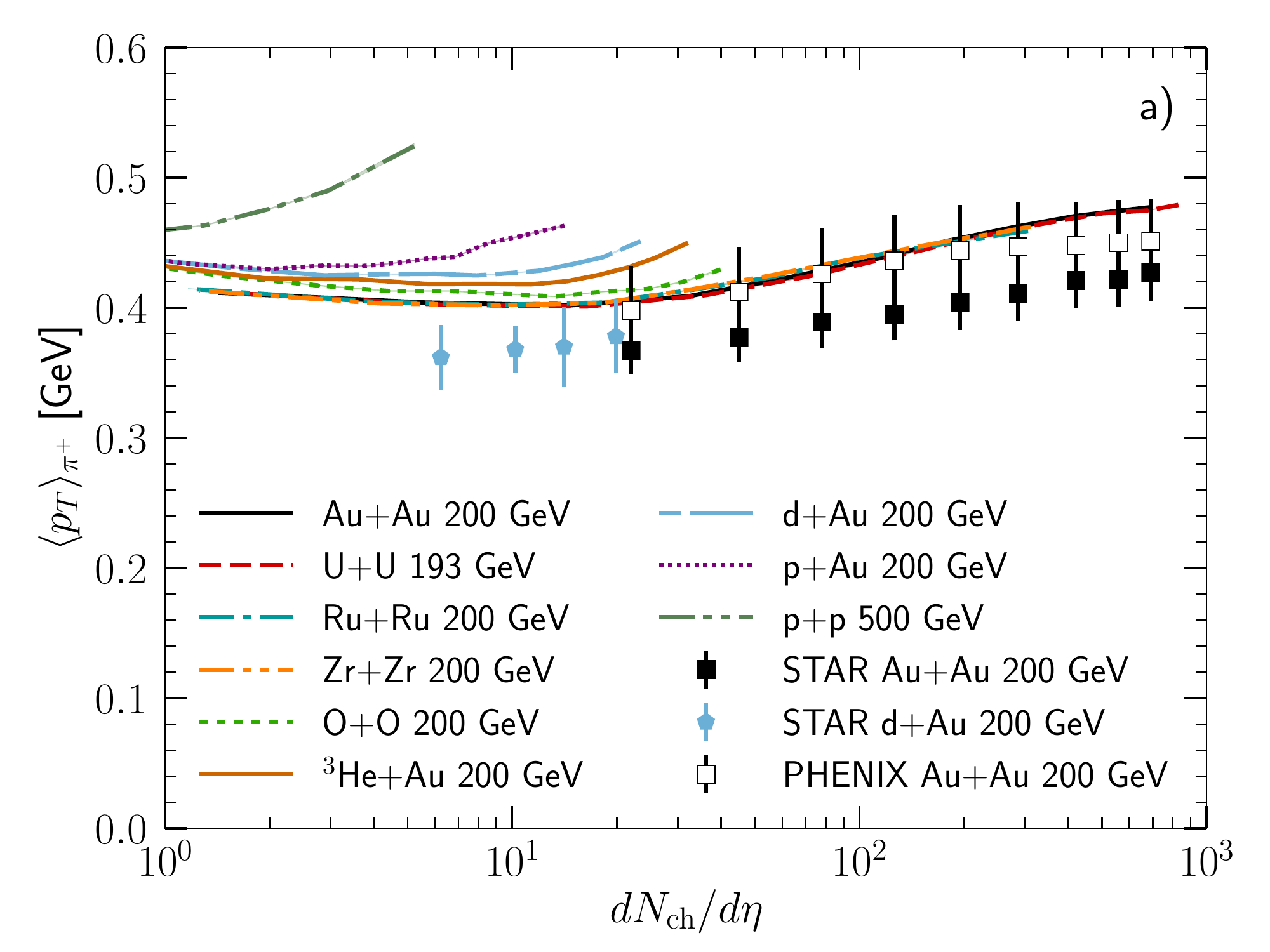}
  \includegraphics[width=0.48\textwidth]{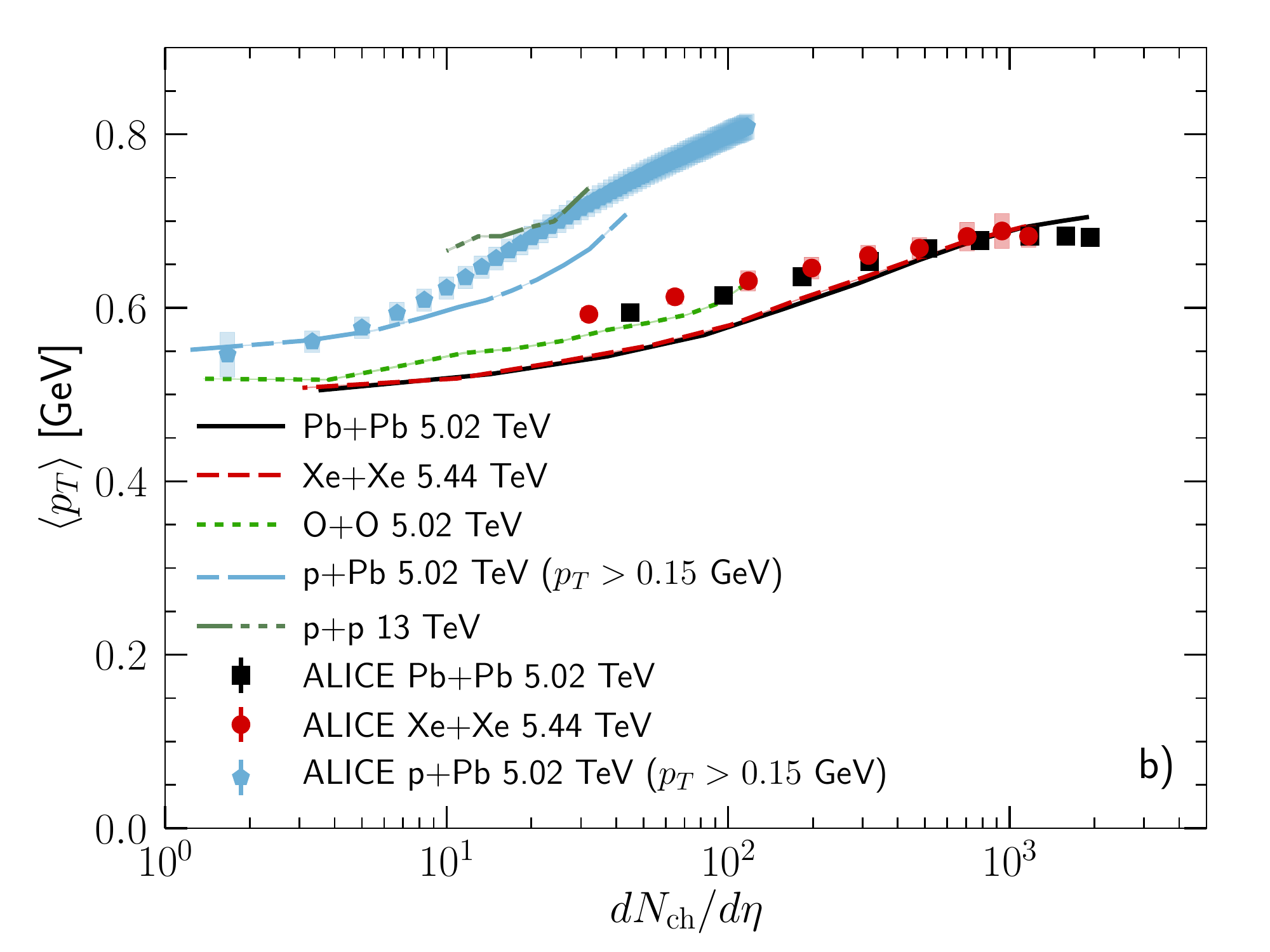}
  \caption{a) Positive pion mean transverse momenta as a function of $dN_{\rm ch}/d\eta$ in various collision systems at RHIC. b) Charged hadron mean transverse momentum as a function of $dN_{\rm ch}/d\eta$ in p+p, p+Pb, O+O, Xe+Xe, and Pb+Pb collisions at LHC. Experimental data from the STAR \cite{Abelev:2008ab}, PHENIX \cite{Adler:2003cb}, and ALICE \cite{Acharya:2018eaq} Collaborations. \label{fig:chmeanpT}}
\end{figure}

The next observable we study is the mean transverse momentum of charged hadrons and identified particles. Within hydrodynamics, the mean transverse momentum constrains the $p_T$-spectrum strongly, such that reproducing $\langle p_T\rangle$ is equivalent to reproducing the $p_T$ differential spectrum \cite{Pratt:2015zsa, Sangaline:2015isa}. As discussed in \cite{Ryu:2015vwa,Ryu:2017qzn}, the mean transverse momentum is very sensitive to the bulk viscosity and puts the biggest constraints on the form and size of $(\zeta/s)(T)$ shown in Fig.\,\ref{fig:coefficients} a). 

In Fig.\,\ref{fig:PIDmeanpT}\,a) we show the identified particle mean transverse momentum as a function of centrality for Au+Au collisions at 200 GeV. We compare to experimental data from the STAR \cite{Abelev:2008ab} and PHENIX \cite{Adler:2003cb} Collaborations. Agreement for pions is better with the PHENIX data, while protons agree better with the STAR data and overestimate the PHENIX data in most centrality bins. Fig.\,\ref{fig:PIDmeanpT}\,b) shows the same observables in 5.02 TeV Pb+Pb collisions at the LHC compared to experimental data from the ALICE Collaboration \cite{Acharya:2019yoi}. Here, agreement with the experimental data is somewhat worse. For pions and kaons central events are well described, but the calculated mean transverse momentum drops faster with centrality than the experimental data. For protons, we underestimate all data points. We chose a lower peak temperature for bulk viscosity compared to the one used in the Refs. \cite{Ryu:2015vwa,Ryu:2017qzn}. The $T_\mathrm{peak} = 0.16$~GeV is constrained by the overall centrality dependence of the mean $p_T$ measurements in 200 GeV Au+Au collisions. A higher peak temperature would lead to overestimation of mean-$p_T$ measurements in peripheral collisions, in which the maximum temperature at the starting time of hydrodynamics could be already below the bulk viscous peak \cite{Schenke:2018fci}. The underestimation of the mean transverse momentum in peripheral Pb+Pb collisions at LHC, which typically start out at a higher initial temperature, indicates that the used shape of $(\zeta/s)(T)$ is not yet the ideal choice. 

In Fig.\,\ref{fig:chmeanpT}\,a) we show the charged hadron mean transverse momentum as a function of charged particle multiplicity for eight different collision systems at RHIC energies and compare to experimental data from STAR and PHENIX where available \cite{Abelev:2008ab,Adler:2003cb}. 
Agreement of the 200 GeV Au+Au data from PHENIX is very good, while data from STAR is overestimated. Our result for the mean transverse momentum in 200 GeV d+Au collisions is significantly larger than our Au+Au result, which is expected in a hydrodynamic framework, where a smaller system with the same approximate amount of entropy exhibits larger gradients and thus more radial flow, which translates to larger final transverse momentum. The STAR data on d+Au collisions does not clearly show this trend, however. 

For the larger systems, Ru+Ru, Zr+Zr, and U+U, the mean transverse momentum is very close to that of Au+Au, while for all smaller systems, p+p, p+Au, d+Au, $^3$He+Au, O+O, we observe the same trend discussed above for d+Au and Au+Au. The smaller in transverse size a system is at a given multiplicity, the larger the mean transverse momentum.

Fig.\,\ref{fig:chmeanpT}\,b) shows the same observable for five different systems at LHC energies, compared to experimental data from the ALICE Collaboration \cite{Acharya:2018eaq} in Pb+Pb, Xe+Xe, and p+Pb collisions. Again, we observe the same trend as discussed for the collision systems at RHIC. As seen in the identified particle mean transverse momentum in Fig.\,\ref{fig:PIDmeanpT} b), we underestimate the mean transverse momentum in Pb+Pb and Xe+Xe collisions, except in the most central centrality classes. In p+Pb collisions, our prediction agrees at the lowest multiplicities, but underestimates the mean transverse momentum at the larger $dN_{\rm ch}/d\eta$. The change from Pb+Pb to p+Pb at a given multiplicity is however reproduced by our model. The result for 13 TeV p+p collisions is the largest, and our prediction for the mean transverse momentum in 5.02 TeV O+O collisions lies between the results for Pb+Pb and p+Pb collisions of the same center of mass energy.

We note again that all parameters were fixed in 200 GeV Au+Au collisions at RHIC, and all other results, in particular those at LHC, are predictions. We thus consider the level of agreement with the experimental data reasonable. It is possible that a modified temperature dependence of the bulk viscosity over entropy density ratio can improve the agreement with the experimentally observed mean transverse momentum at both RHIC and LHC. This will best be explored in a Bayesian analysis including a wide range of available experimental data, such as done in \cite{Bernhard:2016tnd} for different initial state models.

\section{Momentum anisotropy}\label{sec:vn}

\begin{figure}[tb]
  \centering
  \includegraphics[width=0.46\textwidth]{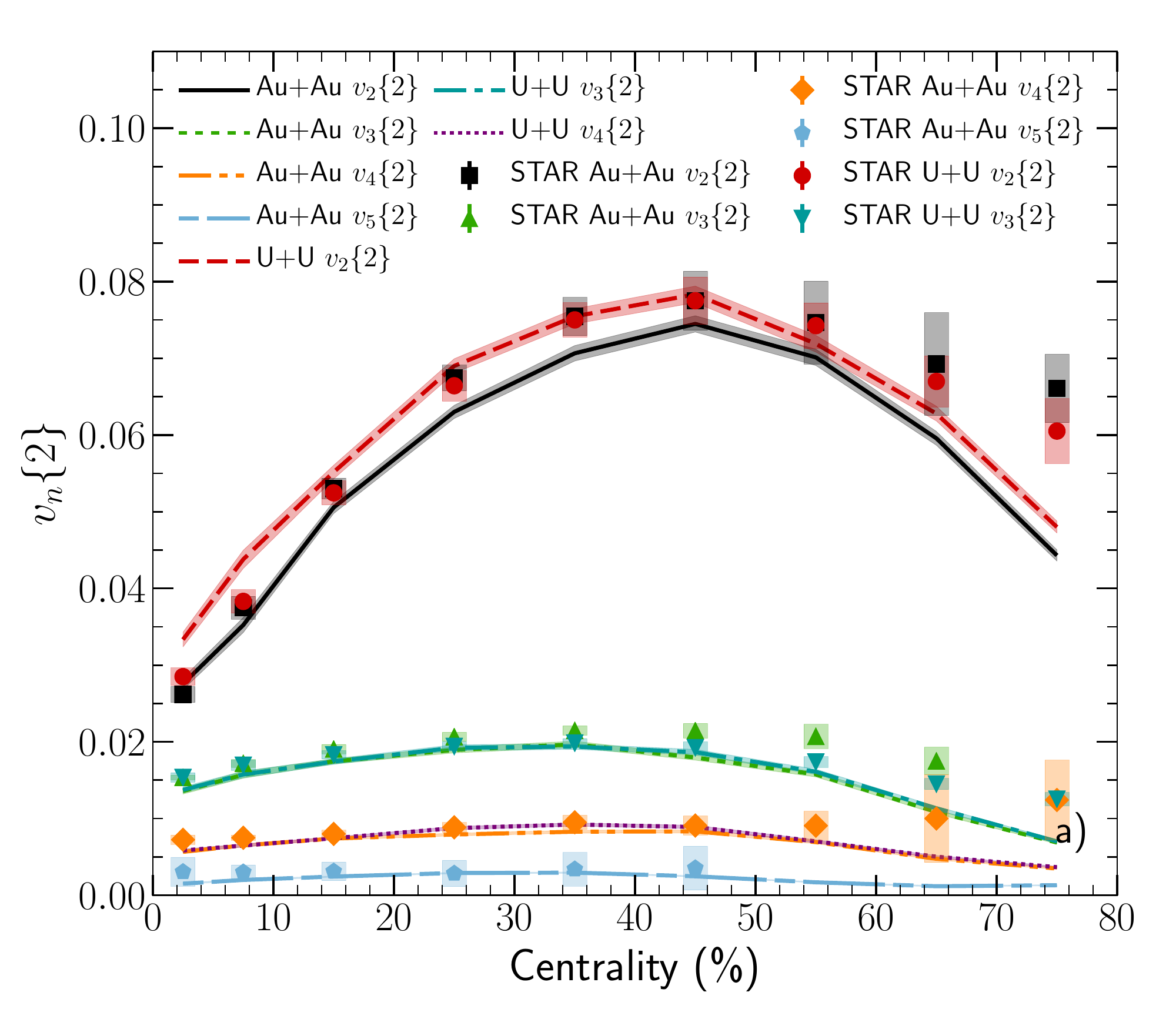}\\\includegraphics[width=0.46\textwidth]{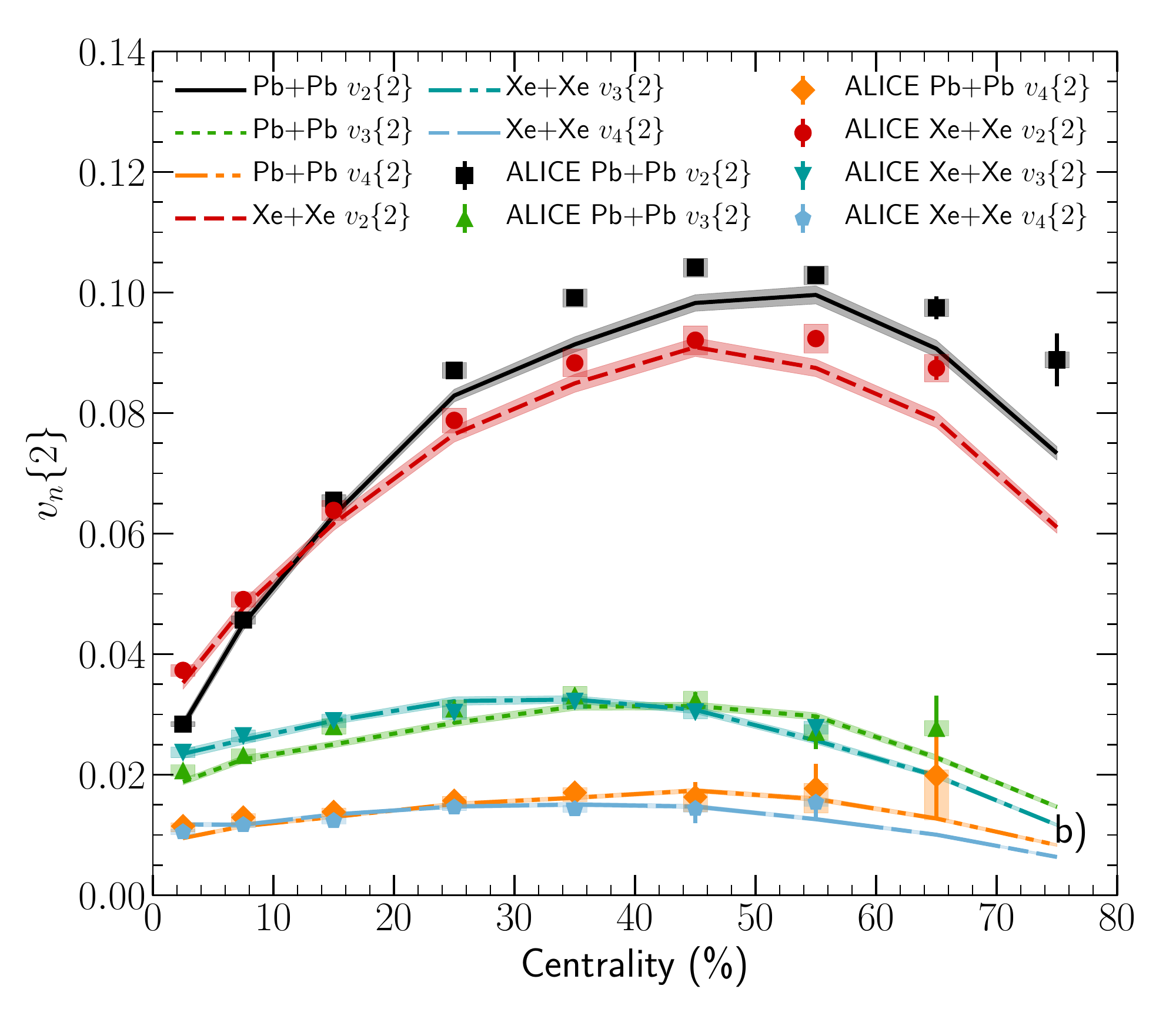}
  \caption{Azimuthal anisotropy coefficients $v_n\{2\}$ for charged hadrons vs.~centrality in 200 GeV Au+Au and 193 GeV U+U collisions at RHIC (a) and 5.02 TeV Pb+Pb and 5.44 TeV Xe+Xe collisions at LHC (b). Experimental data from the STAR \cite{Adamczyk:2016exq,Adamczyk:2017hdl,Adam:2019woz} and ALICE \cite{Adam:2016izf,Acharya:2018ihu} Collaborations. \label{fig:v2}}
\end{figure}

The telltale observable that uncovered the almost perfect fluid nature of the matter produced in relativistic heavy ion collisions is the elliptic flow of low transverse momentum ($p_T\lesssim 3\,{\rm GeV}$) hadrons. A plethora of more complex observables, most of them confirming the fluid dynamic picture to great accuracy have since emerged, among them higher flow harmonics and flow correlations obtained from multi-particle correlation measurements. Here, we will focus on the most basic flow observables $v_n\{m\}$ with the harmonic number $n$ up to 5 and obtained from $m$-particle correlations (with $m$ up to six). We leave the study of more complex, e.g. three or four particle cumulants that correlate different order flow harmonics or event planes, to future work. 

\subsection{Two- and four-particle cumulant flow coefficients}
We compute azimuthal momentum anisotropies of charged hadrons using flow vectors as detailed in Appendix \ref{sec:flowAnalysis}. 

A comparison of the centrality dependence of charged hadron $v_n\{2\}$ obtained from two-particle correlations in 200 GeV Au+Au and 193 GeV U+U collisions to experimental data from the STAR Collaboration \cite{Adamczyk:2016exq,Adamczyk:2017hdl,Adam:2019woz} is shown in Fig.\,\ref{fig:v2} a). Agreement for $v_2\{2\}$ in 200 GeV Au+Au collisions is particularly good for central collisions, while the data is somewhat underestimated in more peripheral events. $v_3\{2\}$ is slightly lower than the data for all centralities and in both systems. Agreement of $v_4\{2\}$ and $v_5\{2\}$ is rather good. The underestimation of the experimental data in more peripheral events could be a result of residual non-flow present in the measurement. Our calculations do not include non-flow from jets and we also eliminate short-range correlations from decays and hadronic scattering by over-sampling. Consequently, we expect better agreement with cumulants of 4 and more particles, which we show below. Nevertheless, overall agreement with the data for $n=1$ to $5$ over the whole centrality range is good for both systems. We see a larger difference between Au+Au collisions and U+U collisions than present in the data. For central collisions, some difference stems from the deformation of the uranium nucleus. For more peripheral events it is more likely an effect from the different lifetime of the two systems.

In Fig.\,\ref{fig:v2} b), we show the same observables for 5.02 TeV Pb+Pb and 5.44 TeV Xe+Xe collisions at the LHC and compare to experimental data from the ALICE Collaboration \cite{Adam:2016izf,Acharya:2018ihu}. Here, $v_3\{2\}$ and $v_4\{2\}$ are generally well described, while $v_2\{2\}$ agrees best in the most central collisions. We even reproduce the difference between Xe+Xe and Pb+Pb collisions well, which is a manifestation of the deformation of the Xe nucleus. In more peripheral collisions we see again that the data is underestimated, more so than in the systems studied at RHIC. 
Despite the larger rapidity gap used compared to the STAR data, part of this discrepancy could be caused by non-flow in the experimental data.

\begin{figure}[tb]
  \centering
  \includegraphics[width=0.48\textwidth]{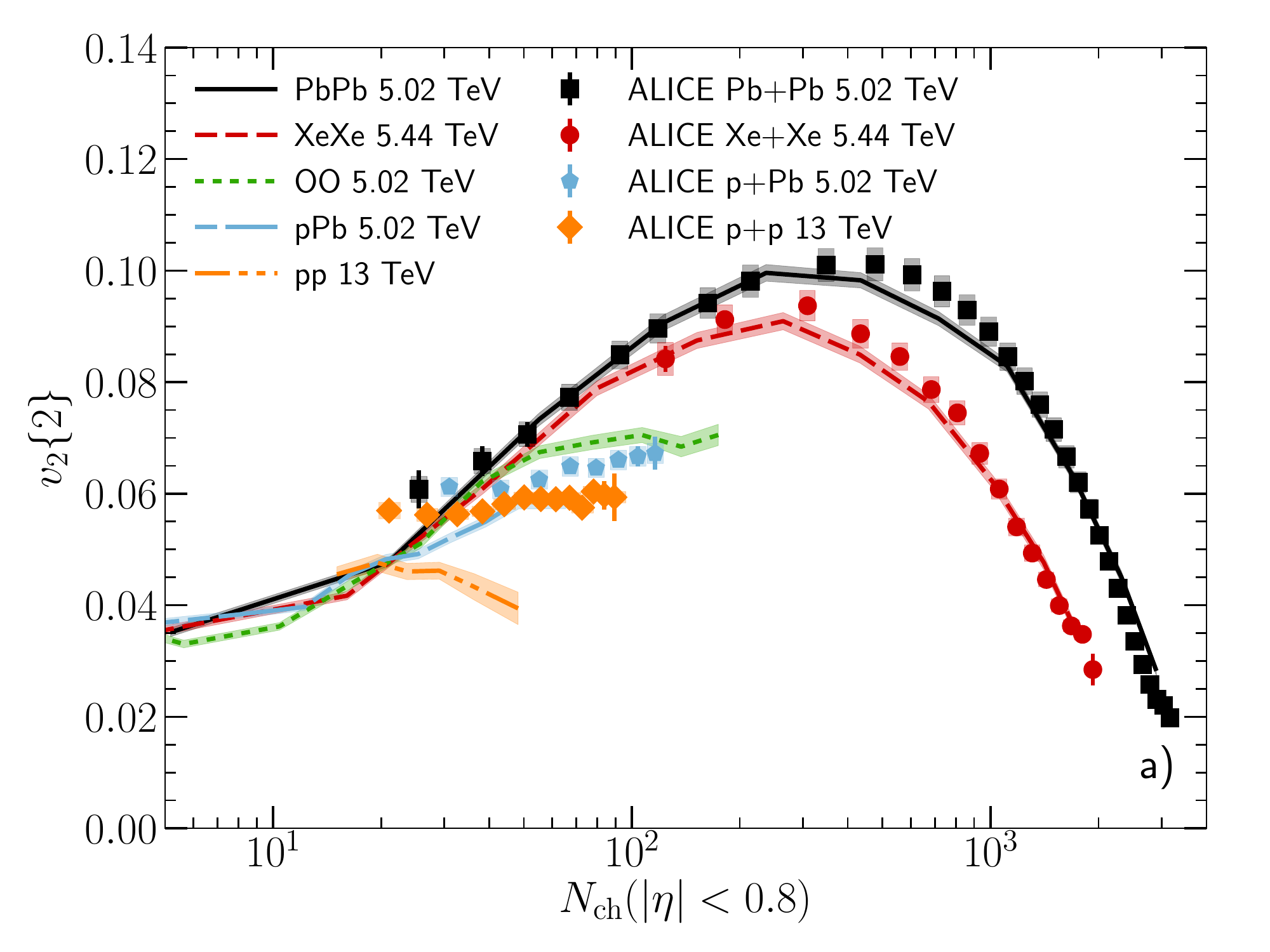}\\\includegraphics[width=0.48\textwidth]{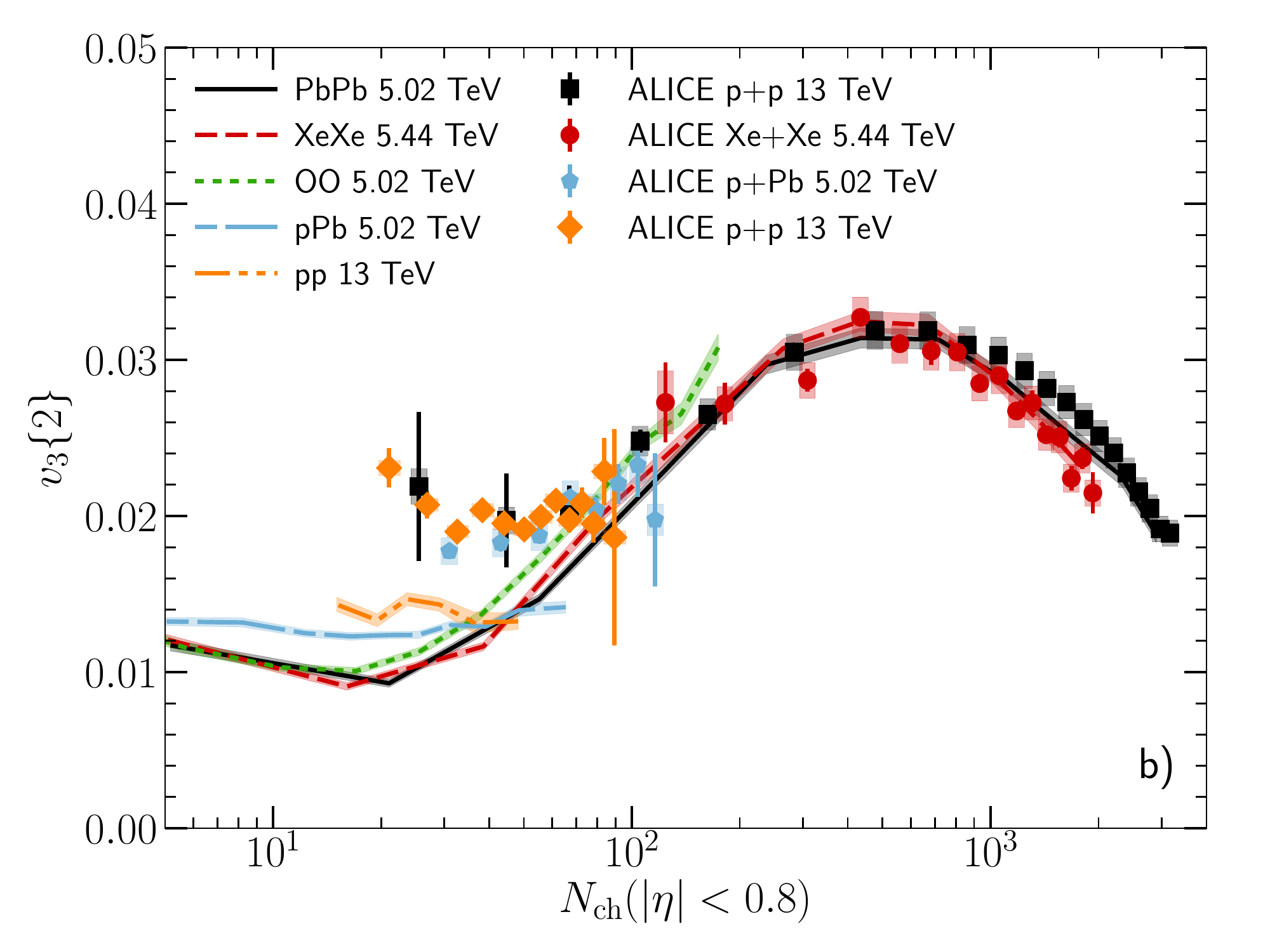}
  \caption{Anisotropy coefficients $v_2\{2\}$ (a) and $v_3\{2\}$ (b) for charged hadrons vs.~charged hadron multiplicity in various collision systems at LHC, compared to experimental data from the ALICE Collaboration \cite{Acharya:2019vdf}. \label{fig:v2v3}}
\end{figure}

We show charged hadron $v_2\{2\}$ as a function of charged hadron multiplicity for more collision systems at LHC in Fig.\,\ref{fig:v2v3} a) and compare to experimental data from the ALICE Collaboration \cite{Acharya:2019vdf}. Apart from the results in Xe+Xe and Pb+Pb collisions, which resemble those shown as a function of centrality in Fig.\,\ref{fig:v2} b), we show predictions for 5.02 TeV O+O collisions, 5.02 TeV p+Pb collisions, and 13 TeV p+p collisions. Given our limited statistics we unfortunately could not push the results for p+Pb collisions to as large a $dN_{\rm ch}/d\eta$ as the experiment could, but our result is close to the experimental data, if slightly low at the largest multiplicity we could compute. For lower $dN_{\rm ch}/d\eta$, we underestimate the experimental data in p+Pb collisions, which could be related to non-flow at the very low multiplicity.

\begin{figure}[tb]
  \centering
    \includegraphics[width=0.48\textwidth]{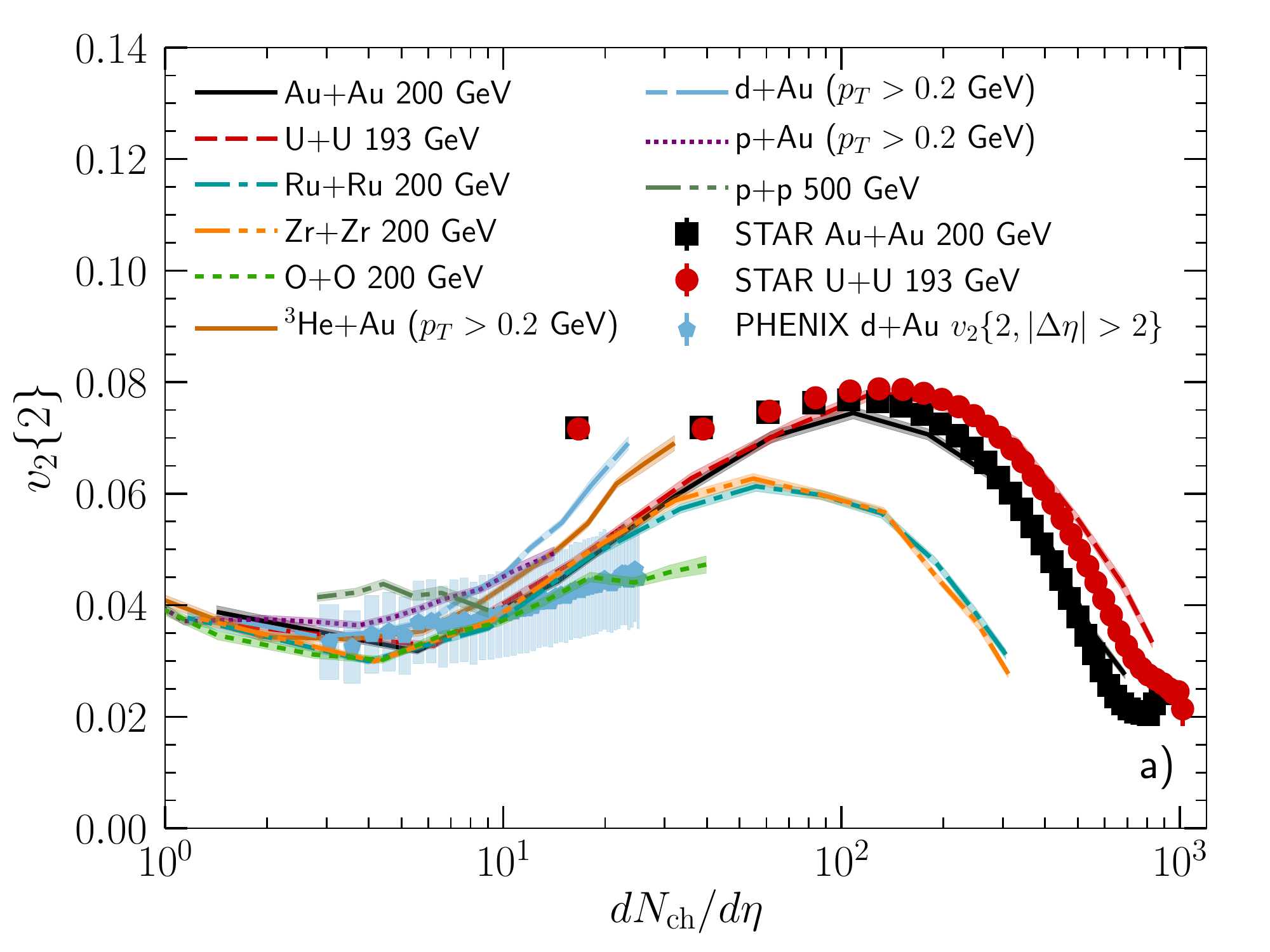}
    \includegraphics[width=0.48\textwidth]{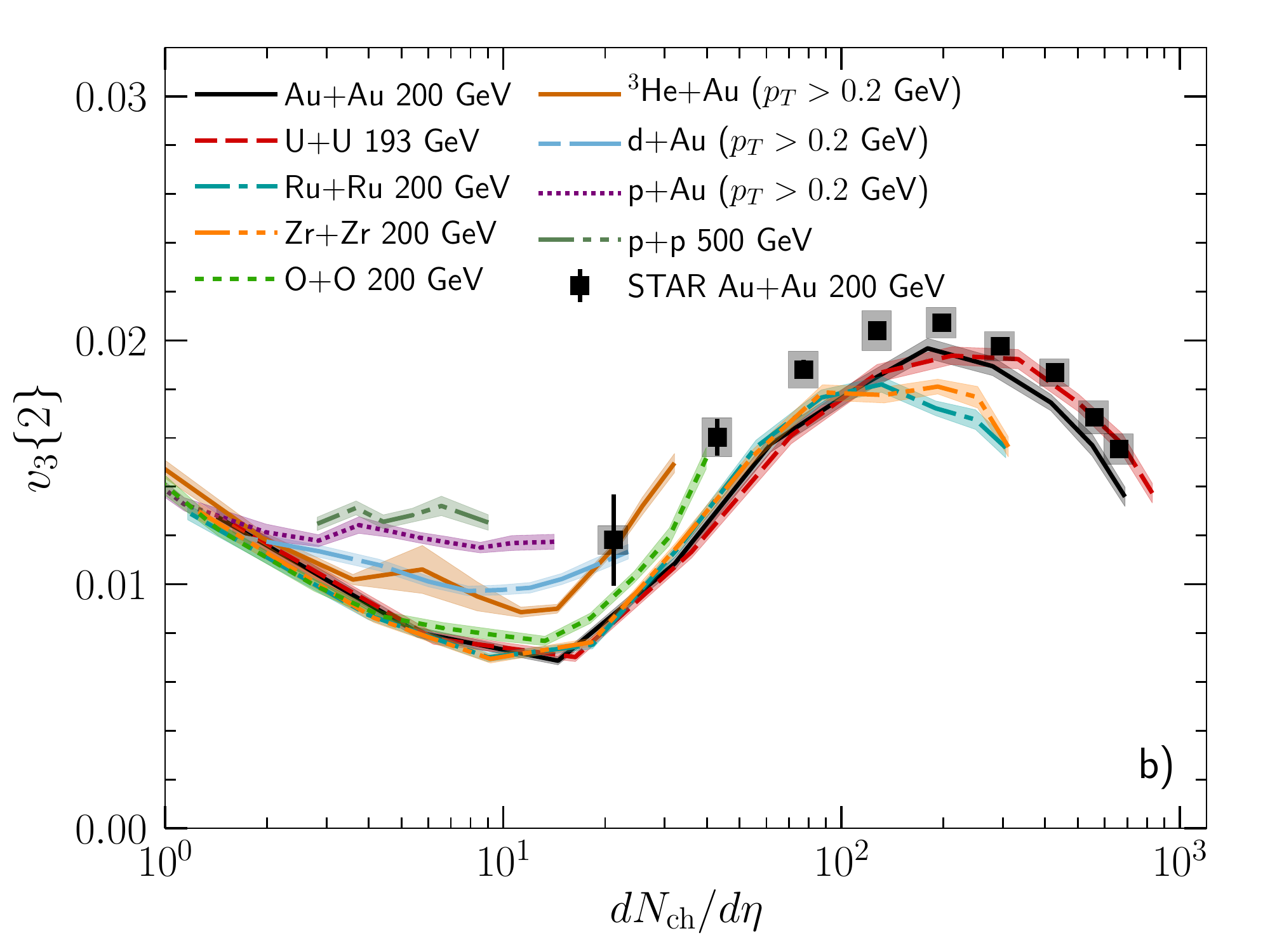}
  \caption{Anisotropy coefficients $v_2\{2\}$ (a) and $v_3\{2\}$ (b) for charged hadrons as functions of multiplicity in various collision systems at RHIC, compared to experimental data from the STAR \cite{Adamczyk:2015obl,Adam:2019woz} and PHENIX \cite{Aidala:2017ajz} Collaborations. \label{fig:v22RHIC}}
\end{figure}

For p+p collisions, we clearly miss both the experimentally observed magnitude, as well as the trend of the centrality dependence. Our result decreases with increasing multiplicity, while the experimental result shows a slight increase. It would not be too surprising if our model missed important physics as the system becomes very small and the multiplicity very low, as is the case in p+p collisions, even though the initial $T^{\mu\nu}$ does include initial state momentum anisotropies from the color glass condensate, as discussed in detail in \cite{Schenke:2019pmk}.

Studying the $v_2\{2\}$ in the range of multiplicity where we have results from all systems, we see that the larger the system, the larger the $v_2\{2\}$. This trend is also seen in the experimental data, and is opposite to the trend seen for the mean transverse momentum, shown in Fig.\,\ref{fig:chmeanpT}.

Results for $v_3\{2\}$ are shown in Fig.\,\ref{fig:v2v3} b) and the comparison with experimental data resembles that for $v_2\{2\}$ quite closely. Agreement is best for more central large systems. We can clearly see that the $v_3\{2\}$ is rather insensitive to the system's average geometry, as it is driven solely by fluctuations. Both in p+p and p+Pb collisions, we underestimate the experimental data quite significantly. This could be due to non-flow contributions, as the disagreement between data and experiment is similar in very low multiplicity Pb+Pb collisions. Interestingly, the calculated $v_3\{2\}$ in p+p collisions does not show the same decrease with multiplicity as did the $v_2\{2\}$.

\begin{figure}[tb]
  \centering
  \includegraphics[width=0.48\textwidth]{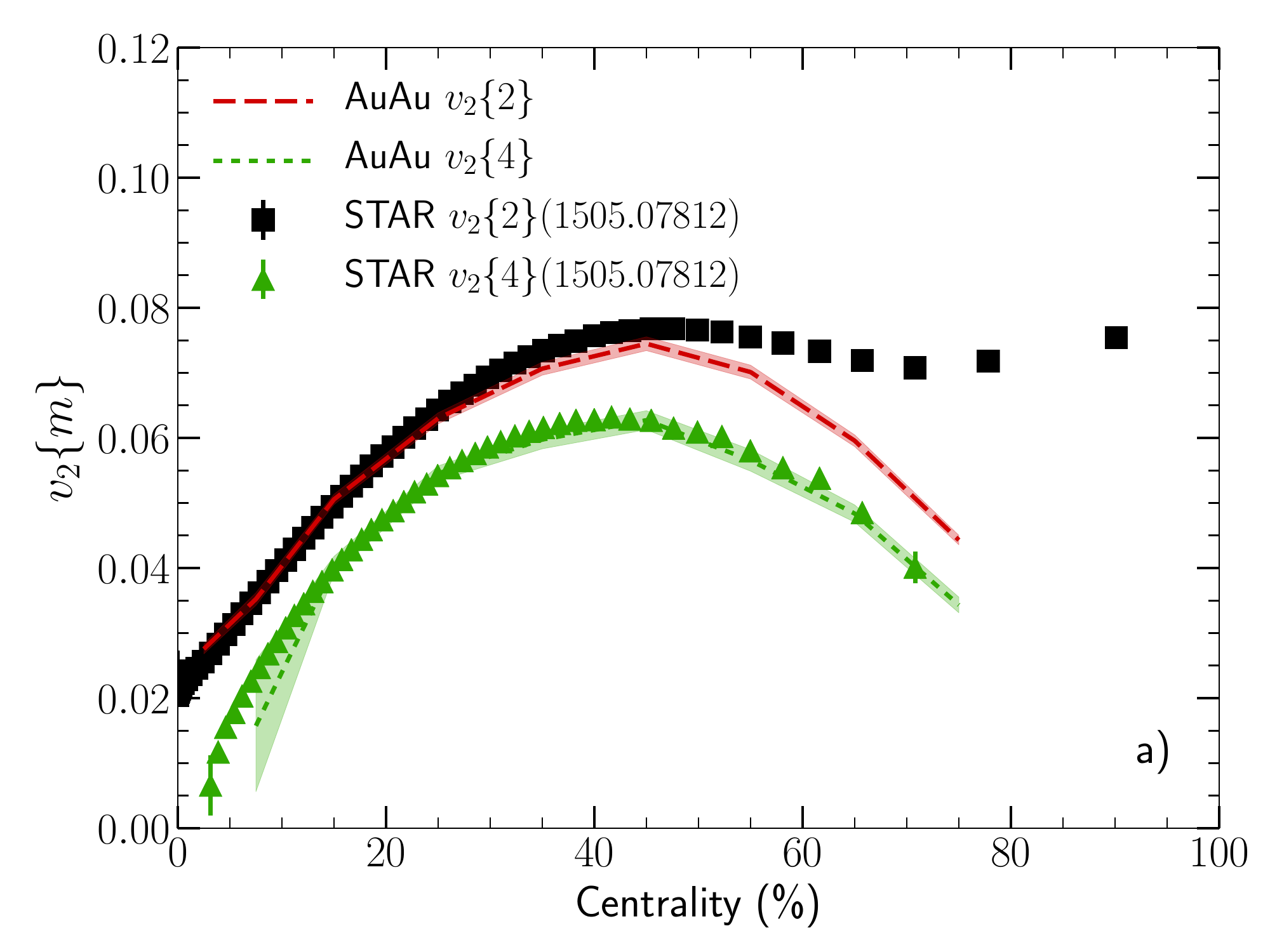}\\\includegraphics[width=0.48\textwidth]{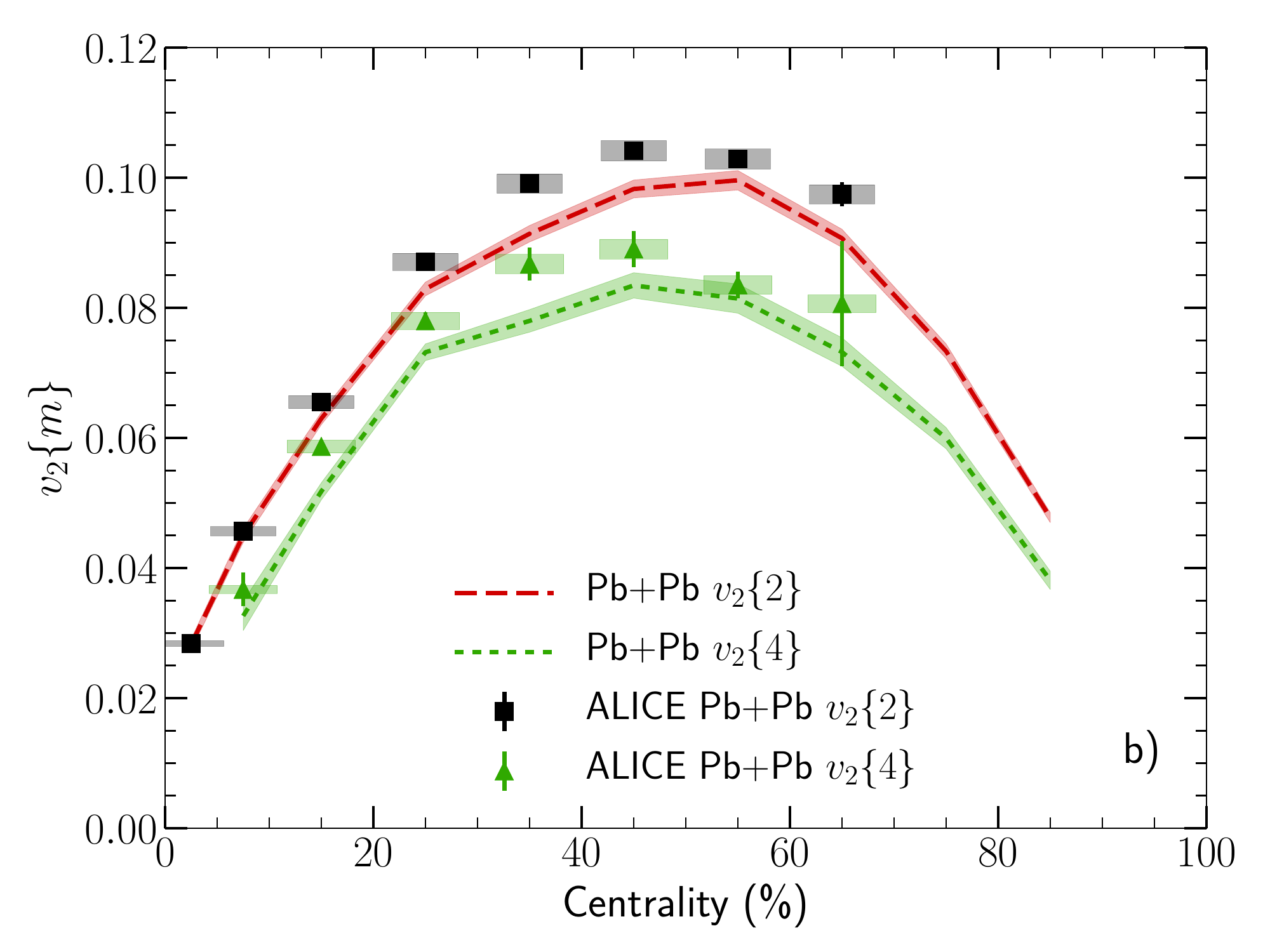}
  \caption{Elliptic anisotropies $v_2\{2\}$ and $v_2\{4\}$ for charged hadrons vs.~centrality in 200 GeV Au+Au collisions at RHIC (a) and 5.02 TeV Pb+Pb collisions at LHC (b). Experimental data from the  STAR \cite{Adamczyk:2015obl} and ALICE \cite{Adam:2016izf} Collaborations. \label{fig:v2m}}
\end{figure}

Fig.\,\ref{fig:v22RHIC} shows the charged hadron $v_2\{2\}$ (a) and $v_3\{2\}$ (b) as a functions of charged hadron multiplicity in a variety of collision systems at RHIC.\footnote{The comparison of $v_2\{2\}$ in Fig.\,\ref{fig:v22RHIC}\,a) (and $v_2\{4\}$ below) vs $dN_{ch}/d\eta$ between STAR data and our calculation is approximate. The efficiency corrected values of $\left< dN_{ch}/d\eta \right>$ from STAR is sensitive to the exact shape of the distribution $P(dN_{ch}/d\eta)$ because of the strong multiplicity dependence of the efficiency. 
We anticipate a small systematic uncertainty in our data-model comparison because of the difference between the IP-Glasma and true experimental $P(dN_{ch}/d\eta)$, published data for which is not available for comparison. 

The values of $dN_{\rm ch}/d\eta$ shown in Fig.\,\ref{fig:v22RHIC}\,b) obtained from the STAR paper~\cite{Adam:2019woz} are estimated using a Monte-Carlo Quark-Glauber model and therefore have some model dependence.} Again we see that triangular flow, which is sensitive to fluctuations only, is very similar in all systems with a slight increase with decreasing system size when comparing at the same $dN_{\rm ch}/d\eta$. For $v_2\{2\}$ this dependence gets more complicated as an interplay of lifetime, radial flow and initial ellipticity takes over. We note that the predicted $v_2\{2\}$ in O+O collisions is below that for Au+Au and U+U collisions, while d+Au and $^3$He+Au collisions exhibit larger $v_2\{2\}$. 
Notably, this is different from our results for LHC energies, shown in Fig.\,\ref{fig:v2v3} a), where $v_2\{2\}$ increases systematically with system size.
Zr+Zr and Ru+Ru collisions have similar $v_2\{2\}$ to that in the larger systems at the smallest $dN_{\rm ch}/d\eta$ but show significantly lower values above $dN_{\rm ch}/d\eta\approx 30$. This is of course expected, as the initial ellipticity has a different $dN_{\rm ch}/d\eta$ dependence for the smaller Ru and Zr nuclei. For $v_3\{2\}$, Zr+Zr and Ru+Ru collisions exhibit much more similar values to Pb+Pb collisions, except at the highest multiplicities.

We show first predictions for 500 GeV p+p collisions and find both $v_2\{2\}$ and $v_3\{2\}$ to be larger than in 200 GeV p+Au collisions (and with that, all other systems), except for the largest multiplicities for $v_2\{2\}$. This motivates measurements of the ridge in high multiplicity events in the upcoming 500 GeV p+p run at RHIC, using the extended pseudorapidity capability of the STAR detector \cite{starbur, fstar}. 

\begin{figure}[tb]
  \centering
    \includegraphics[width=0.48\textwidth]{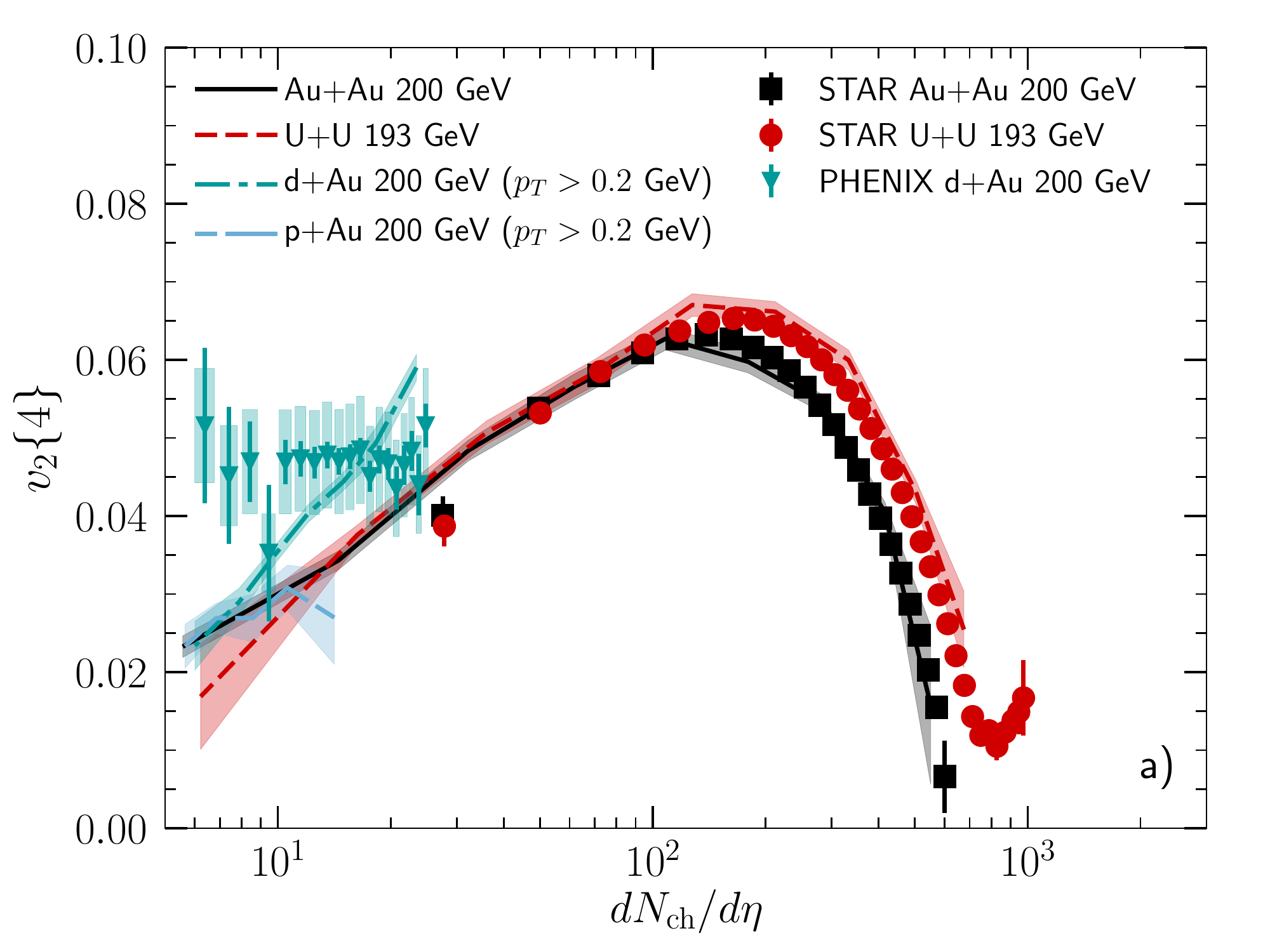}
    \includegraphics[width=0.48\textwidth]{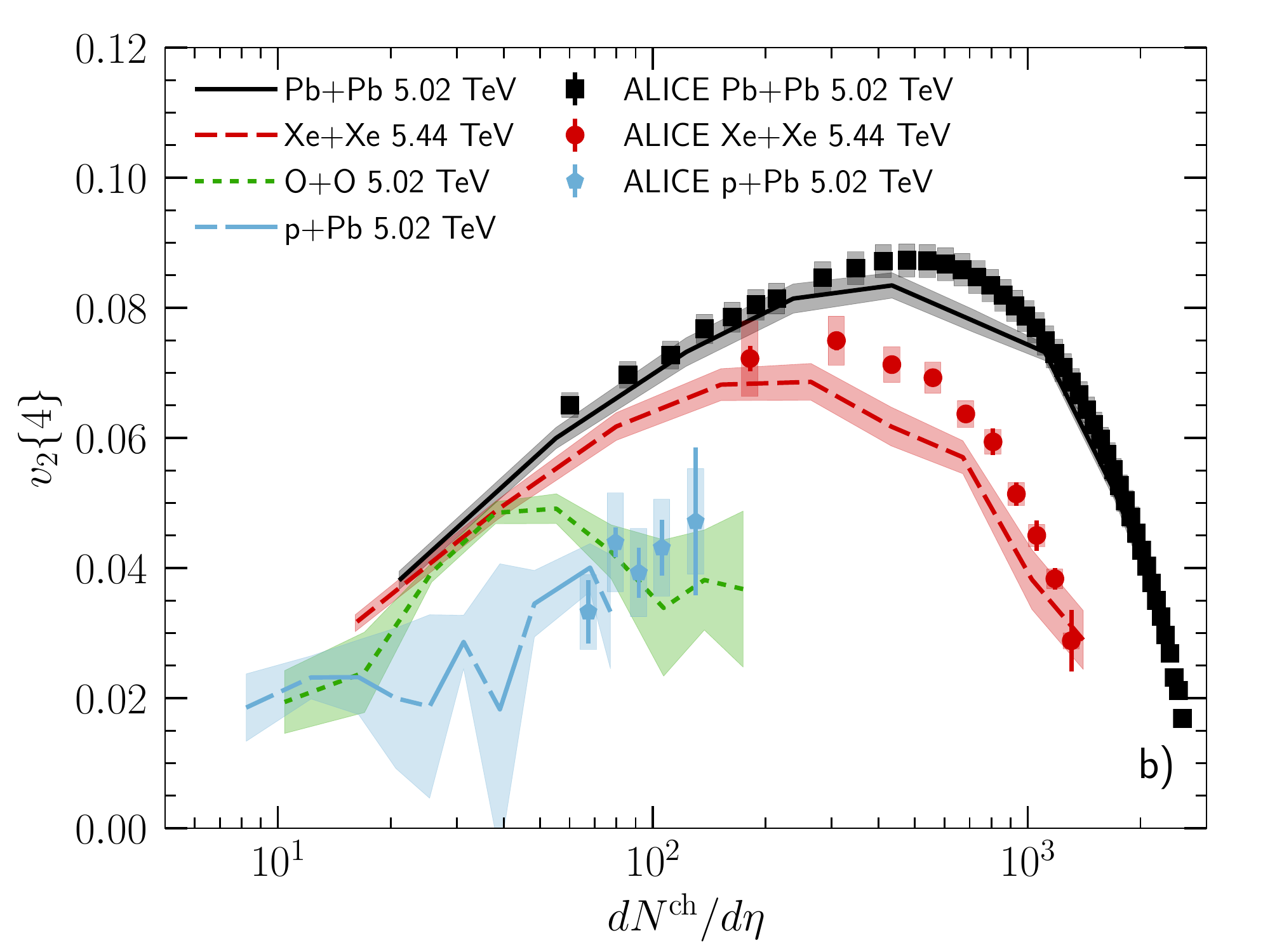}
  \caption{The charged hadron $v_2\{4\}$ in 200 GeV Au+Au, 193 GeV U+U and 200 GeV p+Au and d+Au collisions (a) and 5.02 TeV Pb+Pb, O+O, and p+Pb, and 5.44 TeV Xe+Xe collisions (b) as a function of charged hadron multiplicity, compared to experimental data from the STAR \cite{Adamczyk:2015obl}, PHENIX \cite{Aidala:2017ajz}, and ALICE \cite{Abelev:2014mda,Acharya:2019vdf} Collaborations. \label{fig:v24RHICLHC}}
\end{figure}

As mentioned above, using the correlations of more than two particles can eliminate non-flow and provide a measurement more suitable to be compared to hydrodynamic calculations such as ours.  In Fig.\,\ref{fig:v2m} a), we show both charged hadron $v_2\{2\}$ and $v_2\{4\}$ for 200 GeV Au+Au collisions as functions of centrality and compare to experimental data from the STAR Collaboration \cite{Adamczyk:2015obl}. Indeed, we see that agreement between theory and experiment is better for $v_2\{4\}$ than for $v_2\{2\}$, which is expected as the $\Delta \eta$ gap, another means to eliminate non-flow, employed in this particular $v_2\{2\}$ measurement, is rather small ($|\Delta\eta|>0.1$).

Fig.\,\ref{fig:v2m}\,b) shows the same observables for 5.02 TeV Pb+Pb collisions compared to experimental data from the ALICE Collaboration \cite{Adam:2016izf}. Here, both $v_2\{2\}$ and $v_2\{4\}$ data are underestimated in the more peripheral events. While non-flow could play a role for $v_2\{2\}$, we likely truly underestimate the $v_2\{4\}$ result, which is consistent with us underestimating the mean transverse momentum as shown in Fig.\,\ref{fig:chmeanpT}\,b). Again, we stress that parameters were fixed in 200 GeV Au+Au collisions, making these 5.02 TeV Pb+Pb results predictions. It is likely that a better descriptions of both systems can be found by fine tuning the fluid's transport properties, for example using Bayesian techniques.

\begin{figure}[tb]
  \centering
    \includegraphics[width=0.48\textwidth]{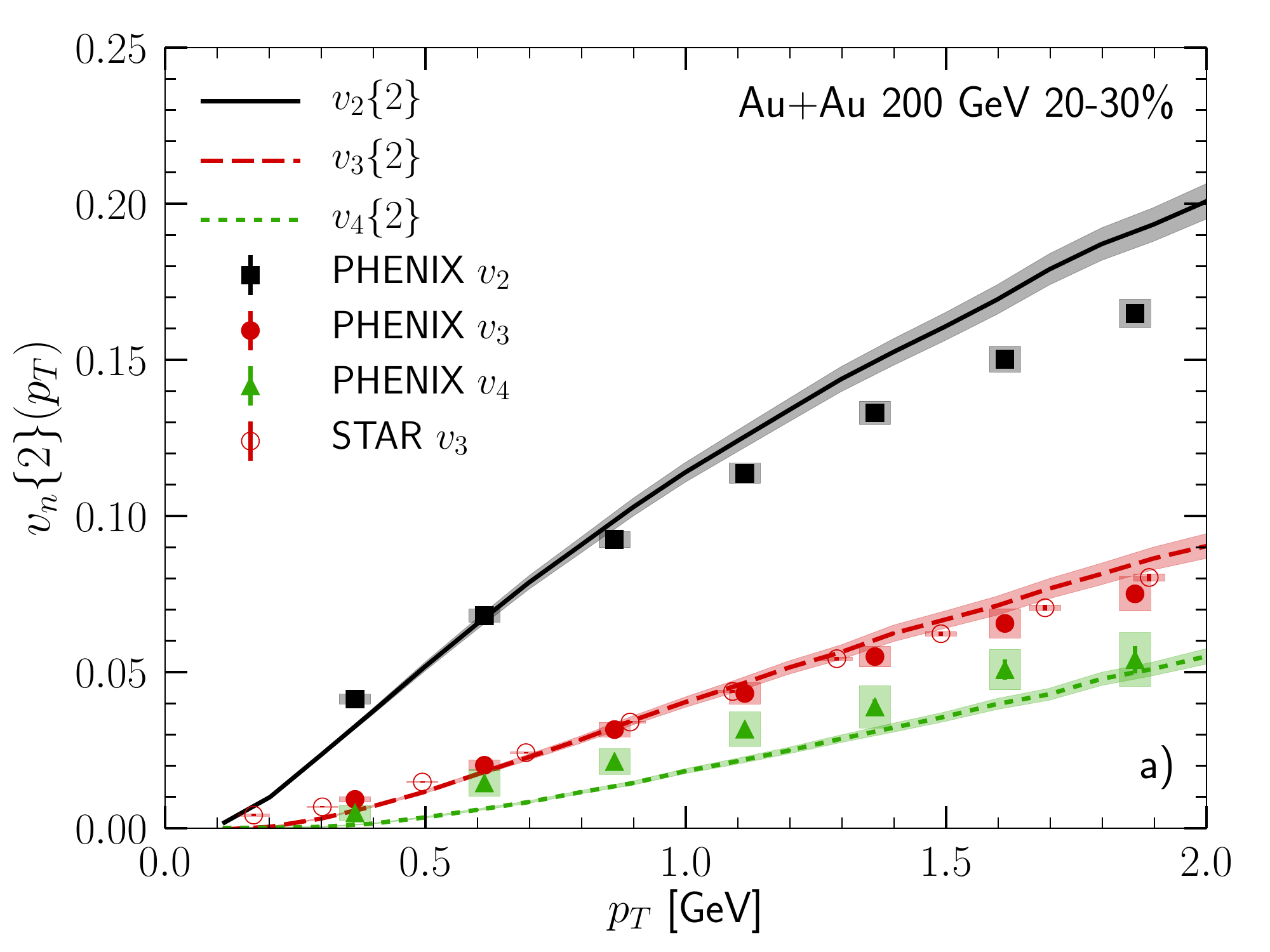}
    \includegraphics[width=0.48\textwidth]{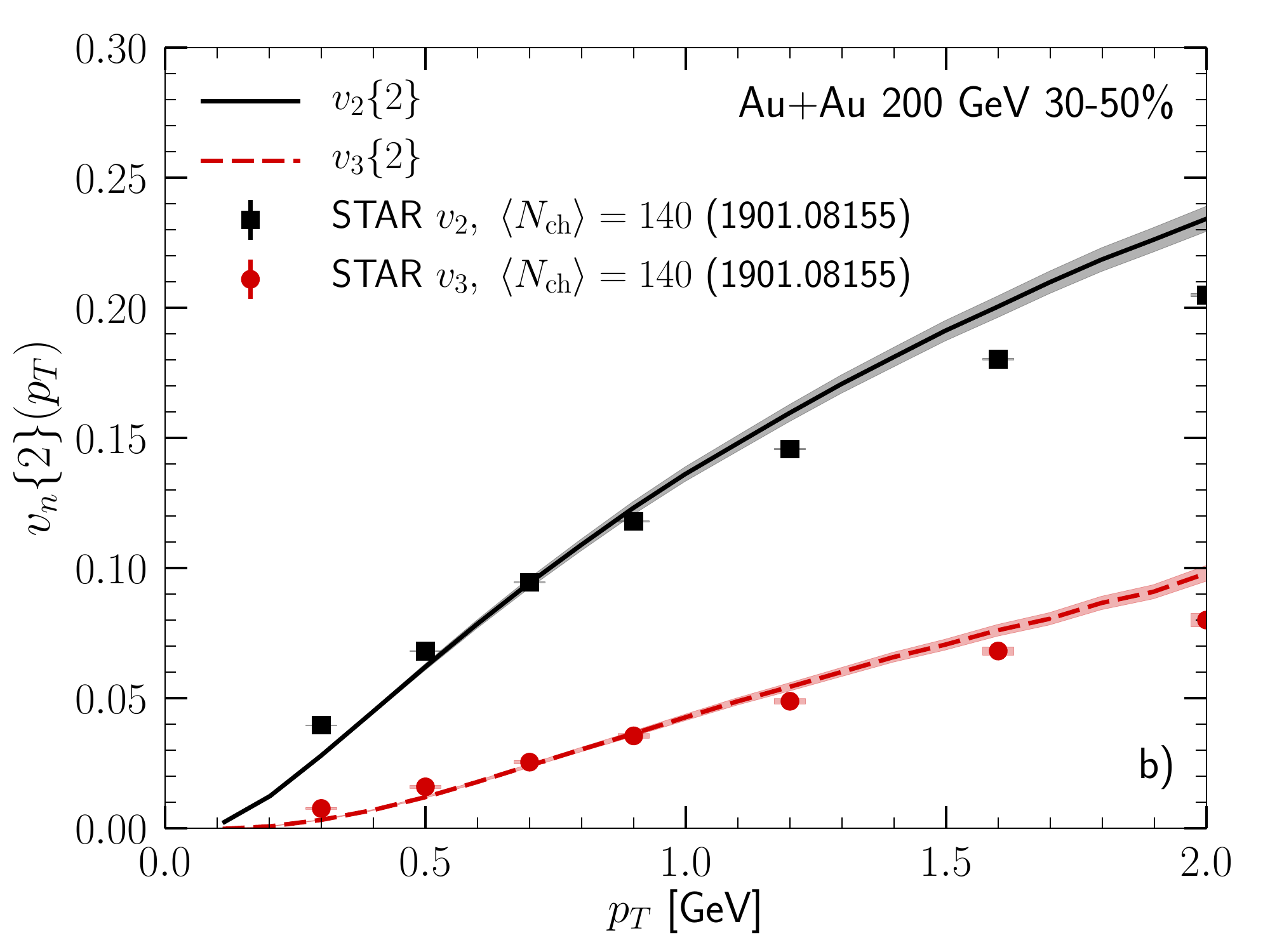}
  \caption{Transverse momentum dependent charged hadron $v_n\{2\}(p_T)$ in 20-30\% central 200 GeV Au+Au collisions compared to experimental data from the PHENIX \cite{Adare:2011tg} and STAR \cite{Adamczyk:2013waa} Collaborations (a) and in 30-50\% central collisions compared to experimental data from the STAR Collaboration \cite{Adam:2019woz} (b). \label{fig:vn2pTRHIC}}
\end{figure}

\begin{figure}[htb]
  \centering
    \includegraphics[width=0.48\textwidth]{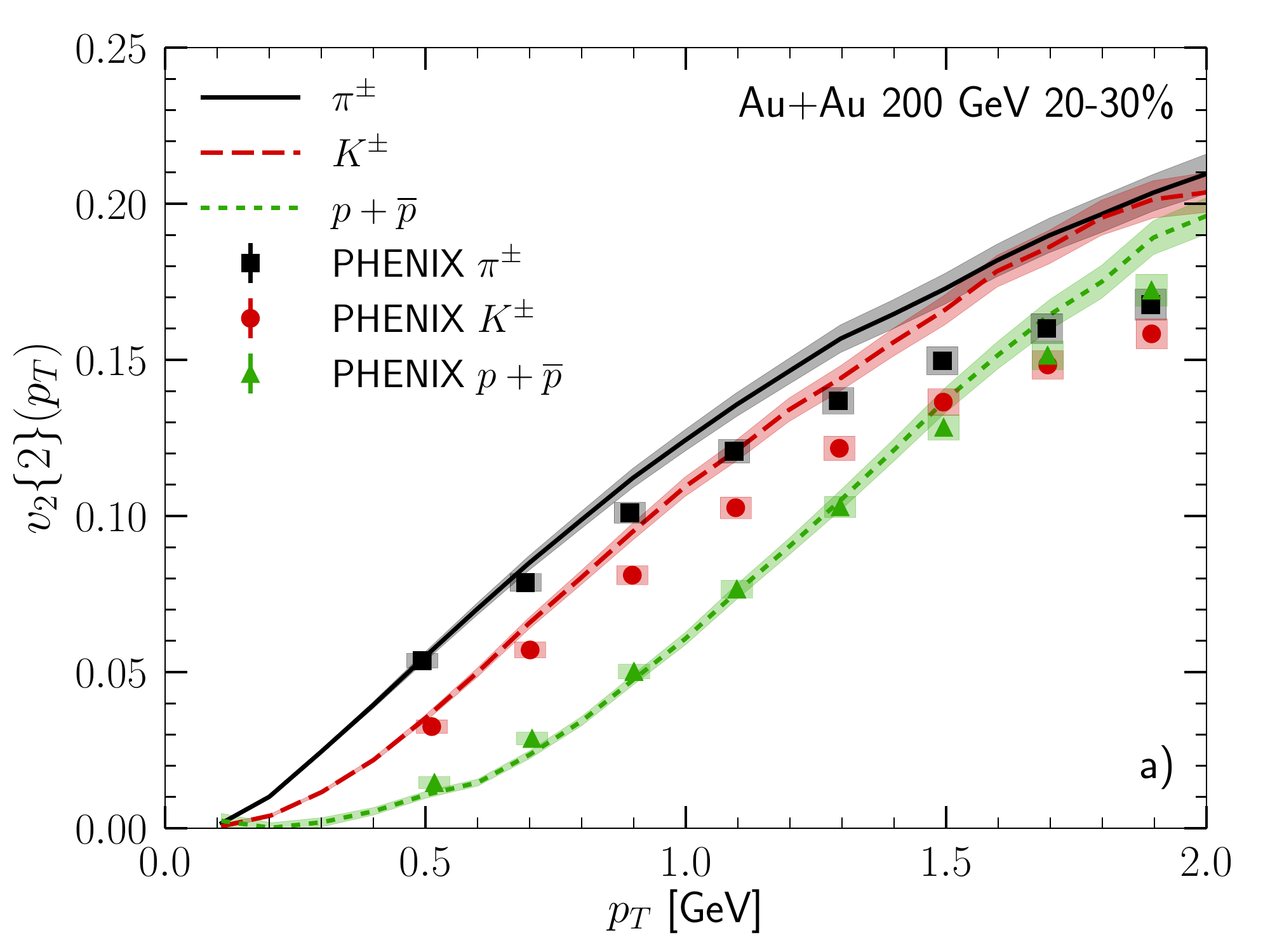}
    \includegraphics[width=0.48\textwidth]{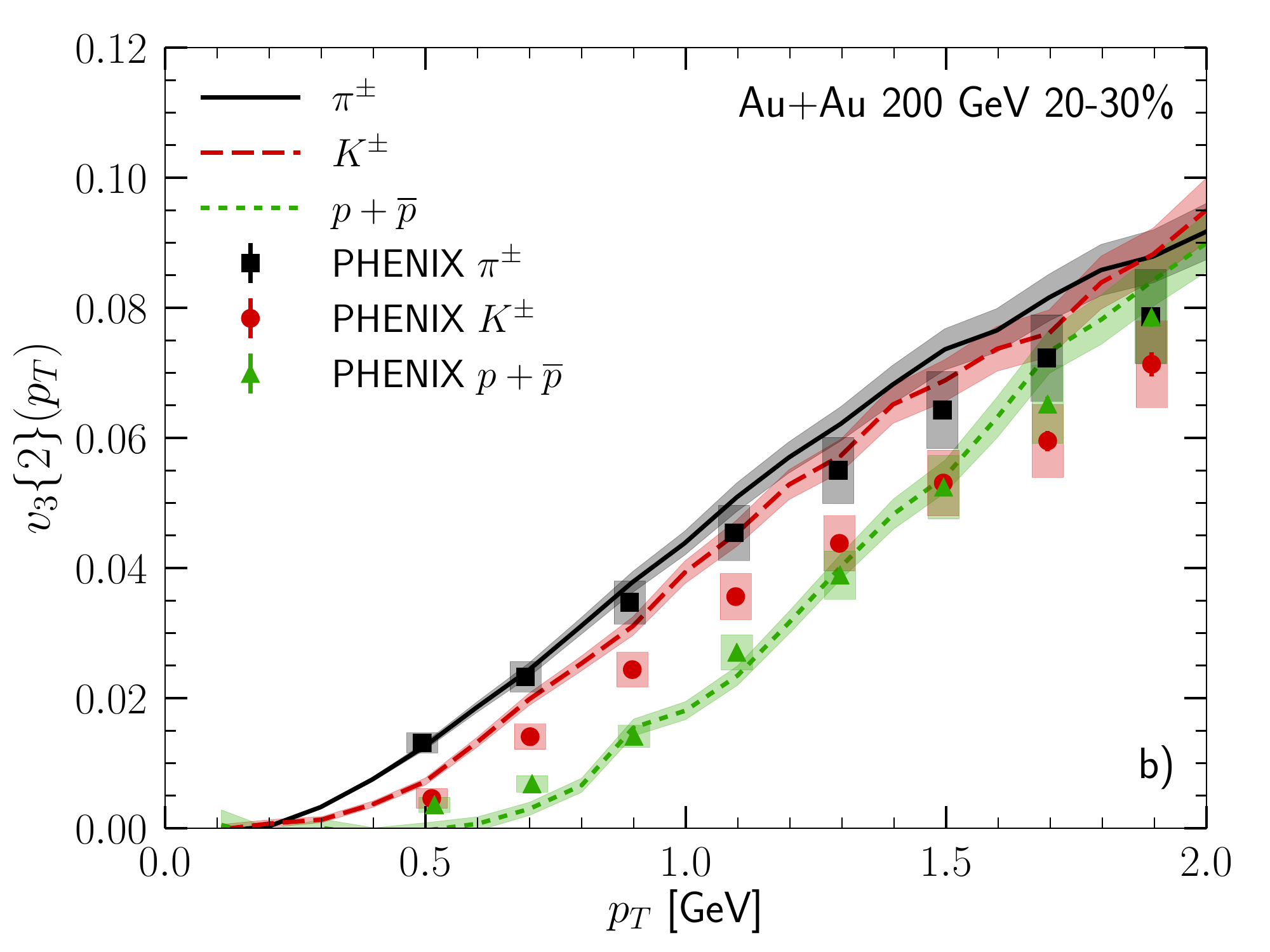}
  \caption{Transverse momentum dependent identified particle $v_2\{2\}(p_T)$ (a) and $v_3\{2\}(p_T)$ (b) in 20-30\% central 200 GeV Au+Au collisions compared to experimental data from the PHENIX Collaboration \cite{Adare:2014kci}. \label{fig:vn2pTRHICPID}}
\end{figure}

We note that the relative difference between $v_2\{2\}$ and $v_2\{4\}$ depends on fluctuations, which are dominated by the initial state. This puts large constraints on the initial state model, as the relative size of $v_2\{2\}$ and $v_2\{4\}$ is not dramatically affected by the details of the medium evolution. We will come back to this point when we study ratios of 4- and 2-, as well as 6- and 4- particle cumulants in Section \ref{sec:ratios}. 

In Fig.\,\ref{fig:v24RHICLHC}\,a), we show the charged hadron $v_2\{4\}$ as a function of multiplicity for 193 GeV U+U, and 200 GeV Au+Au, p+Au, and d+Au collisions, compared to experimental data from the STAR \cite{Adamczyk:2015obl} and PHENIX \cite{Aidala:2017ajz} Collaborations. 
We note that the PHENIX result has some effective lower $p_T$ cut between 0.2 and 0.3 GeV as a result of limited acceptance. In our calculation we used $p_T>0.2\,{\rm GeV}$, the same as for the other systems shown. As for $v_2\{2\}$, we see that the calculated $v_2\{4\}$ in d+Au collisions is larger than that in Au+Au or U+U collisions at the same multiplicity. It is possible that the experimental data confirms this, but the comparison between the d+Au data from PHENIX and Au+Au data from STAR is difficult, because of the possibly different $p_T$ cuts. Given that the p+Au $v_2\{4\}$ is smaller than that for d+Au at the same multiplicity, the large $v_2\{4\}$ in d+Au is likely a result of the larger ellipticity.

\begin{figure}[tb]
  \centering
    \includegraphics[width=0.48\textwidth]{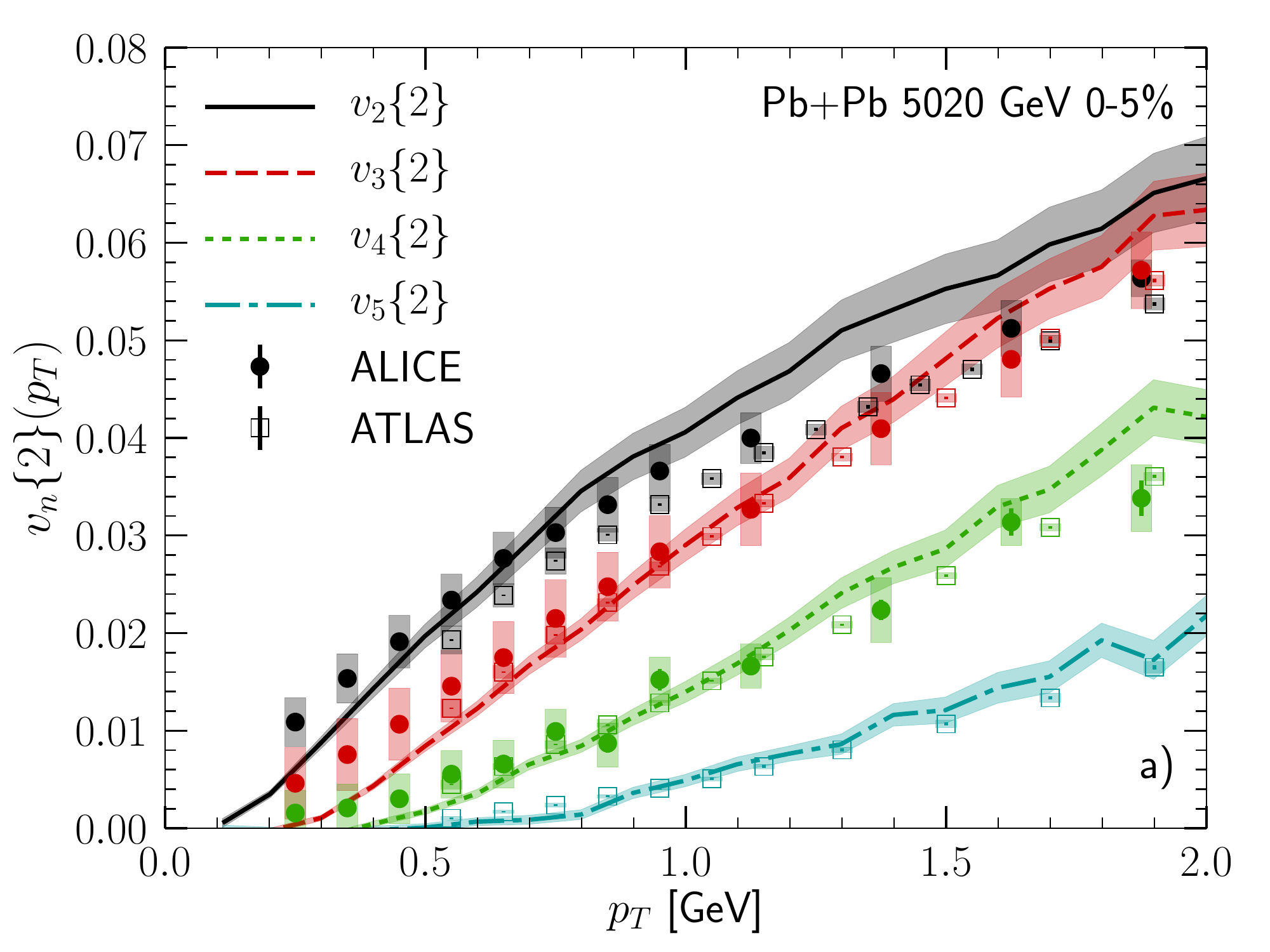}
    \includegraphics[width=0.48\textwidth]{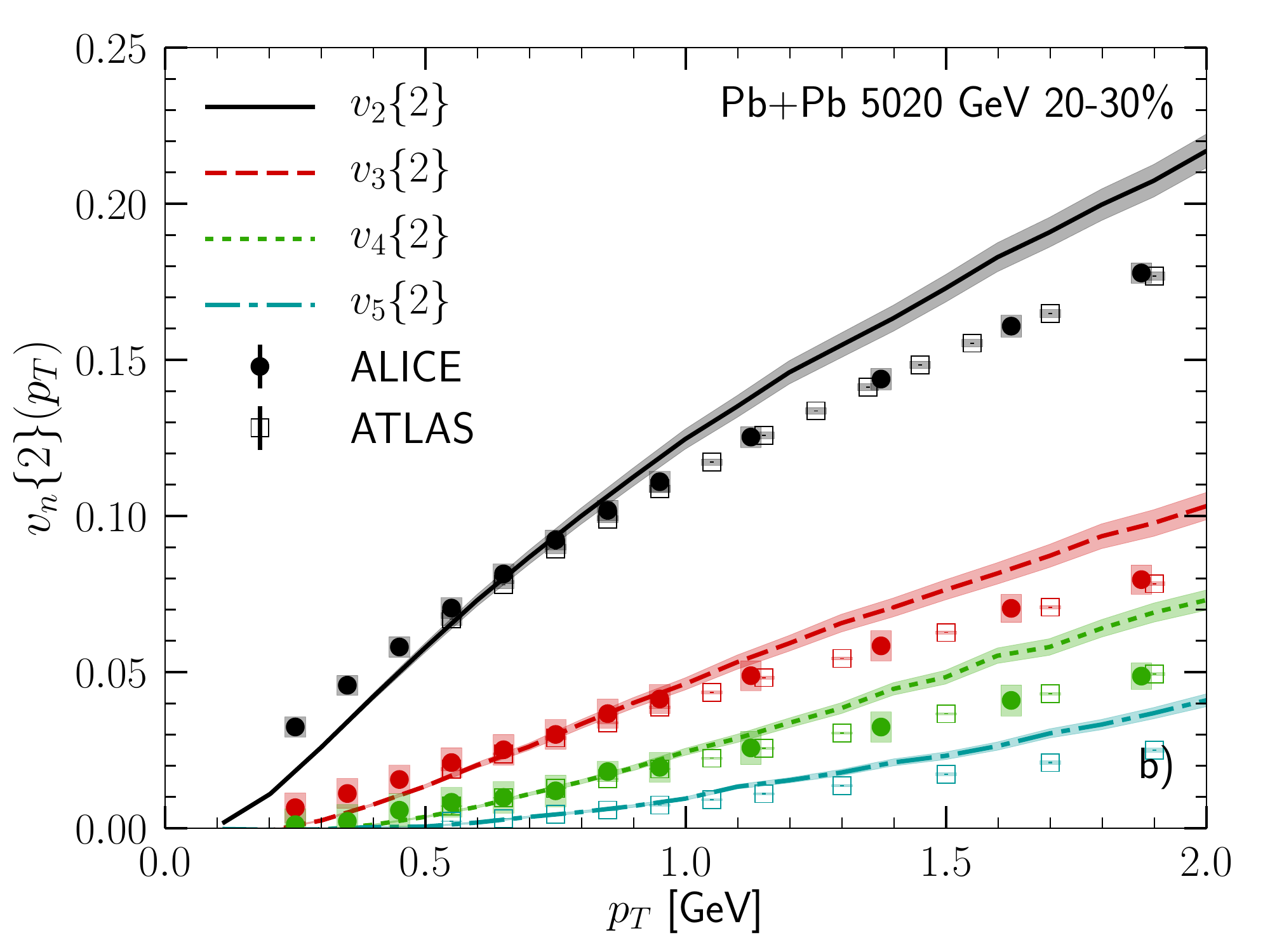}
  \caption{Transverse momentum dependent charged hadron  $v_n\{2\}(p_T)$ for 5.02 TeV Pb+Pb collisions in 0-5\% (a) and 20-30\% (b) centrality classes.  Experimental data from the ALICE \cite{Acharya:2018lmh} and ATLAS \cite{Aaboud:2018ves} Collaborations.\label{fig:vn2pTLHC}}
\end{figure}

\begin{figure}[tb]
  \centering
    \includegraphics[width=0.48\textwidth]{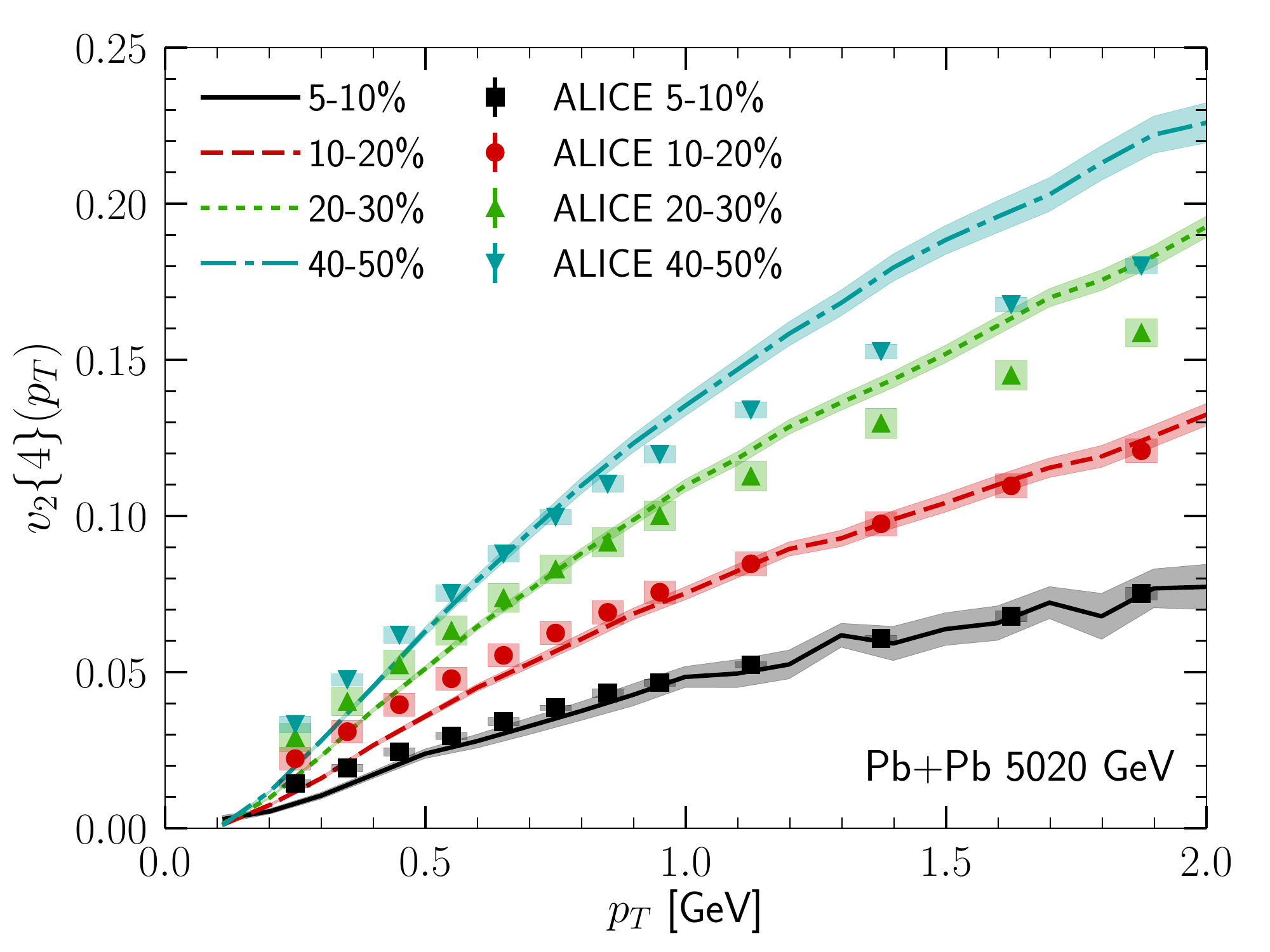}
  \caption{Transverse momentum dependent charged hadron $v_2\{4\}(p_T)$ for four different centrality classes in 5.02 TeV Pb+Pb collisions. Experimental data from the ALICE Collaboration \cite{Acharya:2018lmh}. \label{fig:vn4pTLHC}}
\end{figure}

Fig.\,\ref{fig:v24RHICLHC}\,b) shows charged hadron $v_2\{4\}$ as a function of multiplicity in 5.02 TeV p+Pb, O+O, and Pb+Pb, as well as 5.44 TeV Xe+Xe collisions. As in Fig.\,\ref{fig:v2m}\,b), we see how $v_2\{4\}$ for Pb+Pb collisions underestimates the experimental data for intermediate multiplicities. Xe+Xe collisions show a very similar trend. Because of limited statistics, we cannot reach the highest multiplicities in p+Pb collisions, but where we have results, agreement with the ALICE data is good. O+O collisions lead to results close to p+Pb at the highest multiplicities and close to the larger systems for lower $dN_{\rm ch}/d\eta$.

\subsection{Transverse momentum dependent flow harmonics}

\begin{figure*}[t]
  \centering
    \includegraphics[width=0.95\textwidth]{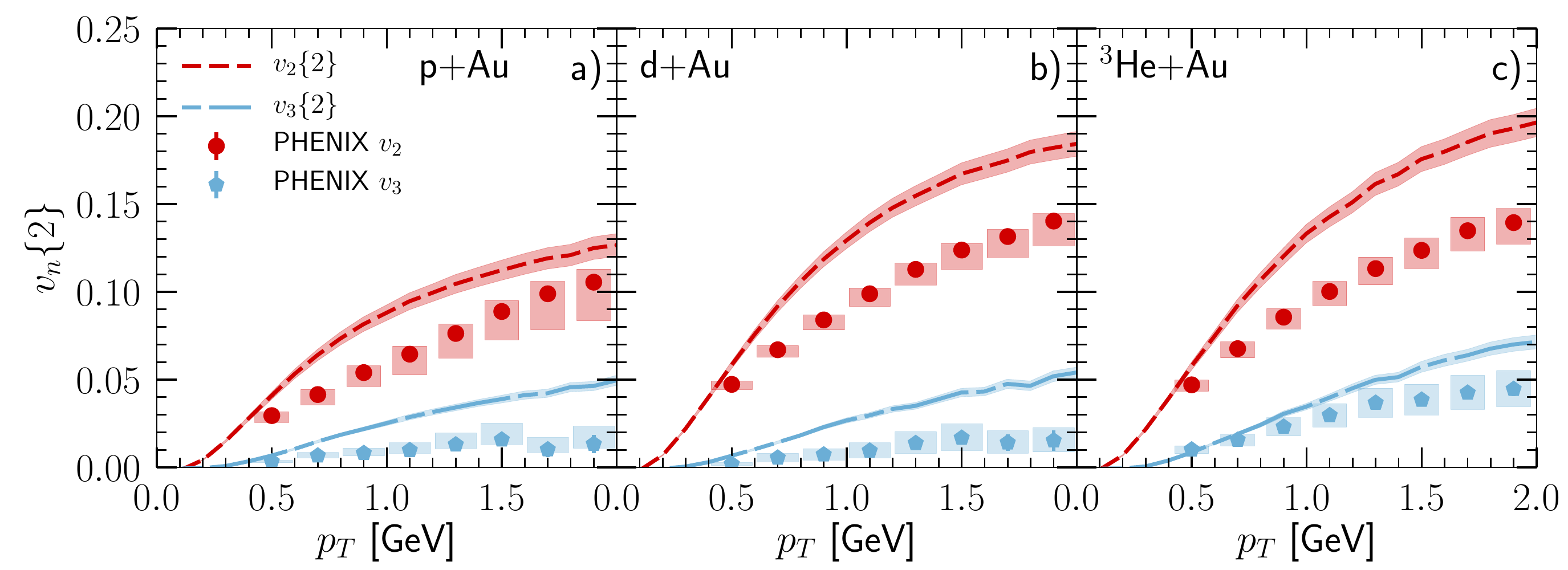}
  \caption{Transverse momentum dependent charged hadron $v_2\{2\}(p_T)$ and $v_3\{2\}(p_T)$ in 0-5\% central collisions for p+Au (a), d+Au (b), and $^3$He+Au (c) collisions at 200 GeV center of mass energy. Experimental data from the PHENIX \cite{PHENIX:2018lia} Collaboration. 
  \label{fig:RHIC_smallsystems}}
\end{figure*}

We now move to flow harmonic measurements differential in transverse momentum, $v_n\{m\}(p_T)$. Their definitions can be found in Appendix \ref{sec:flowAnalysis}. These observables are theoretically less controlled than the previously shown integrated quantities, because viscous corrections to the equilibrium distribution functions on the switching surface (switching to UrQMD) can be large at $p_T>1\,{\rm GeV}$, especially in small systems (see Appendix \ref{sec:deltaf}). In the integrated case, the larger $p_T$ contribute little, making the result more robust. Because we have studied many systems and many centralities, we will present only a selection of results, both for RHIC and LHC energies.

Fig.\,\ref{fig:vn2pTRHIC} a) shows charged hadron $v_2\{2\}$, $v_3\{2\}$, and $v_4\{2\}$ as functions of transverse momentum for 20-30\% central 200 GeV Au+Au collisions, comparing to event-plane $v_n$ measurements from PHENIX \cite{Adare:2011tg} and two particle correlation $v_n\{2\}(p_T)$ measurements from STAR \cite{Adamczyk:2013waa}. The experimental data for $v_2\{2\}(p_T)$ and $v_3\{2\}(p_T)$ is underestimated at low $p_T\lesssim 0.6\,{\rm GeV}$ and overestimated above $p_T\approx 1\,{\rm GeV}$. The $v_4\{2\}(p_T)$ from the PHENIX Collaboration is slightly underestimated by our calculation, seemingly more so than the integrated $v_4\{2\}$ in 20-30\% collisions compared to STAR data (Fig.\,\ref{fig:v2} a)). Fig.\,\ref{fig:vn2pTRHIC} b) shows charged hadron $v_2\{2\}(p_T)$ and $v_3\{2\}(p_T)$ in 30-50\% central 200 GeV Au+Au collisions compared to recent experimental data from the STAR Collaboration \cite{Adam:2019woz}. Agreement with the experimental data is similar to the case of 20-30\% central collisions.

We also show identified particle $v_2\{2\}(p_T)$ and $v_3\{2\}(p_T)$ in 20-30\% central 200 GeV Au+Au collisions in comparison to experimental data for event-plane method $v_n(p_T)$ from the PHENIX Collaboration in Figs.\,\ref{fig:vn2pTRHICPID} a) and b), respectively. While agreement with the experimental data around $p_T\approx 0.5\,{\rm GeV}$ is excellent for pions kaons and protons, only proton $v_n\{2\}(p_T)$ are very well described also for larger $p_T$. Pion and kaon $v_n\{2\}(p_T)$ is overestimated for $p_T \gtrsim 0.7\,{\rm GeV}$.

\begin{figure}[ht]
  \centering
    \includegraphics[width=0.48\textwidth]{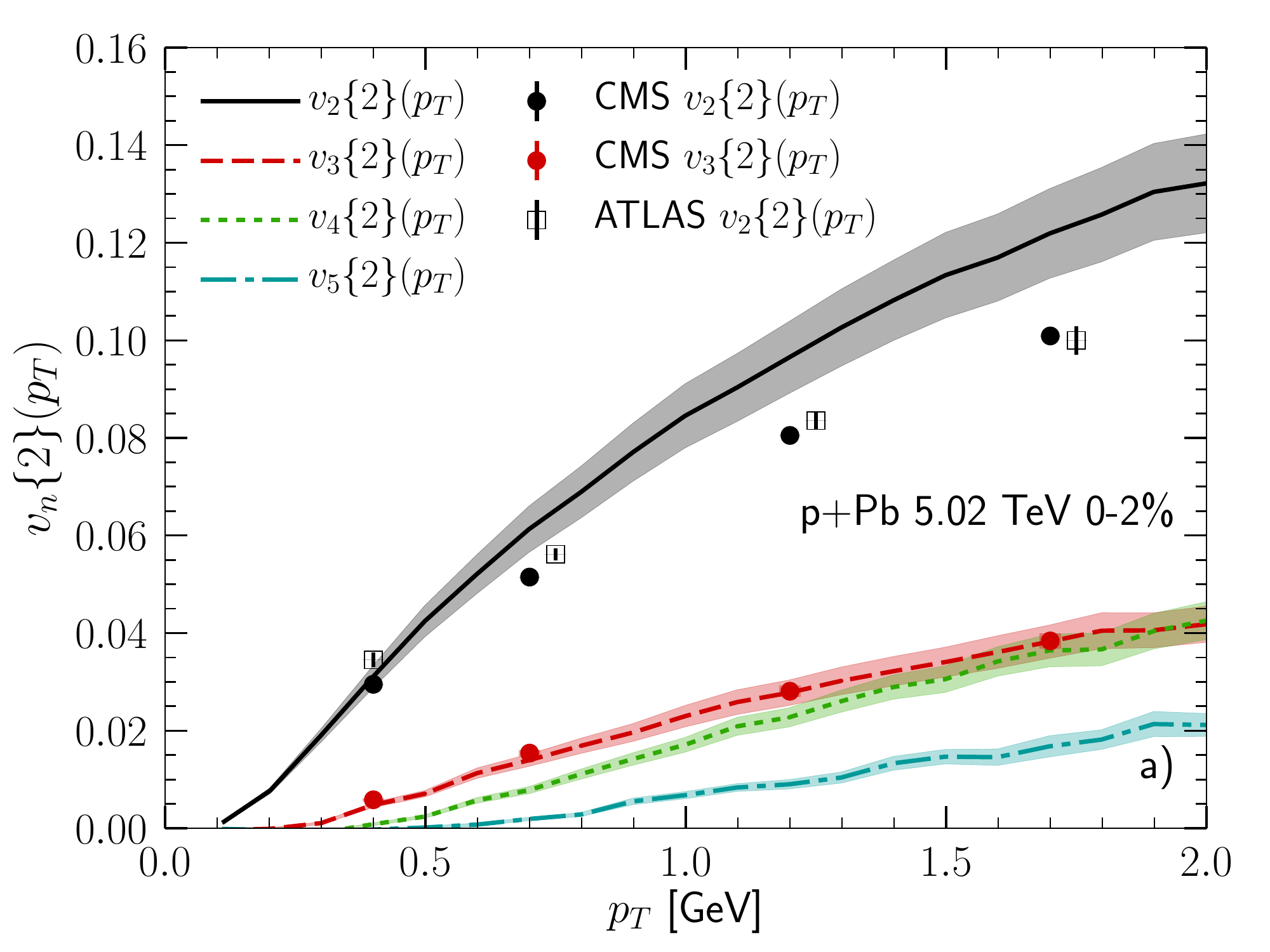}
    \includegraphics[width=0.48\textwidth]{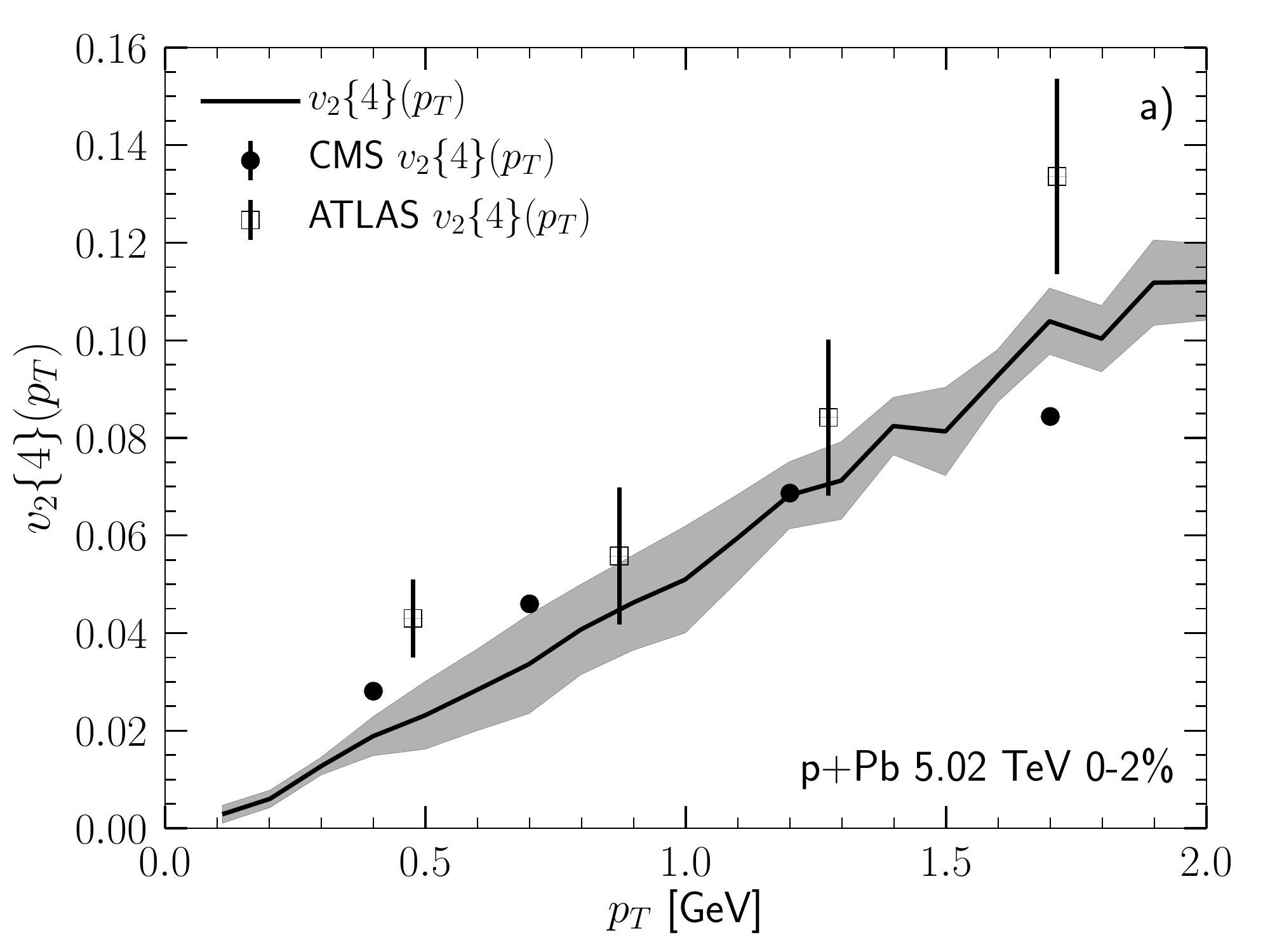}
  \caption{Transverse momentum dependent charged hadron  $v_n\{2\}(p_T)$ (a) and $v_2\{4\}(p_T)$ (b) for 0-2\%  central 5.02 TeV p+Pb collisions.  Experimental data from the CMS \cite{Chatrchyan:2013nka} and ATLAS \cite{Aad:2013fja} Collaborations.\label{fig:pPbvn2vn4pTLHC}}
\end{figure}

\begin{figure}[htb]
  \centering
    \includegraphics[width=0.48\textwidth]{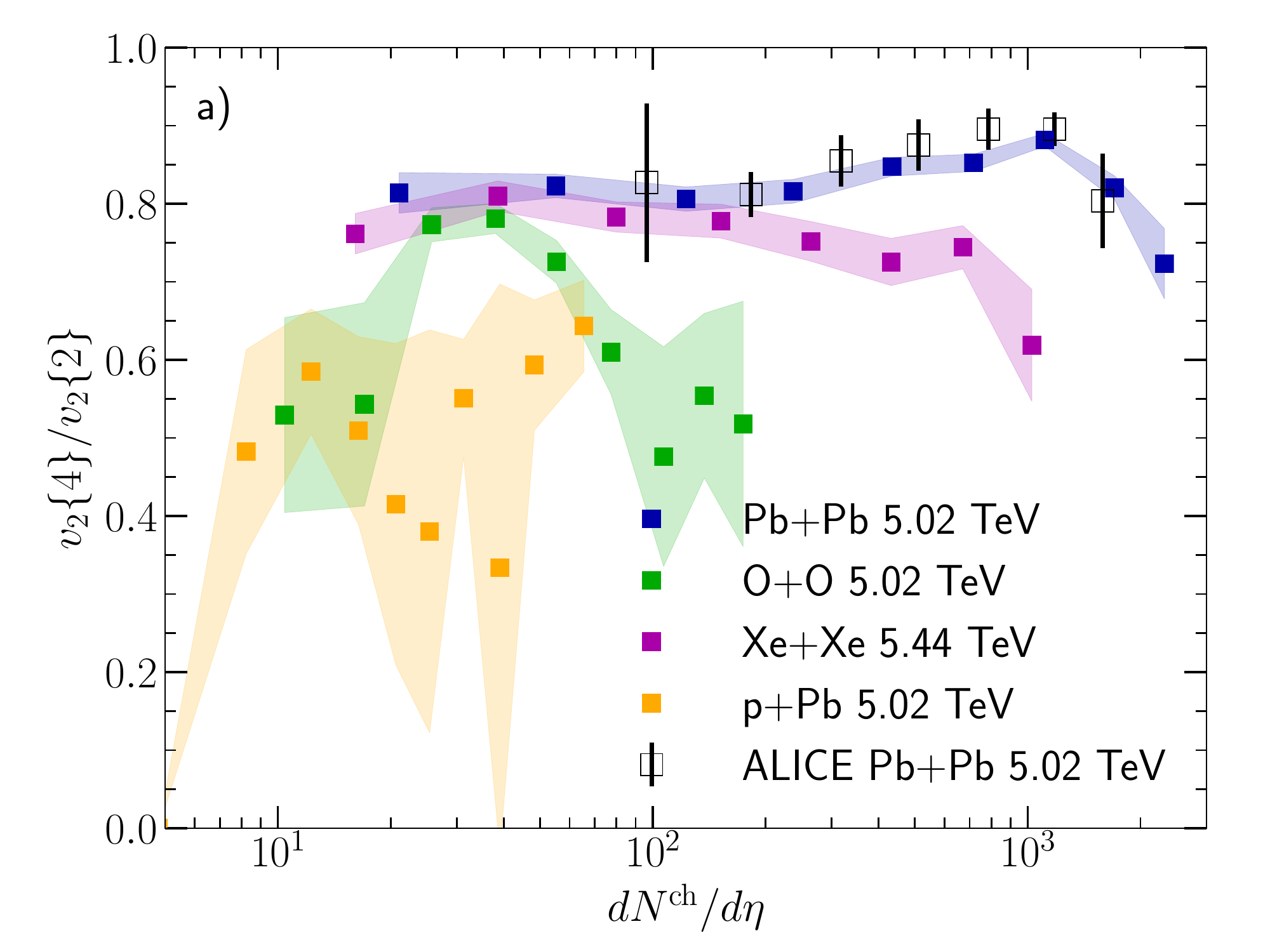}
    \includegraphics[width=0.48\textwidth]{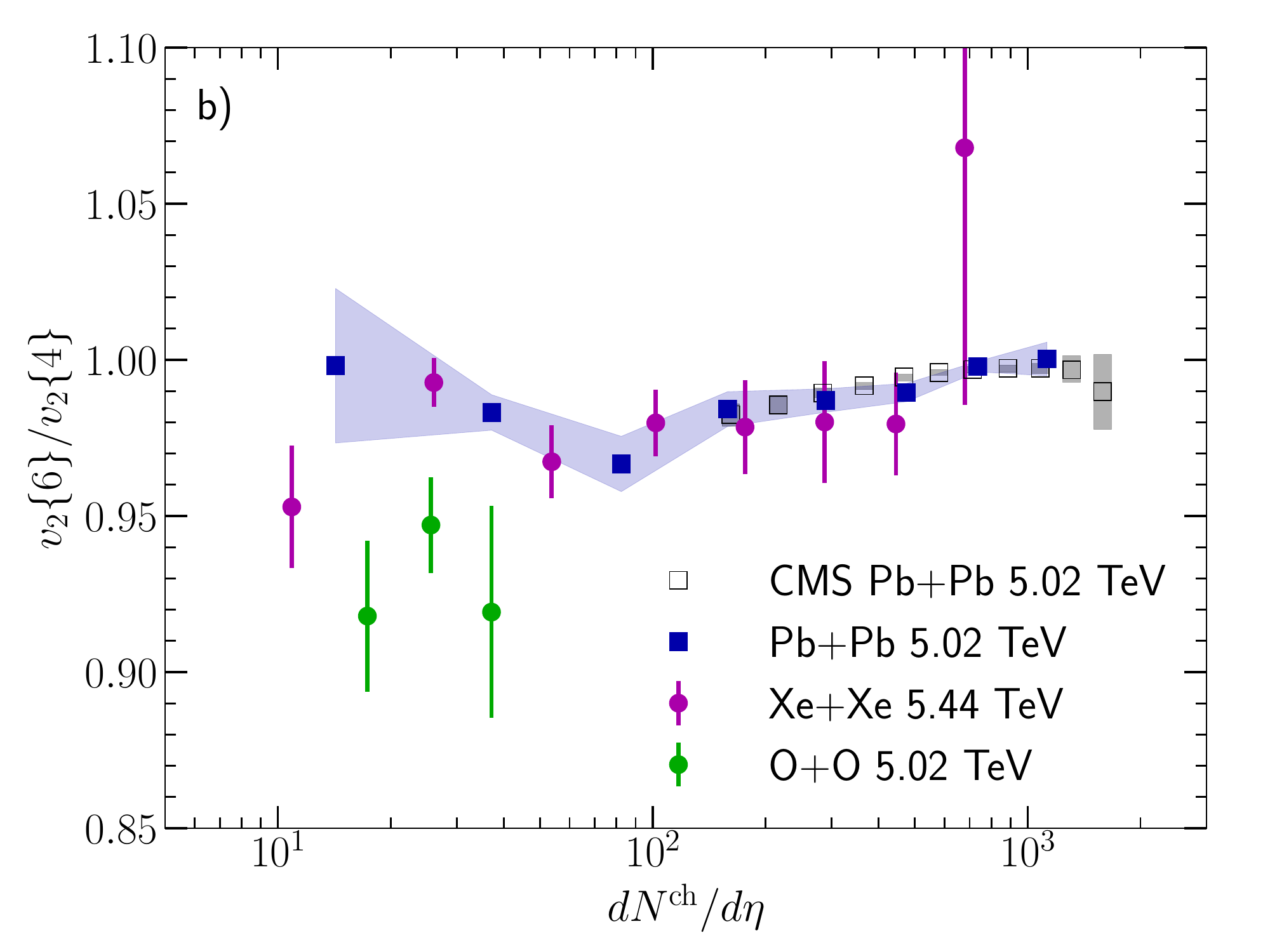}
  \caption{Ratio of fourth and second order charged hadron elliptic flow cumulants in 5.02 TeV O+O and Pb+Pb, 5.44 TeV Xe+Xe, and 5.02 TeV p+Pb collisions as functions of charged hadron multiplicity compared to experimental data from the ALICE Collaboration \cite{Adam:2016izf} (a). Ratio of the sixth and fourth order cumulants compared to experimental data from the CMS Collaboration \cite{Sirunyan:2017fts} (b). \label{fig:v2ratios}}
\end{figure}

In Fig.\,\ref{fig:vn2pTLHC} we show results for charged hadron $v_n\{2\}(p_T)$ in 0-5\% (a) and 20-30\% (b) central 5.02 TeV Pb+Pb collisions. We include results for $n=2-5$ and compare to experimental data from the ALICE \cite{Acharya:2018lmh} and ATLAS \cite{Aaboud:2018ves} Collaborations. In particular in the most central events, agreement with the experimental data for all $n$ is good for most $p_T$ values studied. At the lowest $p_T$ our results tend to underpredict the data, while at larger $p_T$ they slightly overestimate the data, similar to what we observed at RHIC energies. For the 20-30\% central events, this trend is stronger. 

Fig.\,\ref{fig:vn4pTLHC} shows the charged hadron $v_2\{4\}(p_T)$ in 5.02 TeV Pb+Pb collisions in four different centrality classes. Best agreement with the experimental data from the ALICE Collaboration \cite{Acharya:2018lmh} is achieved in the most central collisions. For the more peripheral events, we see the same trend as for $v_2\{2\}(p_T)$, with our result being lower than the experimental data at low $p_T$ and higher at larger $p_T$. This is consistent with our results for the integrated $v_2\{4\}$ in Pb+Pb collisions (see Fig.\,\ref{fig:v2m}), which is dominated by the low $p_T$ region.

Because some parameters, especially the parametrization of the bulk viscosity over entropy density, have changed from our previously presented result \cite{Schenke:2019pmk}, we show the transverse momentum dependent $v_n\{2\}(p_T)$ in 0-5\% central p+Au, d+Au, and $^3$He+Au collisions compared to experimental data from the PHENIX \cite{PHENIX:2018lia} 
Collaboration in Fig.\,\ref{fig:RHIC_smallsystems}. We only find a minor change in our result from what was presented in \cite{Schenke:2019pmk}, and again
for $p_T\gtrsim 0.5\,{\rm GeV}$ the experimental data for $v_n\{2\}(p_T)$ is overestimated in our model.
We note that preliminary STAR data for $v_3\{2\}(p_T)$  \cite{Lacey:2020ime} (not shown here) is well described for $p_T\gtrsim 0.5\,{\rm GeV}$, but underestimated for lower $p_T$. Qualitatively, our calculation shows a weaker system dependence for $v_3\{2\}(p_T)$ than the PHENIX data. It was discussed in \cite{Lacey:2020ime} that initial conditions with subnucleonic fluctuations, such as ours, lead to a weaker system dependence for the eccentricities than initial conditions with only nucleon degrees of freedom, which can describe the PHENIX data rather well \cite{PHENIX:2018lia}.

In Fig.\,\ref{fig:pPbvn2vn4pTLHC}\,a) we show the transverse momentum dependent charged hadron $v_n\{2\}(p_T)$ for $n=1\dots 5$ compared to experimental data from the CMS \cite{Chatrchyan:2013nka} and ATLAS \cite{Aad:2013fja} Collaborations in the case of $n=2$ and $n=3$ (CMS). The comparison of our result for $v_2\{2\}(p_T)$ with the experimental result resembles that for p+Au collisions shown in Fig.\,\ref{fig:RHIC_smallsystems}\,a), as we overestimate the experimental data for $p_T\gtrsim 0.5\,{\rm GeV}$. On the other hand, our result for $v_3\{2\}(p_T)$ agrees very well with the experimental data from the CMS Collaboration. 

Fig.\,\ref{fig:pPbvn2vn4pTLHC}\,b) shows $v_2\{4\}(p_T)$, also compared to experimental data from the CMS \cite{Chatrchyan:2013nka} and ATLAS \cite{Aad:2013fja} Collaborations. At low $p_T<1\,{\rm GeV}$, we underestimate the experimental data, while agreement is reasonable at larger $p_T$, given the uncertainty, also coming from the disagreement between CMS and ATLAS results at large $p_T$. 

\subsection{Ratios of multi-particle cumulants}\label{sec:ratios}

We close our discussion of bulk observables by showing two measures of flow fluctuations, namely ratios of 4th and 2nd order and 6th and 4th order elliptic flow cumulants in Fig.\,\ref{fig:v2ratios} for 5.02 TeV Pb+Pb and O+O, and 5.44 TeV Xe+Xe collisions, and in the case of $v_2\{4\}/v_2\{2\}$ also for 5.02 TeV p+Pb collisions. The ratio $v_2\{4\}/v_2\{2\}$ is a measure of fluctuations of $v_2$, which is mainly driven by fluctuations in the initial eccentricity, a feature of the initial condition. A deviation of $v_2\{6\}/v_2\{4\}$ from one is an indication of non-Gaussianity in the $v_2$ distribution \cite{Giacalone:2016eyu}. The excellent agreement of our results for $v_2\{4\}/v_2\{2\}$ in Pb+Pb collisions with experimental data from the ALICE Collaboration \cite{Adam:2016izf} and $v_2\{6\}/v_2\{4\}$ in Pb+Pb collisions with data from the CMS Collaboration \cite{Sirunyan:2017fts} is an indication that the eccentricity fluctuations in the IP-Glasma model describe those in nature very closely, as has been previously demonstrated in \cite{Gale:2012rq}.

Our result for $v_2\{4\}/v_2\{2\}$ in 5.02 TeV p+Pb collisions has large statistical errors, but is significantly lower than that in Xe+Xe or Pb+Pb collisions at the same multiplicity, indicating increased fluctuations in the small system. For O+O collisions, errors are also large, but one can see that at the highest multiplicities $v_2\{4\}/v_2\{2\}$ in O+O collisions is significantly smaller than in Xe+Xe and Pb+Pb collisions, and closer at lower multiplicities. For $v_2\{6\}/v_2\{4\}$, we also find smaller values in O+O than in the larger systems at the same multiplicity. 

\begin{figure}[htb]
  \centering
    \includegraphics[width=0.48\textwidth]{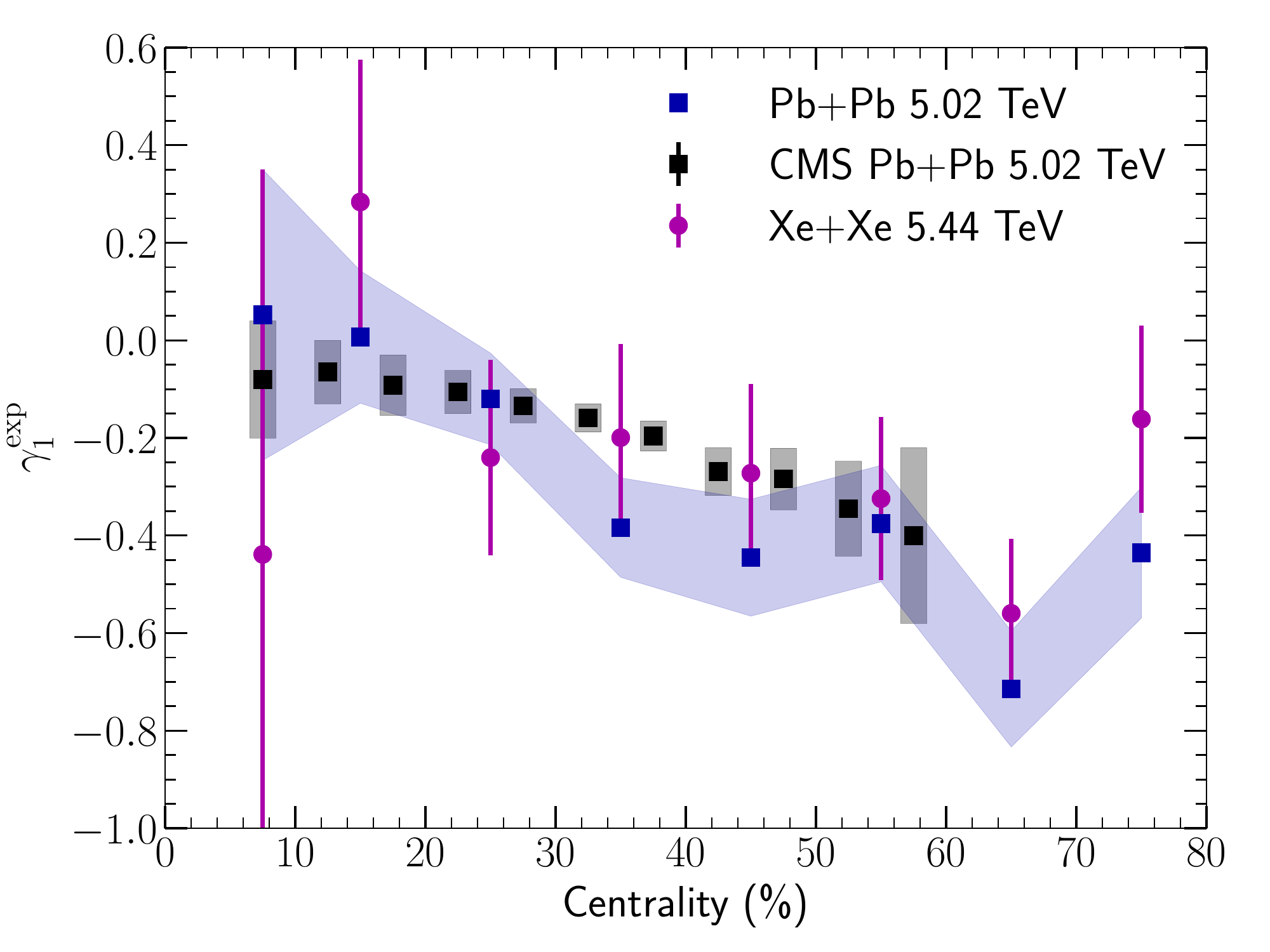}
  \caption{The standardized skewness, obtained from cumulants, $\gamma_1^{\rm exp}$, in 5.02 TeV Pb+Pb collisions and 5.44 TeV Xe+Xe collisions as a function of centrality. Experimental data from the CMS Collaboration \cite{Sirunyan:2017fts}. \label{fig:gamma1}}
\end{figure}

Finally, we compute the standardized skewness of the elliptic flow fluctuations, which was shown to be the main effect that lifts the degeneracy between $v_2\{4\}$ and $v_2\{6\}$ \cite{Giacalone:2016eyu}. It can be approximately obtained from cumulants as
\begin{equation}
    \gamma_1^{\rm exp} = -6\sqrt{2}\, v_2\{4\}^2\frac{v_2\{4\}-v_2\{6\}}{(v_2\{2\}^2-v_2\{4\}^2)^{3/2}}\,.
\end{equation}
We show the result for $\gamma_1^{\rm exp}$ as a function of centrality in 5.02 TeV Pb+Pb collisions compared to experimental data from the CMS Collaboration \cite{Sirunyan:2017fts} in Fig.\,\ref{fig:gamma1}. Considering the relatively large statistical errors, agreement with the experimental data is very good, except for the centrality range 30-50\% where our model slightly underestimates the experimental data. We also show our prediction for 5.44 TeV Xe+Xe collisions in the same figure.

\newpage 

\section{Conclusions}\label{sec:conclusions}
We have presented a detailed discussion of the hybrid model consisting of the IP-Glasma initial state model, \textsc{Music} viscous fluid dynamics, and UrQMD hadronic transport, and discussed its results for bulk and multi-particle correlation observables in a wide range of nuclear collision systems at top RHIC and LHC energies.
This work provides a) baseline calculations and validation of the model and b) the ingredients to reproduce our calculations with the publicly available \cite{ipglasma,iSS,music,iEBE-MUSIC,urqmd} framework. 
All parameters were fixed using 200 GeV Au+Au collisions at RHIC such that all other results are parameter-free predictions of our model. 

Given this constraint, agreement of the model with experimental data on charged hadron and identified particle multiplicity distributions, mean transverse momentum, and anisotropic flow from heavy ion collisions both at RHIC and LHC is very good. Agreement with observables gets worse with decreasing system size, especially for p+p collisions, where we do not describe the experimentally observed momentum anisotropy very well. This is despite our model including both initial state anisotropy from the CGC encoded in the initial energy-momentum tensor from IP-Glasma, and anisotropy generated by fluid dynamics in response to the initial geometry. Of course non-flow contributions that we miss in the calculation and may not be fully eliminated in the experimental data, are a concern in such small systems. 

Achieving better agreement with all experimental data is very likely possible by performing a detailed parameter fit using for example Bayesian techniques \cite{Bernhard:2016tnd,Paquet:2020rxl}, which is beyond the scope of this work. Nevertheless, our results demonstrate that a sophisticated model based on color glass condensate initial conditions and the fluid dynamic evolution of the produced matter, combined with a reasonable treatment of the hadronic transport phase, can describe the whole range of heavy ion collisions at top RHIC energy and above, as well as aspects of small collision systems, with one set of transport coefficients (constant shear viscosity over entropy density $\eta/s=0.12$ and temperature dependent bulk viscosity over entropy density $\zeta/s$ with a maximum of 0.13 at $T=160\,{\rm MeV}$). 

At low temperatures, i.e., in the hadronic transport phase, the shear viscosity over entropy density $\eta/s$ increases with decreasing temperature (see e.g.  \cite{Demir:2008tr,Plumari:2012ep,Rose:2017bjz}). However, our results show that the available experimental data is consistent with a constant $\eta/s$ in the QGP phase. This is in line with previous findings \cite{Denicol:2015nhu}, where no or only a mild increase of $\eta/s$ with temperature in the QGP phase was shown to be preferred by the data.

We also provided predictions for Ru+Ru and Zr+Zr collisions at RHIC, for which data has been taken but not yet published, and O+O collisions at both RHIC and LHC energies, which will potentially be conducted in the future, and could shed more light on small systems as it is a symmetric system whose multiplicity range overlaps with that of the asymmetric small systems studied previously at the same collision energy. We also make first predictions for the upcoming 500 GeV p+p run at RHIC, and find large (compared to the other studied systems at the same multiplicity) $v_2\{2\}$ and $v_3\{2\}$.

Beyond determining QCD transport parameters, this work sets the stage for calculations of more complex observables and studies that tackle specific physics problems, such as pinning down the equation of state \cite{Gardim:2019xjs}, details of the initial state fluctuations \cite{Noronha-Hostler:2015coa,Mantysaari:2017cni}, nuclear deformation \cite{Noronha-Hostler:2019ytn,Giacalone:2020awm}, understanding details of hydrodynamic response \cite{Bhalerao:2014mua,Yan:2015jma,Mazeliauskas:2015vea,Noronha-Hostler:2015dbi}, the effect of pre-equilibrium evolution \cite{Keegan:2016cpi,Kurkela:2018vqr,Kurkela:2018wud,Heinz:2015arc,Gale:2020xlg}, Hanbury-Brown-Twiss source size measurements \cite{Wiedemann:1999qn,Lisa:2005dd,Plumberg:2015mxa}, photon and dilepton production \cite{Paquet:2015lta,Vujanovic:2016anq,Shen:2016zpp,Vujanovic:2019yih,Gale:2020xlg}, the chiral magnetic effect \cite{Kharzeev:1998kz,Voloshin:2004vk,Schenke:2019ruo}, 
and understanding the origins of momentum anisotropies in small systems \cite{Dusling:2015gta,Schlichting:2016kjw,Schenke:2019pmk}. 

To extend our studies to lower beam energies and address the regime of high baryon density and the search for the QCD critical point, it will be important to relax the assumption of boost-invariance. This can be achieved by extending the IP-Glasma initial state to include a rapidity dependence as pioneered in \cite{Schenke:2016ksl} or by using a different initial state model, such as the dynamic string picture used in \cite{Shen:2017bsr}.

\vspace{0.5cm}
\section*{Acknowledgments}
We thank Ron Belmont, Heikki M\"antysaari and Raju Venugopalan for helpful discussions. B.P.S. and P.T. are supported under DOE Contract No. DE-SC0012704. C.S. is supported under DOE Contract No. DE-SC0013460. This research used resources of the National Energy Research Scientific Computing Center, which is supported by the Office of Science of the U.S. Department of Energy under Contract No. DE-AC02-05CH11231 and resources of the high performance computing services at Wayne State University. This work is supported in part by the U.S. Department of Energy, Office of Science, Office of Nuclear Physics, within the framework of the Beam Energy Scan Theory (BEST) Topical Collaboration.

\appendix
\section{Regulation of large viscous corrections}\label{sec:regulator}
We discuss the regulation of viscous corrections that become large compared to the ideal parts of $T^{\mu\nu}$.

Viscous hydrodynamics considers the dissipative tensors $\pi^{\mu\nu}$ and $\Pi \Delta^{\mu\nu}$ as perturbative corrections to the equilibrium part of the energy-momentum tensor $T^{\mu\nu}_\mathrm{ideal} = e u^\mu u^\nu - P \Delta^{\mu\nu}$. However, the size of the viscous stress tensor can be comparable and even larger than $T^{\mu\nu}_\mathrm{ideal}$ in dilute regions or where pressure gradients are very large. Although these fluid cells are typically far outside the particlization (switching) surface and their dynamical evolution does not affect any physical observables, they may cause numerical instability problems during the evolution. To stabilize the simulations, we regulate the ill-behaved shear and bulk viscous tensors. The relative size of the viscous stress tensors compared to the ideal part of the energy momentum tensor can be computed as
\begin{equation}
    r_\pi \equiv \frac{1}{f_\mathrm{s}(e)} \sqrt{\frac{\pi^{\mu\nu} \pi_{\mu\nu}}{T^{\mu\nu}_\mathrm{ideal} T_\mathrm{ideal, \mu\nu}}}\,,
\end{equation}
and
\begin{equation}
    r_\Pi \equiv \frac{1}{f_\mathrm{s}(e)} \sqrt{\frac{3\Pi^2}{T^{\mu\nu}_\mathrm{ideal} T_\mathrm{ideal, \mu\nu}}}\,,
\end{equation}
where the equilibrium part is $T^{\mu\nu}_\mathrm{ideal} T_\mathrm{ideal, \mu\nu} = e^2 + 3P^2$. The energy density dependent regulation strength parameter is defined as,
\begin{equation}
    f_\mathrm{s}(e) = \chi_0 \left[\frac{1}{\exp\left(-\frac{e - e_0}{\xi_0}\right) + 1} -  \frac{1}{\exp \left( \frac{e_0}{\xi_0} \right) + 1} \right] \label{eq:freg}\,.
\end{equation}

We reduce the sizes of the viscous stress tensors as,
\begin{equation}
    \tilde{\pi}^{\mu\nu} = \frac{1}{r_\pi} \pi^{\mu\nu}
\end{equation}
and
\begin{equation}
    \tilde{\Pi} = \frac{1}{r_\Pi} \Pi\,,
\end{equation}
if the respective ratio $r_\pi$ or $r_\Pi$ is greater than one.
This regulation scheme is similar to the one imposed for the net baryon diffusion current in Ref.~\cite{Denicol:2018wdp}.
In this work, we choose $e_0 = 0.02$\,GeV/fm$^{3}$, $\xi_0 = 0.05$, and $\chi_0 = 1$, which stabilize almost all the event-by-event simulations with the IP-Glasma initial conditions. The parameter $e_0$ in Eq.~(\ref{eq:freg}) controls below which energy density regulation strengths stronger than $2/\chi_0$ are imposed on the viscous stress tensors. Our choice of $e_0 = 0.02$\,GeV/fm$^{3}$ is almost 10 times smaller than the switching energy density $e_\mathrm{sw} = 0.18$\,GeV/fm$^3$. The parameter $\chi_0$ controls the maximum allowed size of the viscous stress tensors compared to the size of the equilibrium energy-momentum tensor. Our choice $\chi_0=1$ allows the magnitude of the viscous stress tensors to be at most the same as their equilibrium part. This choice is stronger than usual because the pressure gradients in the IP-Glasma initial conditions are large. We tested that the final flow observables do not vary if a larger $\chi_0 = 5$ is used, although more events are numerically unstable with a larger $\chi_0$.

Recently, explicit conditions for causality in Israel-Stewart like theories of hydrodynamics have been derived \cite{Bemfica:2020xym} and could be implemented to provide constraints on the size of viscous corrections in the future.

\section{Centrality selection}\label{sec:centrality}
We base our centrality selection in this work on the midrapidity charged hadron multiplicity distribution, as is done in experiments. Previous works (e.g. \cite{Gale:2012rq}) have used the gluon distribution to determine centrality classes. We now demonstrate what this choice means for the assignment of events to different centrality classes.

\begin{figure}[tb]
  \centering
    \includegraphics[width=0.5\textwidth]{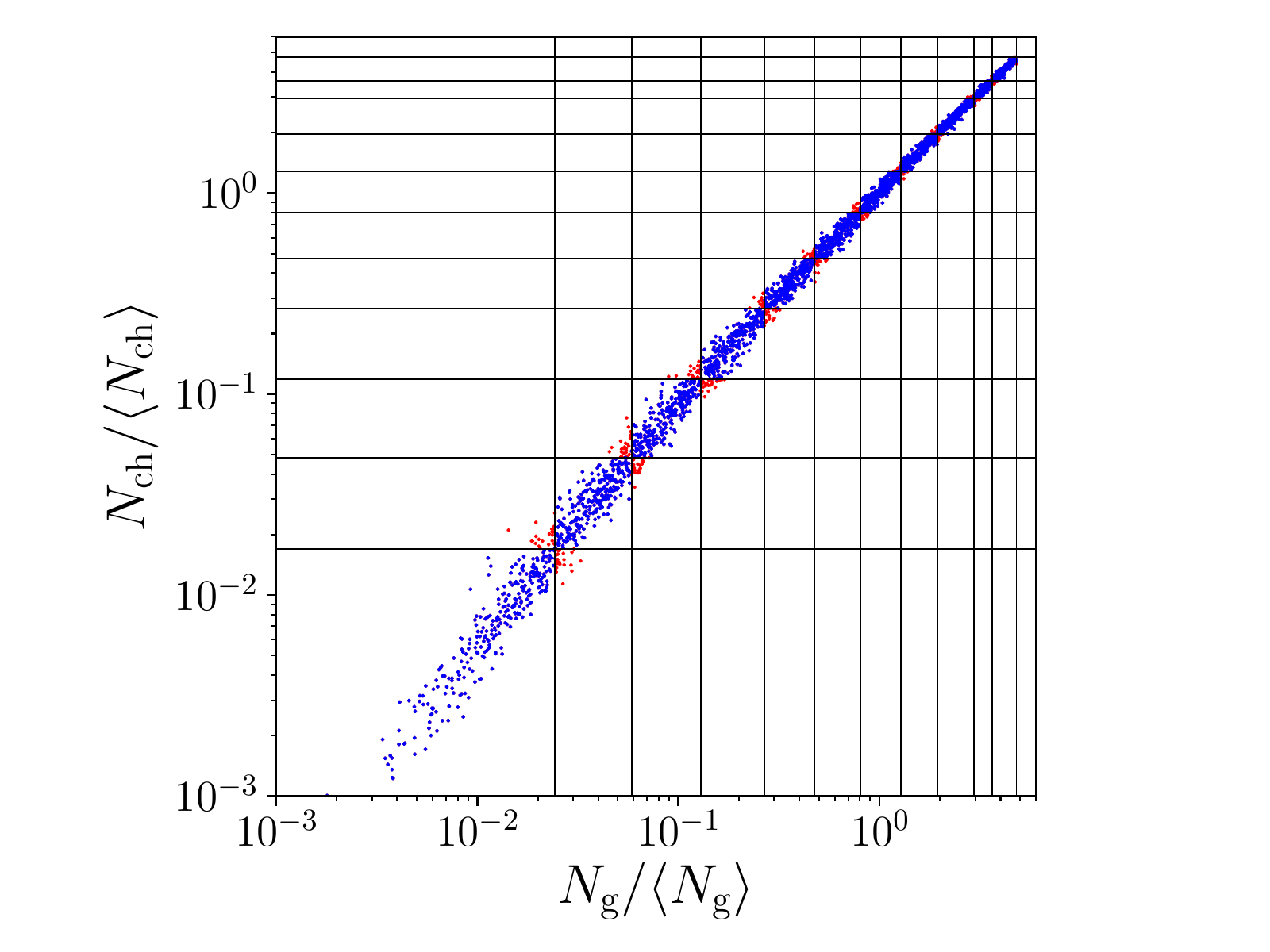}
  \caption{Correlation between the charged particle number $N_{\rm ch}$ (scaled by the mean), obtained after hydrodynamic evolution and UrQMD, and the (scaled) gluon number extracted directly from IP-Glasma. Horizontal (vertical) lines indicate edges of centrality bins defined using the $N_{\rm ch}$ ($N_{\rm g}$) distribution. The red points are those events that move to different centrality classes when one switches the method from using $N_{\rm ch}$ to $N_{\rm g}$ or vice versa. The system shown is Pb+Pb at 5.02 TeV. \label{fig:centrality}}
\end{figure}

Fig.\,\ref{fig:centrality} shows the correlation between the charged hadron number and the gluon number in a scatter plot of all events in $\sqrt{s_{\rm NN}}=5.02\,{\rm TeV}$ Pb+Pb collisions. Horizontal lines indicate edges of centrality bins defined using the $N_{\rm ch}$ distribution. Vertical lines correspond to centrality bin edges when the gluon number distribution is used to define centralities. Red points near the intersections of the lines indicate the events that change centrality class when switching the definition from using $N_{\rm ch}$ to $N_{\rm g}$ or vice versa. The narrower the correlation, the smaller the effect of using gluons instead of charged hadrons. In our case, using the gluon number is a good approximation, as long as centrality bins are not chosen too narrow (here we use 10\% bins, except for the 0-5\% and 5-10\% centralities).

The observables in heavy ion collision systems we have presented in this work are only weakly dependent on the method, and differences were within the statistical errors quoted in most cases. The largest effect was found for higher order cumulants in small systems, with even the sign of $C_n\{4\}$ depending on the centrality selection method.

\section{Flow analysis and multi-particle correlations}\label{sec:flowAnalysis}

At the end of the hybrid simulations, particles are collected to compute the $p_T$-integrated and $p_T$-differential flow vectors $\mathcal{Q}_n$ and $\mathcal{Q}_n(p_T)$, respectively. For every IP-Glasma + hydrodynamics event, we compute the complex flow vectors as
\begin{equation}
    \mathcal{Q}_n \equiv Q_n e^{i n \Psi_n}= \sum_{j} e^{i n \phi_j}\,,
\end{equation}
and
\begin{equation}
    \mathcal{Q}_n(p_T) \equiv Q_n(p_T) e^{i n \Psi_n(p_T)} = \sum_{j \in p_T \mathrm{bin}} e^{i n \phi_j}\,,
\end{equation}
Here the label $j$ runs over final state particles from all oversampled UrQMD simulations and the azimuthal angle is $\phi_j = \mathrm{arctan2}(p_j^y, p_j^x)$.  The magnitudes $Q_0$ and $Q_0(p_T)$ are the total number of particles and the numbers in different $p_T$ bins. The $\mathcal{Q}_n$ and $\mathcal{Q}_n(p_T)$ $(n \ge 1)$ are the event-by-event anisotropic flow vectors.

Using the event-by-event flow vectors, we can compute the multi-particle flow correlation as averages over an ensemble of IP-Glasma + hydrodynamics events in a given centrality bin \cite{Bilandzic:2010jr}. The 2-particle $p_T$-integrated anisotropic flow coefficients are computed as
\begin{equation}
    C_n\{2\} = \frac{{\rm Re}\{\langle \mathcal{Q}_n \mathcal{Q}_n^* - N \rangle\}}{\langle N (N - 1) \rangle},
\end{equation}
where the second term in the numerator subtracts self-correlations, $N = Q_0$ is the total number of particles, and $\mathcal{Q}_n^*$ is the complex conjugate of the $n$-th order flow vector. The scalar-product (SP) $p_T$-differential flow can be computed as
\begin{equation}
    v_n\{\mathrm{2}\}(p_T) = \frac{{\rm Re}\{\langle \mathcal{Q}^\mathrm{POI}_n(p_T) (\mathcal{Q}_n^\mathrm{ref})^* \rangle\}}{\langle Q^\mathrm{POI}_0(p_T) N^\mathrm{ref} \rangle \sqrt{C_n^\mathrm{ref}\{2\}}}.
\end{equation}
In our calculations, the $p_T$-differential flow vector of the particle of interest (POI) is defined in a pseudo-rapidity window $\vert \eta \vert < 0.5$. We choose the reference flow vector using all charged particles in $0.5 < \eta < 2$. Because these two flow vectors do not overlap with each other, there is no self-correlation.

\begin{figure}[tb]
  \centering
    \includegraphics[width=0.48\textwidth]{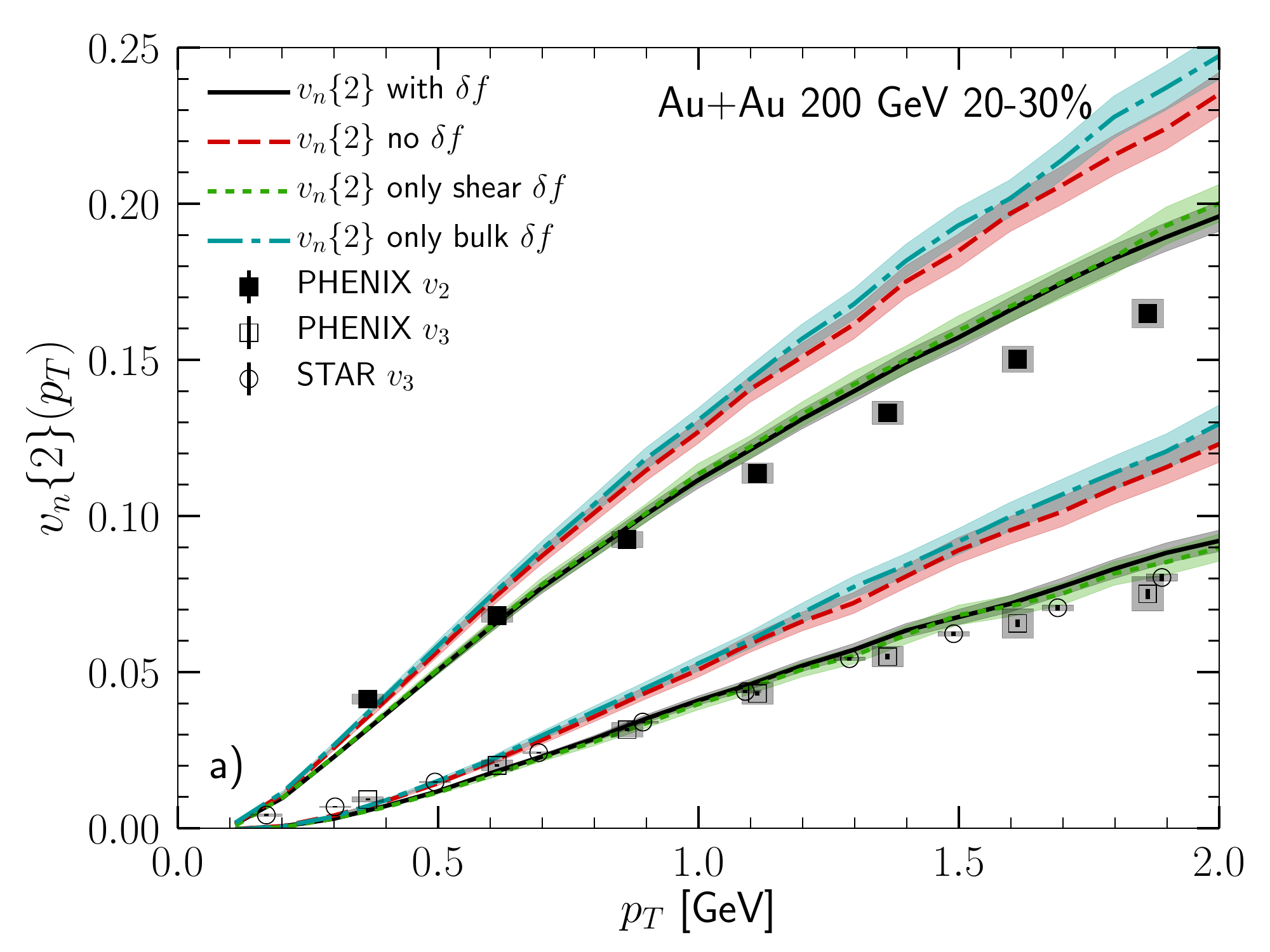}
    \includegraphics[width=0.48\textwidth]{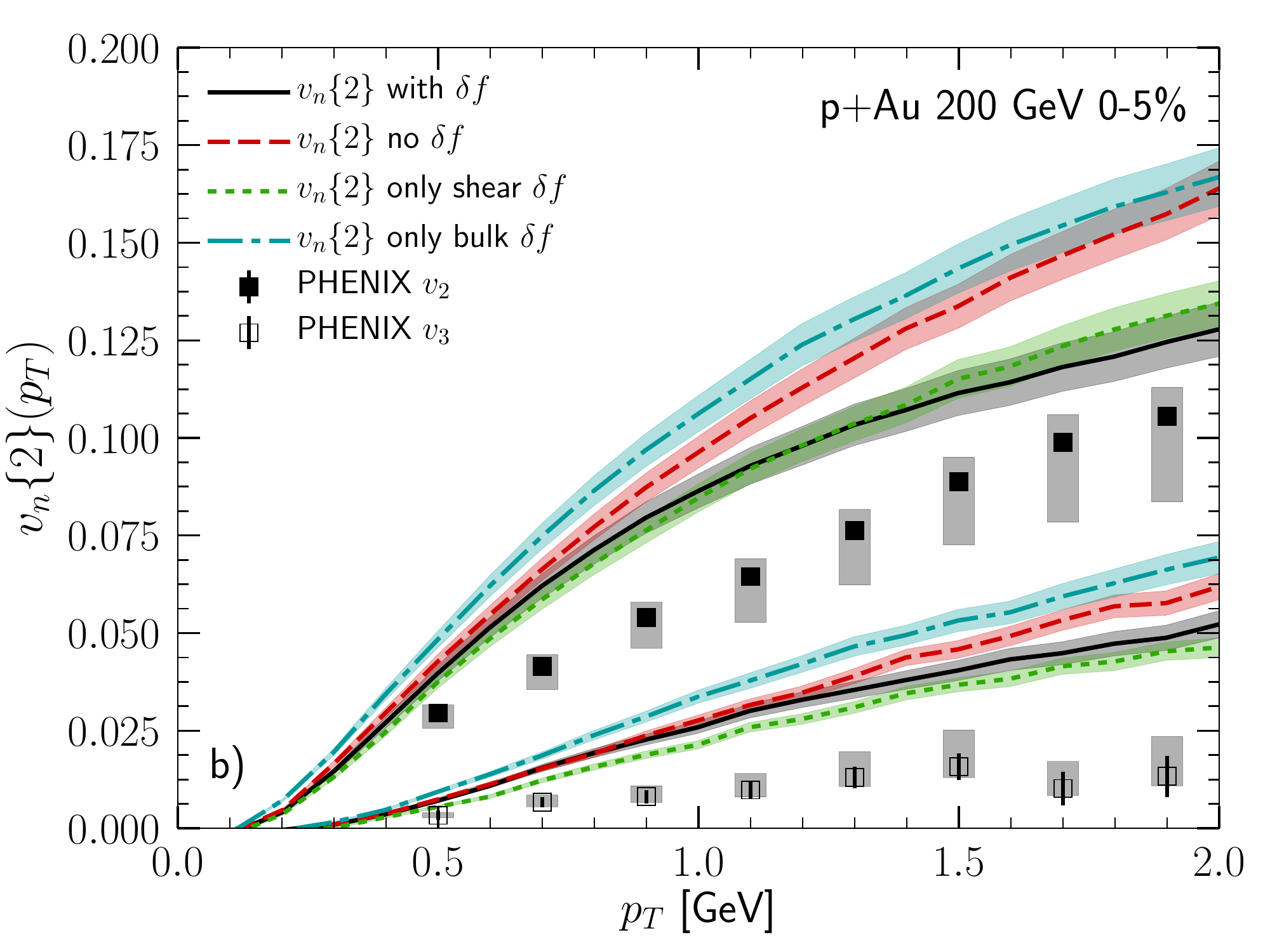}
  \caption{The effect of the off-equilibrium ("$\delta f$") corrections to the distribution function on the transverse momentum dependent charged hadron $v_n\{2\}(p_T)$ in 20-30\% 200 GeV Au+Au collisions (a) and in 0-5\% central 200 GeV p+Au collisions (b) compared to experimental data from the PHENIX \cite{Adare:2014kci,PHENIX:2018lia} and STAR Collaborations \cite{Adamczyk:2013waa}. The upper sets of curves are $v_2\{2\}(p_T)$, the lower sets $v_3\{2\}(p_T)$. \label{fig:vn2pTRHICdeltaf}}
\end{figure}

The $n$-th order 4-particle cumulants of charged hadrons are defined as
\begin{equation}
    C_n\{4\} = \frac{\langle 4 \rangle}{\langle N (N - 1) (N - 2) (N - 3) \rangle} - 2 (C_n\{2\})^2,
\end{equation}
where the 4-particle correlation with self-correlations subtracted is
\begin{eqnarray}
    \langle 4 \rangle &=& \langle (\mathcal{Q}_n \mathcal{Q}_n^*)^2 \rangle - 2 {\rm Re}\{\langle \mathcal{Q}_{2n} \mathcal{Q}_n^* \mathcal{Q}_n^* \rangle \} \nonumber \\
    && - 4 (N - 2) \langle\mathcal{Q}_n \mathcal{Q}_n^* \rangle + \langle \mathcal{Q}_{2n} \mathcal{Q}_{2n}^* \rangle \nonumber \\
    && + 2 N (N - 3).
\end{eqnarray}
If the $C_n\{4\} < 0$, we can compute a real 4-particle cumulant flow coefficient,
\begin{equation}
    v_n\{4\} = (- C_n\{4\})^{1/4}.
\end{equation}

The $p_T$-differential 4-particle flow is defined by choosing one $p_T$-differential flow vector for the POI and correlate it with three reference flow vectors,
\begin{equation}
    v_n\{4\}(p_T) = \frac{-d_n\{4\}(p_T)}{(-C^\mathrm{ref}_n\{4\})^{3/4}},
\end{equation}
where the numerator is
\begin{eqnarray}
    d_n\{4\}(p_T) &=& \frac{\langle 4 \rangle(p_T)}{\langle Q_0^\mathrm{POI}(p_T) N^\mathrm{ref} (N^\mathrm{ref} - 1) (N^\mathrm{ref} - 2) \rangle} \nonumber \\
    && - 2 \frac{{\rm Re}\{\langle \mathcal{Q}^\mathrm{POI}_n(p_T) (\mathcal{Q}^\mathrm{ref}_n)^* \rangle\}}{\langle Q_0^\mathrm{POI}(p_T) N^\mathrm{ref} \rangle} C_n^\mathrm{ref}\{2\} \\
    &=& \frac{\langle 4 \rangle(p_T)}{\langle Q_0^\mathrm{POI}(p_T) N^\mathrm{ref} (N^\mathrm{ref} - 1) (N^\mathrm{ref} - 2) \rangle} \nonumber \\
    && - 2 v_n\{\mathrm{2}\}(p_T) (C_n^\mathrm{ref}\{2\})^{3/2}.
\end{eqnarray}
Here the $p_T$-differential 4-particle correlation function is
\begin{eqnarray}
    \langle 4 \rangle(p_T) &=& {\rm Re}\{\langle \mathcal{Q}^\mathrm{POI}_n(p_T) \mathcal{Q}^\mathrm{ref}_n (\mathcal{Q}^\mathrm{ref}_n)^* (\mathcal{Q}^\mathrm{ref}_n)^* \rangle \nonumber \\
    && - 2 \langle (N^\mathrm{ref} - 1) \mathcal{Q}^\mathrm{POI}_n(p_T) (\mathcal{Q}^\mathrm{ref}_n)^* \rangle \nonumber \\
    && - \langle \mathcal{Q}^\mathrm{POI}_n(p_T) \mathcal{Q}^\mathrm{ref}_n (\mathcal{Q}^\mathrm{ref}_{2n})^* \rangle \}.
\end{eqnarray}
We subtract self-correlation between the three reference flow vectors.

The statistical errors of these multi-particle flow observables are estimated using the jackknife method. The expressions for the six particle cumulants can be found in \cite{Bilandzic:2010jr}.

\section{Effect of off-equilibrium corrections to the particle distribution functions}\label{sec:deltaf}

In this appendix we study the effect the off-equilibrium corrections to the equilibrium distribution functions on the switching surface, discussed in Section \ref{sec:sampling}, have on the transverse momentum dependent anisotropic flow coefficients of charged hadrons.

Fig.\,\ref{fig:vn2pTRHICdeltaf} shows charged hadron $v_2\{2\}(p_T)$ and $v_3\{2\}(p_T)$ in 20-30\% 200 GeV Au+Au collisions (a) and 0-5\% central 200 GeV p+Au collisions (b). We compare the full result shown previously (here using gluon number to select the centrality) to results with shear $\delta f$, bulk $\delta f$, or both set to zero. 

For both systems we can see the following features.
The effect of the bulk $\delta f$ is small by construction as discussed in Sec.\,\ref{sec:music} (the value of $\zeta/s$ is small close to the switching surface), which can be seen by comparing the result for no $\delta f$ to that with bulk $\delta f$ only, as well as the full result to that with shear $\delta f$ only. The effect of shear $\delta f$, which is to reduce the $v_n\{2\}(p_T)$, is significantly larger (except at the smallest $p_T$) and increases with $p_T$, as expected from the $p_T$ dependence of Eq.\,\eqref{eq:sheardf}.

The effect of the shear $\delta f$ in the smaller p+Pb system is only slightly greater than in the studied Au+Au system, with a reduction of $v_2\{2\}(2\,{\rm GeV})$ by approximately 25\% (compared to $\sim 22\%$ for the 20-30\% central Au+Au collisions).

One can further see, more clearly in the case of p+Pb, how the bulk $\delta f$ modifies the shape of the $v_n\{2\}(p_T)$, increasing the result mostly for momenta below approximately $1.5\,{\rm GeV}$. The detailed shape in $p_T$ depends on the choice of $\delta f$ as discussed in detail in \cite{McNelis:2019auj}, resulting in one of the largest systematic uncertainties in this type of calculation. 

\bibliography{spires}
\end{document}